\documentclass[fleqn,usenatbib]{mnras}
\usepackage{float}
\usepackage{newtxtext,newtxmath}
\providecommand{\norm}[1]{\lVert#1\rVert}

\usepackage[T1]{fontenc}
\usepackage{ae,aecompl}

\usepackage{graphicx,epstopdf}	
\usepackage{tabularx}
\usepackage{amsmath}	
\usepackage{amssymb}	
\usepackage[utf8]{inputenc} 
\usepackage{lscape}
\usepackage{threeparttable}
\usepackage{soulutf8}
\usepackage{array}
\graphicspath{ {images/} }
\usepackage[normalem]{ulem}    

\usepackage{subcaption}
\usepackage{caption}

\newcommand{\totalsignals}{25}
\newcommand{\totalstars}{22}
\newcommand{\starswithparams}{19}
\newcommand{\signalswithparams}{21}
\newcommand{\signalswithFPP}{22}
\newcommand{\fppsmallerthanone}{17}
\newcommand{\validated}{12}
\newcommand{\candidates}{12}
\newcommand{\candidatesinsystems}{6}
\newcommand{\validatedbright}{2}
\newcommand{\candidatesbright}{2}
\newcommand{\validatedhabitable}{2}
\newcommand{\candidateshabitable}{3}

\newcommand{\primemissionconfirmed}{2341}
\newcommand{\primemissioncandidates}{2418}
\newcommand{\extendedconfirmed}{409}
\newcommand{\extendedcandidates}{889}

\hypersetup{
    colorlinks=true,
    linkcolor=blue,
    filecolor=magenta,      
    urlcolor=black,
    pdftitle={Sharelatex Example},
    pdfpagemode=FullScreen,
}

\title[Planetary candidates transiting cool dwarf stars]{Planetary candidates transiting cool dwarf stars from Campaigns 12 to 15 of K2}

 \author[Castro Gonz\'alez et al.]{        
    A. Castro Gonz\'alez$^{1}$,
    E. D\'iez Alonso$^{1}$\thanks{E-mail: diezalonso@uniovi.es},
    J. Men\'endez Blanco$^{1}$,
    John H. Livingston$^{2}$,
    \newauthor
    Jerome P. de Leon$^{2}$,
    S. L. Su\'arez G\'omez$^{1,3}$,
    C. Gonz\'alez Guti\'errez$^{1}$,
    F. Garc\'ia Riesgo$^{1}$,
    \newauthor
    L. Bonavera$^{1,4}$,
    F.J. Iglesias Rodr\'iguez$^{1}$,
    R. Mu\~{n}iz$^{5}$,
    Mark E. Everett$^{6}$,
    N. J. Scott$^{7}$,
    \newauthor
    Steve B. Howell$^{7}$,
    David R. Ciardi$^{8}$,
    Erica J. Gonzales$^{9,10}$,
    Joshua E. Schlieder$^{11}$,
    \newauthor
    F. J. de Cos Juez$^{1}$ 
    \\
    $^{1}$Instituto Universitario de Ciencias y Tecnologías Espaciales de Asturias (ICTEA), C.Independencia 13, E-33004 Oviedo, Spain\\
    $^{2}$Department of Astronomy, University of Tokyo, 7-3-1 Hongo, Bunkyo-ku, Tokyo 113-0033, Japan\\
    $^{3}$Departamento de Matemáticas, Universidad de Oviedo, C. Federico García Lorca 18, E-33007, Oviedo, Spain \\
    $^{4}$Departamento de F\'isica, Universidad de Oviedo, C.Federico Garc\'ia Lorca 18, E-33007, Oviedo, Spain\\
    $^{5}$Departamento de Informática, Universidad de Oviedo, Edificio Departamental 1. Campus de Viesques s/n, E-33204, Gijón, Spain\\
    $^{6}$NOIRLab, 950 N. Cherry Ave., Tucson, AZ 85719 USA\\
    $^{7}$NASA Ames Research Center, Moffett Field, CA 94035 USA\\
    $^{8}$NASA Exoplanet Science Institute-Caltech/IPAC 1200 E. California Blvd Pasadena, CA 91125\\
    $^{9}$Department of Astronomy and Astrophysics, University of California, Santa Cruz, 1156 High St. Santa Cruz , CA 95064, USA\\
    $^{10}$National Science Foundation Graduate Research Fellow\\
    $^{11}$Exoplanets and Stellar Astrophysics Laboratory, Mail Code 667, NASA Goddard Space Flight Center, Greenbelt, MD 20771, USA\\
 }
 
\date{Accepted 2020 July 24. Received 2020 June 26; in original form 2019 December 17}

\pubyear{2020}

\begin{document}
\label{firstpage}
\pagerange{\pageref{firstpage}--\pageref{lastpage}}
\maketitle

\begin{abstract}
We analyzed the photometry of 20038 cool stars from campaigns 12, 13, 14 and 15 of the K2 mission in order to detect, characterize and validate new planetary candidates transiting low-mass stars. We present a catalogue of {\totalsignals} new periodic transit-like signals in {\totalstars} stars, of which we computed the parameters of the stellar host for {\starswithparams} stars and the planetary parameters for {\signalswithparams} signals. We acquired speckle and AO images, and also inspected archival Pan-STARRS1 images and Gaia DR2 to discard the presence of close stellar companions and to check possible transit dilutions due to nearby stars.  False positive probability (FPP) was computed for {\signalswithFPP} signals, obtaining FPP  < $1\%$ for {\fppsmallerthanone}. We consider {\validated} of them as statistically validated planets. One signal is a false positive and the remaining {\candidates}  signals are considered as planet candidates. 20 signals have orbital period P$_{\rm orb} < 10$ $d$, 2 have $10$ $d < $ P$_{\rm orb} < 20$ $d$ and 3 have P$_{\rm orb} > 20$ $d$. Regarding radii, 11 candidates and validated planets have computed radius R $<2 R_{\oplus}$, 9 have $2  R_{\oplus} <$ R $< 4  R_{\oplus}$, and 1 has R $>4 R_{\oplus}$. {\validatedbright} validated planets and {\candidatesbright} candidates are located in moderately bright stars ($\rm m_{kep}<13$) and {\validatedhabitable} validated planets and {\candidateshabitable} candidates have derived orbital radius within the habitable zone according to optimistic models. Of special interest is the validated warm super-Earth K2-323 b (EPIC 248616368 b) with T$_{\rm eq} = 318^{+24}_{-43} \, K$, S$_{\rm p} = 1.7\pm 0.2 \, S_{\oplus}$, R$_{\rm p} = 2.1\pm 0.1 \, R_{\oplus} $, located in a m$\rm_{kep}$ = 14.13 star.
\end{abstract}




\section{Introduction}

\begin{table*}
    \caption{Main features of studied campaigns.}
    \label{tab:campaigns}
    \renewcommand{\arraystretch}{1.4}
    \resizebox{18cm}{!} {
    \begin{tabular}{cllllcc}
        \hline
        Campaign & Start date & Stop date & Central coordinates & Area &  Total number of stars & Investigated stars\\
        \hline
        12 & December $15^{th}$ 2016 & March $4^{th}$ 2017 & $\alpha$=23:26:38, $\delta$=-05:06:08 &South Galactic Cap & 29362  & 4928 \\
        13 & March $8^{th}$ 2017 & May $27^{th}$ 2017 & $\alpha$=04:51:11, $\delta$=+20:57:11 &Hyades, Taurus - Auriga & 21543  & 2186 \\
        14 & May  31st 2017 & August $19^{th}$  2017 & $\alpha$=10:42:44, $\delta$=+06:51:06 &Leo and Sextant & 30044  & 6256 \\
        15 & August 23rd 2017 & November $20^{th}$ 2017 & $\alpha$=15:34:28, $\delta$=-20:04:44 &Scorpius & 23398  & 6668 \\
        \hline
    \end{tabular}}
\end{table*}

The Kepler mission \citep{2010Sci...327..977B} was launched in 2009 with the scientific goal of exploring the structure and diversity of planetary systems around main sequence stars. The satellite monitored $\sim$150000 targets in a field located near the Galactic plane which is centered on Cygnus, Lyra and Draco. The main mission finished in 2013 with the loss of two reaction wheels, preventing the necessary pointing precision.

Between 2014 and 2018 the satellite kept operating on its second mission, \textit{K2} \citep{2014PASP..126..398H}, taking advantage of solar photon pressure to compensate for the lack of the reaction wheels. This new scenario implied the observation of fields located in the ecliptic plane. 19 fields were observed for temporal windows spanning $\sim$80 days. In each campaign $\sim$20000 targets were monitored in the standard long cadence mode of 29.5 minutes. Finally, running out of fuel, the satellite was retired on October $\rm 30^{th}$ 2018.

Even though we are in the era of the Transiting Exoplanet Survey Satellite (TESS) \citep{2014SPIE.9143E..20R}, which is expected to find thousands of transiting planets in bright nearby stars, Kepler has been the most successful instrument in detecting planetary candidates. As of June 2020, {\primemissionconfirmed} confirmed and {\primemissioncandidates} candidate planets have been detected during its prime mission \citep[e.g.][]{2011ApJ...736...19B,2013ApJS..204...24B,2014ApJS..210...19B} and {\extendedconfirmed} confirmed and {\extendedcandidates} candidates during its extended mission  \citep[e.g.][]{2015ApJ...800...59V,2016ApJS..226....7C,2018AJ....156...78L,2018AJ....156..277L,2018MNRAS.476L..50D,2018MNRAS.480L...1D,2019MNRAS.489.5928D}. Despite the large amount of confirmed and candidate planets detected by the Kepler's first and second mission, a lot of planetary signals still remain undetected, waiting to be identified for future confirmation and characterization.

Our understanding of planetary systems has totally changed due to satellite discoveries, with the high occurrence rate of small planets ($R_{p} < 4 R_{\oplus}$) around low-mass stars  one of its most important results   \citep{2012ApJS..201...15H,2013ApJ...767...95D,2015ApJ...807...45D,2015ApJ...798..112M,2015ApJ...814..130M}.

Low-mass stars (0.1 $M_{\odot}$ < M < 0.6 $M{_\odot}$) account for 70$\% $ of the stellar population in the Galaxy \citep{2006AJ....132.2360H}, so they have a significant impact on the overall statistics of planets. The advantages of detection are multiple for these stars as well. First, as the dimming in brightness due to transits is proportional to the planet - stellar radius ratio $(R_{p}/R_{*})^{2}$ \citep{2002ApJ...580L.171M}, transiting planets orbiting around these stars are easier to detect. Second, low-mass stars are more suitable for radial velocity follow-up, favoring the confirmation of their planets and mass measurements. Third, the signal amplitude in transmission spectroscopy is higher \citep{2002ApJ...568..377C}, so planets orbiting around cool stars are good targets for atmospheric characterization. Also, habitable zones (HZ) around these cool stars are closer, facilitating the detection and characterization of planets in the habitable zone.
In this paper we present the results of our search of planetary candidates transiting low-mass stars from campaigns 12, 13, 14 and 15 of K2, with 23 new transit-like signals in 21 low-mass stars and 2 signals in an early K-type dwarf star, of which {\validated} were statistically validated as planets, 1 is a false positive and {\candidates} remain as candidates. In Section \ref{candidatesselection} we describe the selection process of the researched stars and the methodology to detect planetary candidate signals. In Section \ref{stellarcaract} we describe the procedure followed in order to obtain the stellar parameters of the planetary candidate hosts, and also the data acquisition and analysis of the speckle and AO images. In Section \ref{candidates_caract} we describe how we obtained the parameters of planetary candidates and the studied validation indicators. In Section \ref{res_dis} we discuss the results obtained, while the conclusions of the work are summarized in Section \ref{conclusions}.


\section{Selection of target stars and searching for planetary candidates}
\label{candidatesselection}

Our search focused on cool stars from campaigns 12, 13, 14 and 15 of K2, which were the last four public campaigns at the time of doing the analysis. Previous campaigns have been studied to inform the community about planetary candidates \citep[e.g.][]{2016MNRAS.461.3399P,2016ApJS..222...14V,2018AJ....155..222D,2018AJ....156...22Y,2019ApJS..244...11K} but, although various planets have been confirmed in campaigns 12 to 15 \citep[e.g.][]{2017Natur.542..456G,2018MNRAS.476L..50D,2018AJ....155..222D,2018MNRAS.480L...1D,2019MNRAS.489.5928D}, according to the \textit{NASA Exoplanet Archive}\footnote{\url{https://exoplanetarchive.ipac.caltech.edu/index.html}}, only 13 candidates have been published in these campaigns by \cite{2019RNAAS...3...43Z} (8 from C12 and 5 from C13. See Figure~\ref{fig:campaigns}). Table~\ref{tab:campaigns} summarizes the main features of each campaign (start and stop observing dates, central coordinates of each field, sky region, number of monitored targets by K2 and the number of selected and studied cool stars in this work).

To select cool stars we started from the \textit{Ecliptic Plane Input Catalog (EPIC)} \citep{2016ApJS..224....2H}, which provides coordinates, photometry and kinematics for all the K2 targets, based on multiple all-sky catalogues. The catalogue also provides stellar properties (temperatures, radii, masses, surface gravities, metallicities, densities, distances and extinctions) computed from colours, proper motions, spectroscopy and stellar population models. For the cross-matching process we used the online query tool available at the \textit{Mikulski Archive for Space Telescopes (MAST)}\footnote{\url{https://archive.stsci.edu/k2/epic/search.php}}.

With a precision of $\sim$2-3$\%$ in T$_{\rm eff}$, the EPIC catalogue misclassifies $56-72\%$ of subgiants as dwarfs, and 9$\%$ of dwarfs as subgiants \citep{2016ApJS..224....2H}. It also underestimates radii for M dwarfs up to 20$\%$, as it relies on Padova stellar models \citep{2008A&A...482..883M} with systematically smaller radii for M dwarfs. For these reasons, in the selection process we only imposed the condition T$_{\rm eff}$ < 4100 K to obtain a set of M and late K-type stars. Later, we obtained independent and more accurate stellar parameters than those collected in EPIC for all the stars with promising transit-like signals presented in this work (see Section~\ref{stellarcaract} and Table \ref{tab:starparams}). The selection process resulted in 6668 cool stars from campaign 15, 6256 from campaign 14, 2186 from campaign 13 and 4928 from campaign 12 (total amount of 20038 selected cool stars).

\begin{figure*}
\centering
    \includegraphics[scale = 0.47]{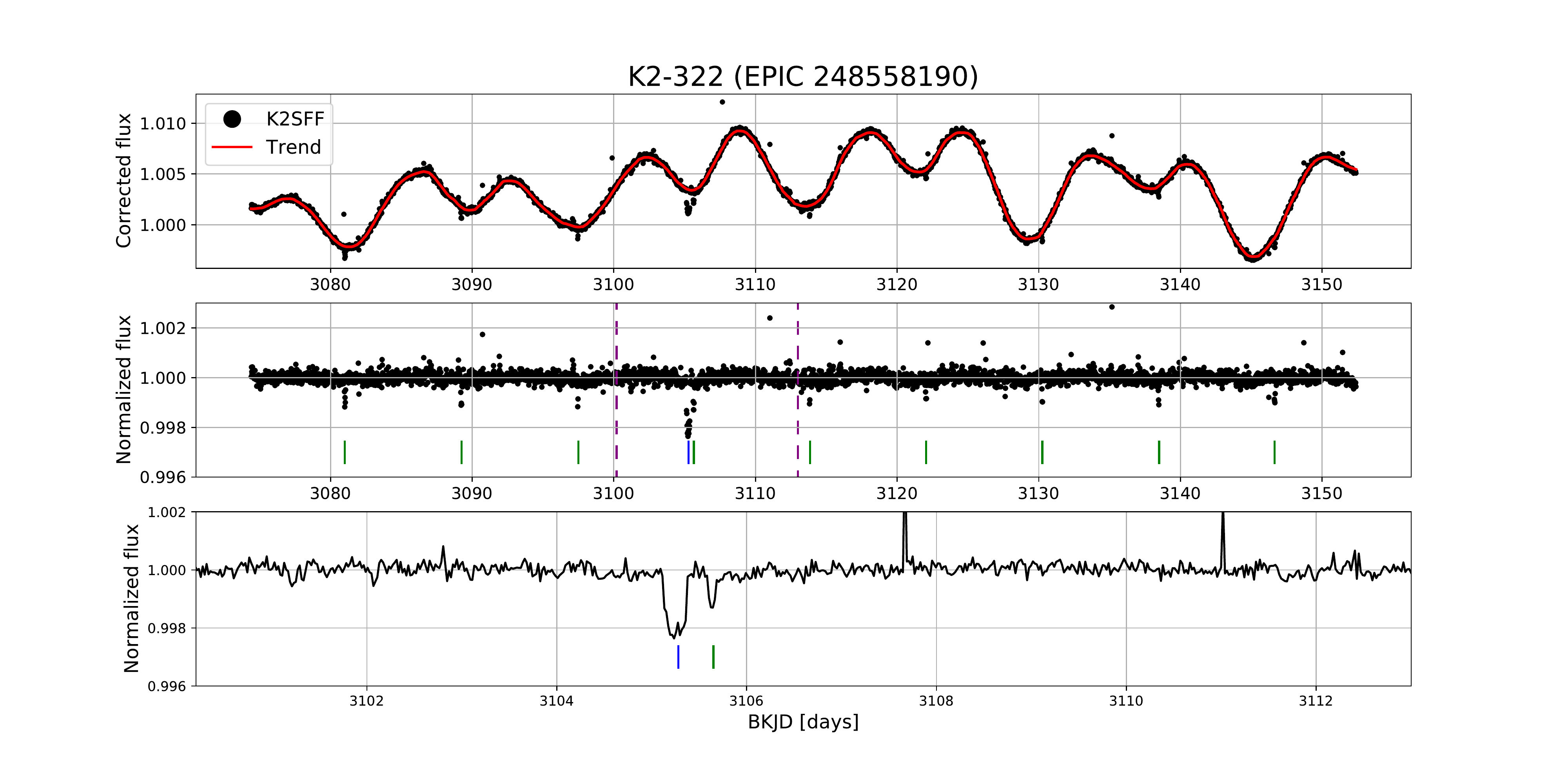}
    \caption{K2 long cadence photometry of the star K2-322. Top: K2SFF photometry after removal of telescope systematics, revealing a clear rotation modulation, along with the trend line (in red) corresponding to a moving median of 1.5-day boxcar size (72 data points). Mid: detrended light curve showing the location of the 8.2-day periodic dips of the validated planet K2-322 b (green lines). The location of a single transit event, most likely due to another planet in the system is also highlighted (blue line). Bottom: zoom in (within the vertical dashed lines) over the normalized light curve showing more clearly the single transit event along with a transit of the validated planet K2-322 b.}
    \label{fig: lc_example}
\end{figure*}


We downloaded from MAST\footnote{\url{https://archive.stsci.edu/hlsp/k2sff}} the K2 Self Flat Fielding (K2SFF) corrected light curves for all the stars of our sample. The K2SFF pipeline \citep{2014PASP..126..948V} eliminates most of the systematic photometric errors caused by the spacecraft drift during the K2 mission. The photometry of the 20038 filtered stars was analysed with the \texttt{lctools} software \citep{2019arXiv191008034S} to search for planetary transit-like signals. The software performs a moving median detrending process, after which the \textit{BLS (Box-fitting Least Squares)} algorithm \citep{2016ascl.soft07008K} is applied in order to find periodic dips that may have been caused by planetary transits. With an imposed lower value of SNR = 6 for the signal-to-noise ratio of candidate signals  \citep[computed from equation 14 of][]{2016A&C....17....1H}, and constrained to find a maximum of 3 different signals per light curve, \texttt{lctools} detected a total of 13309 signals in the four studied campaigns, resulting in a mean value of $\sim$ 0.66 signals per star. Figure \ref{fig: lc_example} shows an example of the corrected K2SFF photometry for a star from this work (K2-322, EPIC 248558190) showing clear modulations due to stellar rotation, along with the normalized light curve after removing the stellar variability. The dips due to the validated planet K2-322 b can be easily seen (highlighted in green). This target also shows the presence of a single transit event (highlighted in blue), most likely due to a long period planet (see individual discussion in Section \ref{res_dis}).

All signals were visually inspected to reject those with obvious non-transit origin. Spurious signals were those most commonly discarded. Furthermore stellar rotations, characterized by periodic modulations with periods ranging from hours to multiple days, eclipsing binaries, with clear primary and secondary minima and showing depths typically deeper than transits ($\geq1\%$), and irregular variable stars. While performing the visual inspection, we also excluded transit-like signals with less than three transits in the K2 light curve, so candidates with orbital periods longer than half the temporal spanning of the campaigns ($\sim$ 40 days) were omitted from this work. However, these candidates are being studied for a specific follow-up.

After the cleaning process, we finished with a list of 25 new promising candidate signals in 22 stars (see Section~\ref{res_dis}, and Tables~\ref{tab:starparams} and~\ref{tab:planet_params}). Furthermore, among the previous published confirmed and candidate systems belonging to the four studied campaigns (30 confirmed and 12 candidates), 8 of them (7 confirmed and 1 candidate) satisfies the selection condition  T$_{\rm eff}$ < 4100 $K$ adopted in this work. These 8 systems were also detected, showing the completeness of our signal search and the posterior vetting.


\section{Stellar characterization of planet candidate hosts}
\label{stellarcaract}
\subsection{Stellar parameters}
To obtain the stellar parameters (radii, masses, effective temperatures and surface gravities) for the stars of our final list of candidate hosts, we used the \texttt{isochrones}  package\footnote{\url{https://github.com/timothydmorton/isochrones}}, an interpolation tool for the fitting of stellar models to photometric or spectroscopic parameters \citep{2015ascl.soft03010M}. The software does trilinear interpolation in mass-age-[Fe/H] space for any given set of model grids, thus being able to predict the value for any physical or photometric property derived by the models. The package uses the MESA Isochrones Stellar Tracks (MIST) models \citep{2015ApJS..220...15P,2016ApJS..222....8D,2016ApJ...823..102C}, which are included in the \texttt{isochrones} package. These models are computed with the Modules for Experiments in Stellar Astrophysics (MESA) code \citep{2011ApJS..192....3P}. The current MIST release consists of masses ranging from 0.1 to 300 $\rm M_{\odot}$, log(age) ranging from 5 to 10.3, and solar-scaled abundances adopted from \citet{2009ARA&A..47..481A}, with metallicities ranging from [Fe/H] = -4 to 0.5.
We acquired parallaxes and their uncertainties from Gaia DR2 \citep{2018A&A...616A...1G} and the wide band \textit{BVHJKgri} photometry and their uncertainties from the EPIC catalogue \footnote{EPIC catalogue adopts \textit{gri} photometry from SDSS (Sloan Digital Sky Survey) DR9 \citep{2012ApJS..203...21A}. In cases where it is not available, we adopted \textit{gri} photometry from SDSS DR15 \citep{2019ApJS..240...23A} or Pan-STARRS DR1 \citep{2018AAS...23143601F}.} (see Table \ref{tab:stars-photo}). Comparing with the EPIC catalogue, the isochrones-derived T$_{\rm eff}$ can be estimated to within $8\%$ for the cool stars of our sample, considering the difference as:
\begin{equation}
\quad \quad \quad \quad \quad  \quad \quad \quad \quad  \frac{\norm{T_{EPIC}-T_{isochrones}}}{\norm{T_{EPIC}}}
\end{equation}
which results in a value of 0.08062 relative error. Although straightforward, the use of this error makes it possible to give a general measure of the difference of both estimations, adjusted in consideration of the EPIC values. 

The \texttt{isochrones}-derived parameters are model-dependent, so when available, they should be replaced by those obtained by asteroseismology or spectroscopy. Regarding uncertainties, we note that some of them are well below 1$\%$ in T$_{\rm eff}$ and $\sim$1$\%$ in mass and radius, which are not creditable.  We emphasize that these tabulated uncertainties only take into account the measurement error from the observations and the dispersion due to the MCMC analysis, and not the inherent uncertainties in the models themselves, so therefore they may be underestimated.

It is worth to mention that 4 stars characterized with \texttt{isochrones} have a final estimated effective temperature T$_{\rm eff}>$ 4100 K, the highest T$_{\rm eff}$  imposed to cross match the EPIC catalogue.
The final adopted stellar parameters and their uncertainties are presented in Table~\ref{tab:starparams}.

Two stars (EPIC 201663913\footnote{Posterior analysis showed that the signals present at EPIC 201663913 are a contamination from EPIC 201663879.} and K2-325) have been characterized from acquired spectra by \cite{2019AJ....158...87D}. In both cases, the published stellar parameters were adopted instead of those derived with \texttt{isochrones}.

We also checked our stellar parameters comparing with those recently published in \citet{2020ApJS..247...28H}, which provides significant improvement over EPIC catalogue due to added spectroscopic constraints. The comparison indicates a good agreement, computing a 4.7$\%$ relative error for the stellar radii, in contrast with the 43$\%$ relative error when comparing with EPIC. The radii discrepancy with EPIC is also consistent with \citet{2019AJ....158...87D}, where it was computed $\sim$40$\%$.

\subsection{Speckle and IR/AO imaging and identification of nearby contaminating stars}

We acquired speckle and AO images for 11 stars\footnote{All the images are available at ExoFOP-K2 \url{https://exofop.ipac.caltech.edu/k2/}} of our final sample to rule out the presence of close or blended sources. Not only could close sources  imply a misidentification in the origin of the signal, but also the stellar characterization with \texttt{isochrones} may be unreliable due to photometric contamination between both sources. We also inspected Pan-STARRS1 \citep{2016arXiv161205560C} images of the 22 stars to discard close or blended sources.

The speckle images were acquired  with the NESSI instrument at the 3.5 m WIYN telescope (Kitt Peak, Arizona) and DSSI at the 8 m GeminiS telescope (Cerro Pach\'on, Chile). The AO images were acquired with NIRC2 at the 10 m Keck telescope (Maunakea, Hawaii).

NESSI \citep{2018PASP..130e4502S} is a dual-channel imager using high-speed readout EMCCD detectors and a dichroic to split optical light at $\sim$700 nm. Speckle images are obtained in a shutterless stream of 1000 images per set, each image being 40 ms in duration. Depending on the target brightness, 1-3 or more sets are obtained in a row, each providing a blue and red image. All of the observations presented here used a blue filter having central wavelength/bandpass values of 562/44 nm and a 832/40 nm red filter. The data files are passed through our standard Fourier analysis pipeline \citep{2009AJ....137.5057H} in which the average power spectrum for each image is computed and summed. We next deconvolve the speckle transfer function through division by the power spectrum of a point source standard star (taken spatially and temporally close to the target star) and compute a weighted least-square fit of a fringe pattern to the result. During this step, pixels in the Fourier plane that have low signal-to-noise and low-frequency values judged to be in the seeing disk are set to zero. In order to determine the highest probability quadrant location of the companion star, we compute a reconstructed image via bispectral analysis \citep{1983ApOpt..22.4028L}. Resolved star systems produce a characteristic interferometric fringe pattern from which the separation, position angle, and delta magnitude can be determined through a modeling procedure. Details of our data reduction techniques and error assessments are given in \citet{2011AJ....141...45H} and \citet{2011AJ....142...19H}.

The Differential Speckle Survey Instrument \citep[DSSI;][]{2009AJ....137.5057H} was installed as a visitor instrument on the Gemini South 8m telescope during a run of 6 nights during 2018 March and April.  DSSI employs two high-speed EMCCD cameras to obtain simultaneous optical speckle image sets in two filters for diffraction-limited spatial resolution.  Speckle image reconstruction is used to detect or constrain the presence of any additional sources within $\simeq1\arcsec$ of a target star. Each target star was observed with filters having central wavelength/bandpass values of 692/40~nm and 880/50~nm.  The star was centered in a $256\times256$~pixel ($2.8\arcsec\times2.8\arcsec$) window of each detector.  The data consisted of sets of images, where each set is a data cube of 1000 frames with frame exposure times of 60~ms.  Along with each science target, a nearby, bright ($V\approx5$) calibrator star was observed with one 1000-frame data set.  These calibration stars were chosen to be single stars, in order to represent the instrumental PSF of the speckle data (their point source nature is confirmed during data reduction).

The data were reduced as described in \citet{2011AJ....142...19H} to produce a set of high-level data products.  These include, for each filter, a reconstructed (diffraction-limited) image of the field surrounding each target star and a contrast curve giving the relative limiting magnitude of any undetected point source as a function of angular separation from the target. Briefly, to construct the images, the autocorrelation function of each frame is calculated, then co-added over the image set.  These are transformed into power spectra, and the power spectrum of each science target is divided by that of its point source to find the modulus of the power spectrum of the science target.  A set of bispectral subplanes are also found for each image set and used according to the algorithm of \citet{1990JOSAA...7.1243M} to provide phase information that, when combined with the power spectrum modulus, correct the phase of the image.  The result is filtered with a Gaussian having a FWHM corresponding to the diffraction limit of the telescope and this is then inverse-transformed to produce the reconstructed image.

The contrast curves are measured from the reconstructed images by measuring, within a set of concentric annuli centered on the target star, the standard deviation, $\sigma$, of local minima and maxima in the image background.  These local minima and maxima are effectively the noise threshold against which the signal (i.e. peak) of any neighboring star must be detected.  The contrast curve is defined based on a smoothed fit over the $5\sigma$ values from each annulus with magnitude differences measured relative to the peak of the target star.

The Keck Observatory observations were made with the NIRC2 instrument on Keck-II behind the natural guide star AO system.  The observations were made on 2018~Nov~22 UT in the standard 3-point dither pattern that is used with NIRC2 to avoid the left lower quadrant of the detector which is typically noisier than the other three quadrants. The dither pattern step size was $3\arcsec$ and was repeated twice, with each dither offset from the previous dither by $0.5\arcsec$.  The observations were made in the narrow-band $Br-\gamma$ filter $(\lambda_o = 2.1686; \Delta\lambda = 0.0326 \,\, \mu$m) with an integration time of 10 seconds with one coadd per frame for a total of 90 seconds on target.  The camera was in the narrow-angle mode with a full field of view of $\sim10\arcsec$ and a pixel scale of approximately $0.099442\arcsec$ per pixel.

The AO data were processed and analyzed with a custom set of IDL tools.  The science frames were flat-fielded and sky-subtracted. The flat fields were generated from a median average of dark subtracted flats taken on-sky. The flats were normalized such that the median value of the flats is unity.  The sky frames were generated from the median average of the dithered science frames; each science image was then sky-subtracted and flat-fielded. The reduced science frames were combined into a single combined image using a intra-pixel interpolation that conserves flux, shifts the individual dithered frames by the appropriate fractional pixels, and median-coadds the frames. The final resolution of the combined dither was determined from the full-width half-maximum of the point spread function; 0.059\arcsec.

The sensitivities of the final combined AO image were determined by injecting simulated sources azimuthally around the primary target every $45^\circ $ at separations of integer multiples of the central source's FWHM \citep{2017AJ....153...71F}. The brightness of each injected source was scaled until standard aperture photometry detected it with $5\sigma $ significance. The resulting brightness of the injected sources relative to the target set the contrast limits at that injection location. The final $5\sigma $ limit at each separation was determined from the average of all of the determined limits at that separation and the uncertainty on the limit was set by the rms dispersion of the azimuthal slices at a given radial distance.

The speckle and AO images rule out the presence of close companions except for EPIC 250001426 and EPIC 246163416, which have a secondary source situated at 0.23\arcsec and 0.65\arcsec respectively. Also, analysis of the Pan-STARRS1 images for the stars without speckle or AO images rule out the presence of close companions with $\Delta$mag < 5 at angular separation > 2.5\arcsec, with the exception of EPIC 245944983 (a binary star with angular separation of 4\arcsec \, between components), EPIC 246331418 (also a binary star with the secondary component at angular separation of 3\arcsec) and EPIC 249391469, with a faint source with angular separation of 5.85\arcsec (see individual discussion in Section \ref{res_dis} for more details). 

In Table \ref{FPP} we summarize the {\totalsignals} signals belonging to the {\totalstars} stars, for which the limits on the presence of nearby stars have been obtained by speckle, AO or archival Pan-STARRS1 images. All the speckle images, with their extracted contrast curves, and the images from Pan-STARRS1, are presented in Appendix~\ref{Appendix} (Figures \ref{fig:AO_1} and \ref{fig:AO_2}, Figures \ref{fig:pans_1} to \ref{ap:pans_last} respectively) and discussed in Sections~\ref{candidates_caract} and~\ref{res_dis}.

Five stars showing transit-like signals have radial velocity measurements from Gaia DR2 (summarized in Table \ref{RV}). From the radial velocities, distances,  proper motions and coordinates of these five stars, we checked their membership to any moving group from \cite{2013ApJ...762...88M}, using the code from \cite{david_rodriguez_2016_192159}, finding no matches.

\begin{table}
\caption{Proper motions and radial velocities for the five stars of our sample with $V_{r}$ available from Gaia DR2. Distances from Gaia DR2 for the entire sample are listed in Table \ref{tab:starparams}.}
\begin{tabular}{cccc}
\hline
EPIC & $\mu_{\alpha} [mas/yr]$ & $\mu_{\delta} [mas/yr]$ & $V_{r}[km/s]$ \\ \hline
201663879 & $-2.85 \pm 0.07$ & $-22.73 \pm 0.05$ & $8.63 \pm 1.83$ \\
245944983 & $109.47 \pm 0.05$ & $25.50 \pm 0.03$ & $25.38 \pm 1.70$ \\
248480671 & $50.92 \pm 0.07$ & $-109.19 \pm 0.06$ & $-43.89 \pm 0.89$ \\
248558190 & $-16.30 \pm 0.04$ & $-2.64\pm 0.03$ & $4.18\pm 1.42$ \\
248782482 & $107.70 \pm 0.07$ & $-38.22 \pm 0.06$ & $-4.21 \pm 1.26$ \\ \hline
\end{tabular}
\label{RV}
\end{table}


\section{Candidates characterization and statistical validation}
\label{candidates_caract}

We used the \texttt{pyaneti} package\footnote{\url{https://github.com/oscaribv/pyaneti}} \citep{2019MNRAS.482.1017B} to characterize the planetary candidates. The code uses a Bayesian approach combined with an MCMC sampling to infer the candidate parameters. From the phase-folded light curve of the transit, the orbital period, the computed stellar mass, radius and  effective temperature as fixed parameters, and the orbital radius, planet radius and the quadratic limb darkening coefficients as uniform priors, the package fits the \cite{2002ApJ...580L.171M} transit model, considering the 29.5 minutes long cadence of K2 following \cite{2010MNRAS.408.1758K}.  As photometry data alone does not properly constrain the eccentricity, we assumed $e=0$ in our fit, a reasonable assumption due to tidal circularization of orbits for close-in planets \citep{2011MNRAS.415.2349R}.

The \cite{2002ApJ...580L.171M} model considers a star limb darkening parametrized by a quadratic law with two coefficients, which were left as free parameters (with a uniform prior between 0 and 1) in the fitting process. For three candidates of our sample (EPIC 246909566 b, in the coolest star with $\rm T_{eff}$ = 3057 $K$, EPIC 248775938 b, in the hottest low-mass star with $\rm T_{eff}$ = 4663 $K$, and EPIC 248782482 b, in an intermediate $\rm T_{eff}$ star with $\rm T_{eff}$ = 3852 $K$), we also computed the planetary parameters fixing the quadratic limb-darkening coefficients to those tabulated in \citet{2018A&A...618A..20C}. The planetary parameters obtained are equal (within the uncertainties) than those obtained with the coefficients as free parameters in the \texttt{pyaneti} fit, so the coefficients were left as free parameters.

Appendix~\ref{Appendix} (Figures \ref{fig:phase_folded1}, \ref{fig:phase_folded2} and \ref{fig:phase_folded3}) shows the phase-folded transits and the best fit provided by \texttt{pyaneti}. The main derived parameters for our candidates are presented in Table~\ref{tab:planet_params}. 

To obtain the false positive probability (FPP) of our candidate list we used the \texttt{vespa} package\footnote{\url{https://github.com/timothydmorton/VESPA}} \citep{2015ascl.soft03010M}, which implements the procedure described in \cite{2012ApJ...761....6M}. The code computes the probabilities of planetary and non-planetary scenarios from  very  limited  follow-up  observations and Galaxy model simulations of stellar population. Using the \texttt{TRILEGAL} Galaxy model \citep{2005A&A...436..895G}, \texttt{vespa} considers different false positive scenarios caused by eclipsing binaries (EBs), background EBs (BEBs) and hierarchical triple systems (HEBs). \texttt{vespa} computes the stellar parameters from broadband photometry and parallaxes using \texttt{isochrones} and compares a large amount of simulated situations to the observed phase-folded light curve. Although \texttt{vespa} considers the most frequent false positive scenarios involving
eclipsing binaries, the package does not take into account other possible false positive scenarios as instrumental artifacts in the data or other astrophysical phenomenons, such as star spots.

The package was run from the following inputs: coordinates, parallaxes from Gaia DR2, wide band BVHJKgri photometry, orbital period, transit phase-folded light curve, ratio $R_{p}/R_{*}$ derived from \texttt{pyaneti}, maximum allowed depth of potential secondary eclipse (fixed with a value of $10^{-4}$) and limits on the presence of nearby background stars derived from speckle and Pan-STARRS1 images. 

The resulting FPPs for the candidates presented in this work and the multiplicity-corrected $\rm FPP_{2}$ for those candidates belonging to a multi-planetary system  are summarized in Table \ref{FPP} and discussed in Section \ref{res_dis}.

\begin{figure*}
\includegraphics[scale = 0.6]{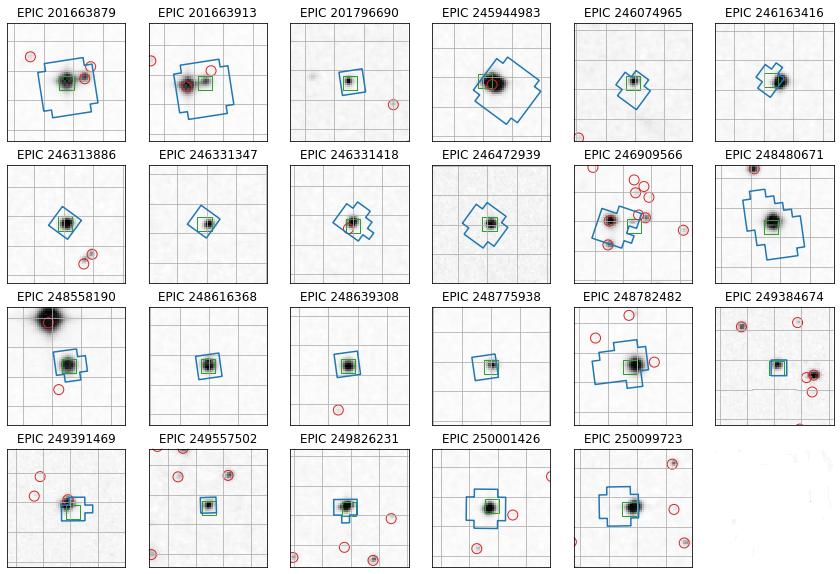}
\caption{1'x 1' cut-out images from the DSS2 survey showing the target (green square) on the center and nearby Gaia sources (red circles). The photometric aperture (blue polygon) from the K2SFF pipeline is superposed. North is up and East is left.}
\label{fig:DSS2_aperture}
\end{figure*}

FPP computed by \texttt{vespa} could not be reliable if faint stars within the best aperture could be the sources of the signals. Also the full width at half-maximum (FWHM) of the point spread function (PSF) varies for Kepler between 3.1\arcsec and 7.5\arcsec \footnote{According to Kepler documentation: \url{https://keplergo.arc.nasa.gov/DataAnalysisProducts.shtml}}, so targets outside the best aperture (but close to its edge) could also result in contamination. To discard this scenarios, we checked exhaustively all the targets, searching for nearby sources in Pan-STARRS1 and DSS2 images, Gaia DR2 and EPIC. The DSS2 images for the {\totalstars} stars are shown in Figure \ref{fig:DSS2_aperture} along with the K2SFF apertures and Gaia DR2 nearby sources\footnote{Figure \ref{fig:DSS2_aperture} has been made using Astropy \citep{exoplanet:astropy13} and Astroquery \citep{2019AJ....157...98G}.} (see the individual discussion in Section \ref{res_dis}).

The relationship between the observed ($\delta$') and true ($\delta$) transit depth considering dilution from a secondary star within the best aperture and $\rm \Delta m$ magnitudes fainter than the brighter star is:

\begin{equation}
\rm \quad \quad  \quad \quad \quad \quad \quad \quad \quad \delta' =\frac{\delta}{\gamma}= \frac{\delta}{1+10^{\,0.4 \, \Delta m}}
\end{equation}

Assuming a conservative value of $\delta$ = 1 (100 $\%$ dip in the faint secondary star), a value of $\delta$' > $\gamma^{-1}$ could not be originated in the secondary star, discarding the secondary star as the source of the signal. In all the cases with secondary stars within the best aperture or outside but at a distance $\lesssim$ 6\arcsec of its edge, we applied the criteria. If the secondary star is discarded as the origin of the signals (and the \texttt{vespa} FPP is < $1\%$) the signal is validated. Otherwise the signal is considered as a candidate.

For each target, we also visually inspected the phase-folded light curves created using small and large apertures from the K2SFF photometric pipeline. In some cases, a nearby star a few pixels away from the target can be substantially excluded from the photometry using a small enough aperture. We find no apparent difference between the transit depths using a large and a small aperture except for EPIC 249391469 (see individual discussion in Section \ref{res_dis} for details).

In addition, for signals with FPP < 1$\%$,  we only consider as validated planets those who fulfill the above criteria and also satisfies the condition SNR > 10, so signals with 6 < SNR < 10 are considered as candidates.


\begin{table}
    \centering
    \begin{tabular}{ccccc}
        \hline \hline
        ID & C & FPP & $\rm FPP_{2}$ & Notes \\
        \hline 
        EPIC 249384674 b &15& 1.3 $\times$ $10^{-3}$ &   7.2 $\times$ $10^{-5}$ & (2) \\
        EPIC 249384674 c &15& 4.2 $\times$ $10^{-4}$ &  2.3 $\times$ $10^{-5}$ & (2) \\
        \textcolor{red}{EPIC 249391469 b} &15& 2.0 $\times$ $10^{-3}$      & --        & (4, 5, 6) \\
        EPIC 249557502 b &15&  7.1 $\times$ $10^{-5}$ & --  & (2) \\
        EPIC 249826231 b &15&  2.2 $\times$ $10^{-3}$         & --       & (2) \\
        \textcolor{red}{EPIC 250001426 b} &15& --  & --             & (2, 5) \\
        \textcolor{red}{EPIC 250099723 b} &15& 5.6 $\times$ $10^{-1}$                  & --  & (2) \\
        EPIC 201663879 b &14& 1.2 $\times$ $10^{-3}$    &   3.0 $\times$ $10^{-5}$          & (1) \\  
        \textcolor{red}{EPIC 201663879 c} &14& 5.0 $\times$ $10^{-1}$              &  2.4 $\times$ $10^{-2}$ &(1) \\
        EPIC 201796690 b  &14& 1.2 $\times$ $10^{-3}$      & --          & (4) \\
        EPIC 248480671 b &14& 2.9 $\times$ $10^{-5}$ & --   & (3) \\
        EPIC 248558190 b&14& 7.7 $\times$ $10^{-5}$ & --   & (1) \\
        EPIC 248616368 b &14& 1.1 $\times$ $10^{-5}$ & -- & (1) \\
        EPIC 248639308 b &14& 3.9 $\times$ $10^{-6}$& --  & (1) \\
        \textcolor{red}{EPIC 248775938 b} &14& 2.0 $\times$ $10^{-1}$               & --  & (4) \\
        \textcolor{red}{EPIC 248782482 b}&14& 1.3 $\times$ $10^{-4}$ & --   & (4, 5) \\
        \textcolor{red}{EPIC 246909566 b} &13& 5.2 $\times$ $10^{-2}$                & --   & (4) \\
        
        \textcolor{red}{EPIC 245944983 b} &12& -- & --& (4) \\
    
        EPIC 246074965 b &12& 6.3 $\times$ $10^{-5}$  & --   & (4) \\
        \textcolor{red}{EPIC 246163416 b} &12& -- & --   & (1) \\
        \textcolor{red}{EPIC 246313886 b} &12& 6.0 $\times$ $10^{-3}$            & --      & (4,7) \\
        \textcolor{red}{EPIC 246331347 b} &12& 2.0 $\times$ $10^{-2}$               & --   & (4) \\
        \textcolor{red}{EPIC 246331418 b} &12& 9.5 $\times$ $10^{-4}$        & 3.4 $\times$ $10^{-5}$           & (4, 5) \\
        \textcolor{red}{EPIC 246331418 c}&12& 2.2 $\times$ $10^{-3}$        &  7.9 $\times$ $10^{-5}$           & (4, 5) \\
        EPIC 246472939 b &12& 1.0 $\times$ $10^{-3}$       & --           & (4) \\
        \hline
    \end{tabular}
    \caption{False positive probabilities (FPP) from \texttt{vespa} and multiplicity-corrected probabilities ($\rm FPP_{2}$) for all our candidates and validated planets except for EPIC 245944983 b (binary star), EPIC 250001426 b and EPIC 246163416 b (both with close stars detected in speckle images). Signals without computed FPP, with FPP $>1\%$ or with FPP $<1\%$ but not validated due to nearby companions within (or close to the edge of) the best aperture or due to not fulfilling the condition SNR > 10 are highlighted in red. (1): with limits on the presence of nearby background stars from speckle images from NESSI (WIYN telescope). (2): from speckle images from DSSI (GeminiS telescope). (3): from AO images at 2168 nm from NIRC2 (Keck telescope). (4): from Pan-STARRS1 images. (5): stars with FPP $<1\%$ but not discarded contaminant sources detected within (or outside, but close to the edge of) the best aperture. (6): False positive. The signal comes from a nearby contaminant star. (7): Candidate status due to not fulfilling the imposed condition SNR > 10.}
    \label{FPP}
  
\end{table}


\section{Results \& Discussion}
\label{res_dis}

In Tables \ref{tab:stars-photo} and \ref{tab:starparams} we present respectively the photometry and \texttt{isochrones}-computed stellar parameters for the candidate host stars. Analysis of the candidate  signals (see below) reveals that in fact one of them, detected in EPIC 201663913, is originated in the close early K-type star EPIC 201663879.

In Table \ref{tab:planet_params} we present the computed candidate parameters for {\signalswithparams} of the {\totalsignals} signals\footnote{We have only tabulated the period, epoch and transit depth of the signal from the binary star EPIC 245944983, from EPIC 250001426 and  EPIC 246163416, both with close stars detected in speckle images, and also from EPIC 249391469, whose signal comes from Gaia DR2 6239523158128298496 (see individual discussion).} detected in these {\totalstars} stars. {\candidatesinsystems}  candidates belong to three multi-planetary candidate systems (EPIC 49384674, EPIC 201663879 and EPIC 246331418), with two candidates in each star.

We performed the false positive analysis with \texttt{vespa} for {\signalswithFPP} signals. Due to photometric contamination, we do not computed FPP on the signal from the binary star EPIC 245944983, from EPIC 250001426 and from EPIC 246163416 (both with close stars detected in speckle images). The analysis concludes that {\fppsmallerthanone} candidates have FPP < $1\%$, which is the common adopted threshold to statistically validate a planet \citep{2014ApJ...784...45R,2015ApJ...809...25M,2016ApJ...822...86M,2019A&A...627A..66H}, but considering the nearby contaminant stars within the photometric aperture, 3 of these {\fppsmallerthanone} candidates with  FPP < $1\%$ are considered as candidates instead of validated planets; 1 more signal is also considered a candidate for not satisfying the condition SNR > 10 and 1 signal is a false positive (see individual discussions in Section \ref{res_dis}). 5 signals have  $1\%$ < FPP < $56\%$. Considering the false positive threshold as FPP > $90\%$ \citep{2015ApJ...809...25M}, these 5 signals are also considered as planetary candidates, along with the remaining 3 signals from stars with close companions detected in speckle and Pan-STARRS1 images. We end up with {\validated} validated planets, {\candidates} candidates and 1 false positive.

For the candidates in multi-planetary systems presented in this work we recalculated the false positive probabilities according to the low probability of multiple false positive signals \citep{2011ApJS..197....8L}. When two planet candidates are detected within the same aperture, these candidates have much lower FPPs than single candidates. Following \citet{2012ApJ...750..112L}, for each candidate belonging to a multi-planetary system we computed the multiplicity-corrected false positive probability as $\rm FPP_{2}$ = 1-$\rm P_{2}$, where
\begin{equation}
\rm \quad \quad  \quad \quad \quad \quad \quad \quad \quad P_{2} \approx \frac{X_{2} P_{1}}{X_{2} P_{1}+(1-P_{1})}
\end{equation}
$\rm X_{2}$ is the \textit{multiplicity boost} for systems with two planet candidates and $\rm P_{1}$ = 1-FPP, being FPP the false positive probability computed with \texttt{vespa} without considering the candidate multiplicity. The \textit{multiplicity boost} can be estimated as the ratio between the fraction of analysed targets with planet candidates and the fraction of planet candidate hosts with more than one candidate. \citet{2012ApJ...750..112L} estimated $ \rm X_{2} \sim$  25 for the prime Kepler mission but \citet{2016ApJ...827...78S} argues that $ \rm X_{2}$ can not be assumed to be the same as that for K2, given the different Galactic backgrounds and pointing characteristics of the observations. The campaigns studied in this work have too few candidates to estimate $\rm X_{2}$ properly (e.g. C14 and C15 have 0 published candidates), so we assumed the $\rm X_{2}$ estimated for the adjoining well-studied campaigns C1, C2, C3 and C4. Using the catalog of \citet{2019ApJS..244...11K}, we computed $\rm X_{2} \sim$ 28 for C3 (adjoining C12),  $\rm X_{2} \sim$ 26 for C4 (adjoining C13), $\rm X_{2} \sim$ 40 for C1 (adjoining C14) and $\rm X_{2} \sim$ 18 for C2 (adjoining C15). The FPPs for the  validated planets and candidates (highlighted in red) are summarized in Table \ref{FPP}. For the planet candidates ($\rm 1\%<FPP<90\%$), the main possible false positive scenarios are individually discussed below.

It is worth to mention that follow-up observations have revealed a false positive origin for several K2 validated planets. \citet{2017ApJ...847L..18S} identified three K2 validated Jupiter-sized planets as confirmed low-mass stellar secondaries. To explain the initial misclasification the authors state different possible causes as the indistinguishable radius distribution of the smaller stars and gas giants planets, the difficulty in detecting a secondary eclipse in eccentric orbits and a poor characterization of the host star. Another source of misidentification can be the presence of an unnoticed background eclipsing binary within the photometric Kepler aperture, as we discuss below for our targets with background stars in the Pan-STARRS1 images and Gaia DR2. \citet{2017A&A...606A..75C} identified three K2 validated super-Earth-sized planets as background eclipsing binaries acquiring ground-based high-resolution images in which the binaries were left out of the photometric aperture. Although planet validation techniques are useful tools to get a quick approach to the goodness of planet candidates, there exist the possibility of misclasifications, so that detailed follow-up is necessary to confirm them.

\begin{figure}
    \includegraphics[width=9cm]{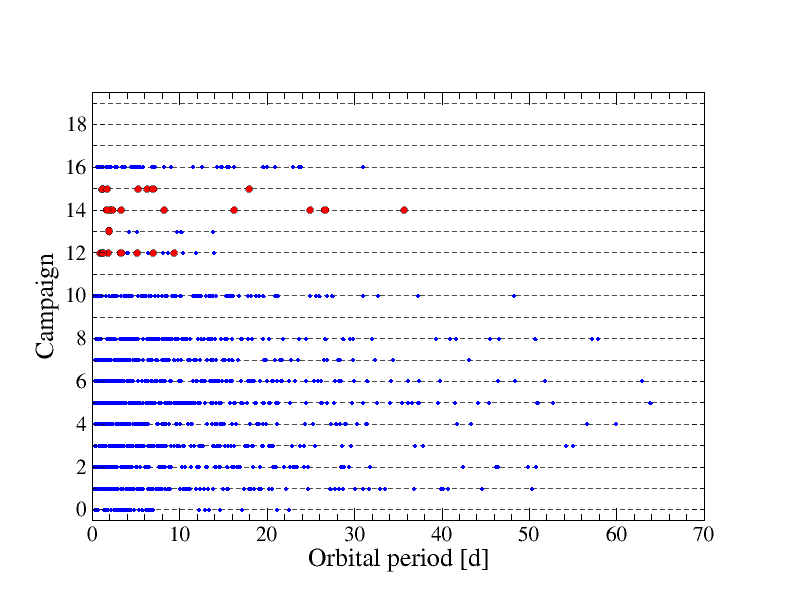}
    \centering
    \caption{Orbital period distribution by Campaign of published K2 candidates and confirmed planets (blue) and the new candidates and validated planets presented in this work (red).}
    \label{fig:campaigns}
\end{figure}

In Figure \ref{fig:campaigns} we plot the orbital period distribution of the new validated planets and candidates presented in this work and the previous published candidates and confirmed planets for each K2 campaign \footnote{In all figures presented in this section we plot the data available at the NASA Exoplanet Archive (\url{https://exoplanetarchive.ipac.caltech.edu/}) for the previous published candidates and confirmed planets in all K2 campaigns.}. As mentioned in Section \ref{candidatesselection}, only 8 candidates have been published in C12 and 5 in C13 by \cite{2019RNAAS...3...43Z}, while C14 and C15 do not have published candidates. 

\begin{figure}
    \centering
    \includegraphics[width=9cm]{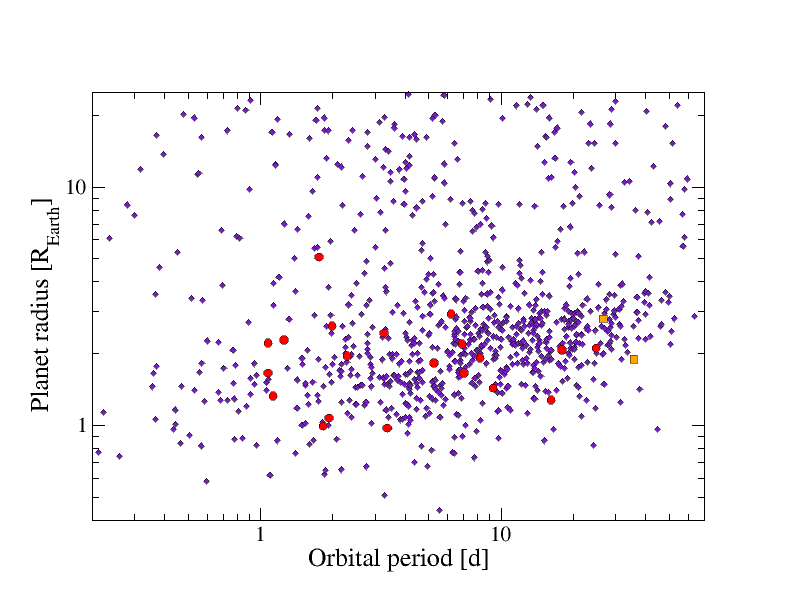}
    \caption{Planet radius versus orbital period for published K2 candidates and confirmed planets (violet) with R$_{p}$ < 25$R_{\earth}$ and candidates and validated planets from this work (red). Orange  squares represent the validated planet b and candidate c in the early K-type star EPIC 201663879 (T$_{\rm eq}$ = 5440 K), also from this work.}
    \label{fig:RpvsPorb}
\end{figure}

Figure \ref{fig:RpvsPorb} plots the planet radii versus orbital periods for all the K2 previously published candidates and confirmed planets with $R<25R_{\earth}$, and the new validated planets and candidates presented in this work. 20 of them have orbital period P$_{\rm orb} < 10$ $d$, 2 have 10 $d$ < P$_{\rm orb}$ < 20 $d$ and 3 have P$_{\rm orb} >$ 20 $d$. Regarding radii, 11 candidates and validated planets have R $<2 R_{\oplus}$, 9 have $2 R_{\oplus} < $R$ < 4 R_{\oplus}$, and 1 has R $>4 R_{\oplus}$.

Figures \ref{fig:campaigns} and \ref{fig:RpvsPorb} show that the number of candidates and planets decreases as the orbital period increases, especially after 40 days. Note that in this work we are only considering signals with at least three transits in the K2 light curve, which constrains periods to last less than 40 days. Figure \ref{fig:RpvsPorb} also shows a greater abundance of candidates with small radii. These trends are related to the short temporal span of the K2 observing windows and the intrinsic distribution of planets (66$\%$ have P$_{\rm orb} < 10 \, d$ and 50$\%$ have R$ < R_{\oplus}$).

\begin{figure}
    \centering
    \includegraphics[width=9cm]{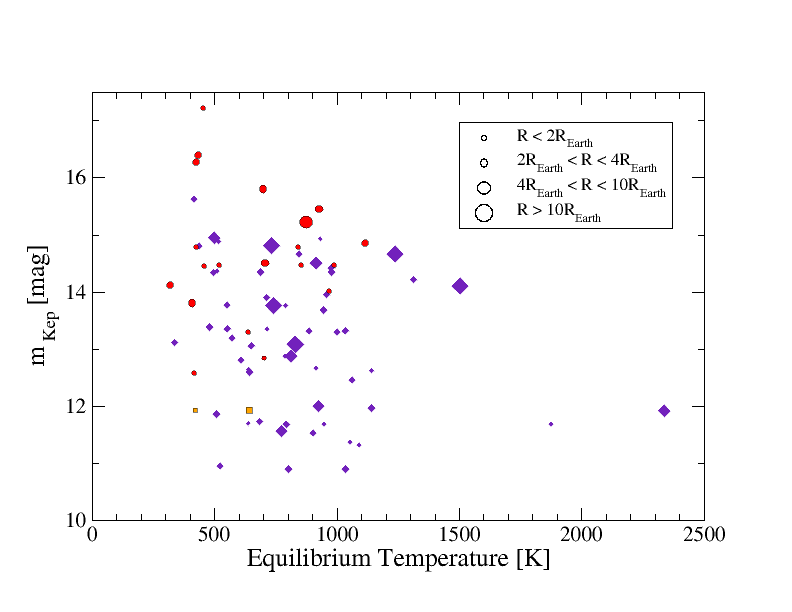}
    \caption{Kepler magnitude m$\rm_{Kep}$ of candidate hosts stars versus T$\rm_{eq}$ of candidate planets for published K2 (violet) and this work (red) candidates. Orange squares represent the validated planet b and candidate c in the early K-type star EPIC 201663879 (T$\rm_{eq}$=5440 K), also from this work.}
    \label{fig:magKep_vs_Teq}
\end{figure}

Figure \ref{fig:magKep_vs_Teq} shows the Kepler magnitude of the hosts stars versus the equilibrium temperature for all the confirmed planets, published K2 candidates and our validated and candidate planets. Two validated planets and two candidates are located in moderately bright stars ($k_{p}$ < 13). EPIC 201663879 ($k_{p}$ = 11.93) has two super-Earths (b, validated, T$_{\rm eq}$ = 641 K, R$_{\rm p}$ = 2.8 $R_{\oplus}$, and c, candidate, T$_{\rm eq}$ = 420 K, R$_{\rm p}$ = 1.9 $R_{\oplus}$). EPIC 248480671 ($k_{p}$ = 12.84) hosts a validated super-Earth (T$_{\rm eq}$ = 701 K, R$_{\rm p}$ = 2.0 $R_{\oplus}$), and EPIC 248782482 ($k_{p}$ = 12.59) with a super-Earth candidate (T$_{\rm eq}$ = 417 K, R$_{\rm p}$ = 1.3 $R_{\oplus}$). These are of special interest as they are well-suited for future radial velocity follow-up and atmospheric characterization.

\begin{figure}
    \centering
    \includegraphics[width=9cm]{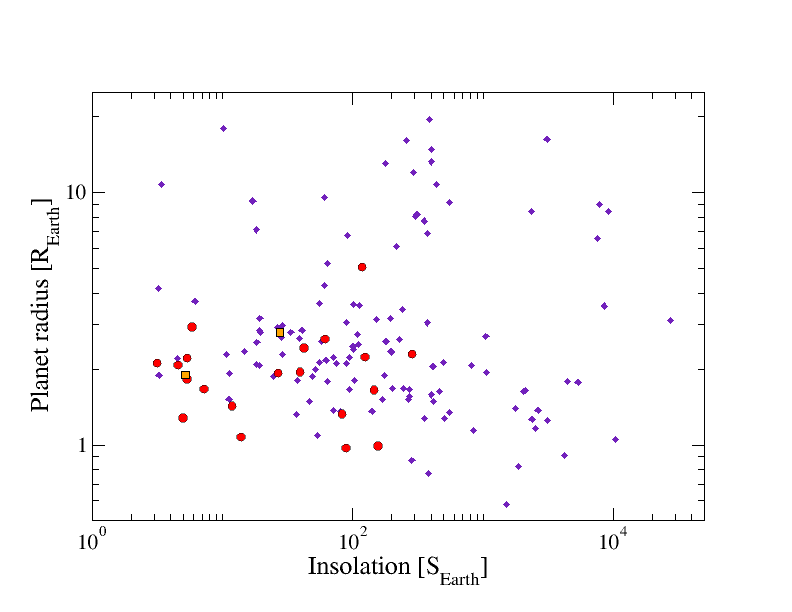}
    \caption{Planet radius versus insolation for published K2 candidates and confirmed planets (violet) with R$_{p}$ < 25$R_{\earth}$ and validated planets and candidates from this work (red). Orange squares represent the validated planet b and candidate c in the early K-type star EPIC 201663879 (T$_{eq}$ = 5440 K), also from this work.}
    \label{fig:RpvsS}
\end{figure}

Most of our validated planets and candidates receive large amounts of insolation (see Figure \ref{fig:RpvsS}), computing S$_{\rm p}$ > 5 $S_{\oplus}$ for 18 of them. The habitable zone (HZ) is defined as the range of planet-star distances in which a planet could keep liquid water on its surface. As the presence of an atmosphere has a great impact on the surface temperature of a planet, several climate models in which different atmospheric constrains are considered provide different boundaries for the habitable zone.

\citet{1993Icar..101..108K} and \citet{2013ApJ...765..131K} HZ boundaries exclude all our validated and candidate planets from the habitable zone. However, \citet{2013ApJ...778..109Z} state that the inner edge of the HZ for hot desert worlds, with low relative humidity and high albedos could be much closer to the star that previously thought. In particular, considering the optimistic proposed constrains of planets with 1$\%$ relative humidity and a terrestrial-like albedo (A = 0.8), 2 validated planets and 3 candidates presented in this work would be within the Zsom's habitable zone, as detailed in the individual discussions (see Table \ref{insolacion}).

\begin{table}
    \centering
    \begin{tabular}{ccccc}
        \hline
        Candidate  &C& Name&  $\rm d_{in}$ [$AU$] & a [$AU$] \\
        \hline
        EPIC 250099723 b &15& --&  0.1208 & 0.1417 \\
        EPIC 201663879 c &14& -- & 0.3038 & 0.3501 \\
        EPIC 248616368 b &14& K2-323 b & 0.0922 & 0.1275 \\
        EPIC 248782482 b &14& --& 0.0910 & 0.0999 \\
        EPIC 246074965 b &12& K2-325 b & 0.0411 & 0.0419 \\

        \hline
    \end{tabular}
    \caption{Validated planets and candidates within the habitable zone according to the optimistic Zsom model, considering 1$\%$ relative humidity and a terrestrial-like albedo (A = 0.8). The fourth column shows the distance to the host star  of the inner edge of the habitable zone according to the Zsom model. The fifth column presents the computed orbital radius.}
    \label{insolacion}
\end{table}

\begin{figure}
    \centering
    \includegraphics[width=9cm]{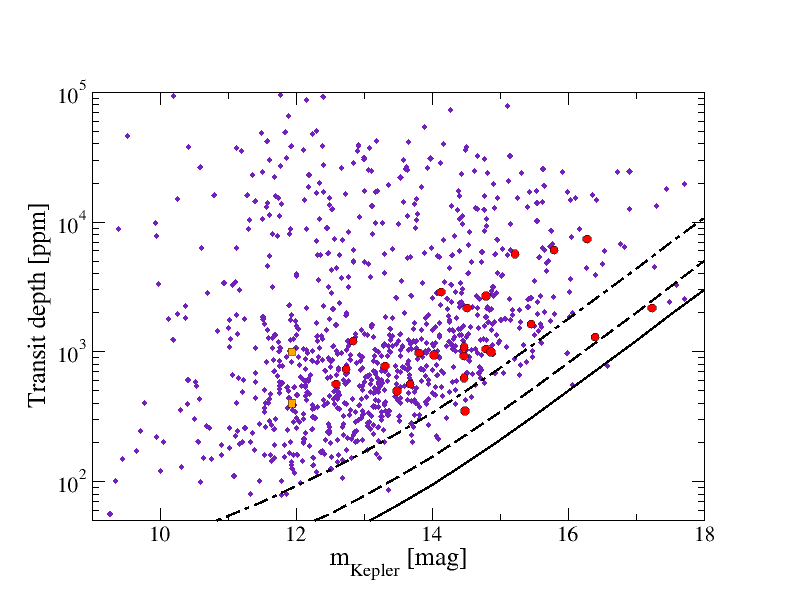}
    \caption{Transit depth versus m$\rm_{Kep}$ of candidate hosts stars for published K2 candidates and confirmed planets (violet) with transit depths < $10 \%$ and validated planets and candidates from this work (red). Orange squares represent the validated planet b and candidate c in the early K-type star EPIC 201663879 (T$_{eq}$ = 5440 K), also from this work. The lines show the limits in transit depth detection for transiting planets with orbital period 1.3 days (solid line), 3.7 days (dashed line) and 17 days (dash-dotted line). }
    \label{fig:Depth_vs_magkep}
\end{figure}

Figure \ref{fig:Depth_vs_magkep} plots the transit depth in logarithmic scale versus stellar magnitude in the Kepler band. Our validated planets and candidates follow the same upward trend that previously K2 published candidates and confirmed planets. This trend is related with the minimum signal-to-noise ratio SNR = 6 imposed in our transit search. In the figure we plot the transit depth detection limits for the typical orbital periods of the validated and candidate planets of our search (1.3, 3.7 and 17 days) with the selected SNR.

Of special interest  are the warm super-Earth K2-323 b (validated, T$_{\rm eq} = 318^{+24}_{-43} \, K$, S$_{\rm p} = 1.7\pm 0.2 \, S_{\oplus}$, R$_{\rm p} = 2.1\pm 0.1 \, R_{\oplus}$), located in a m$\rm_{kep}$ = 14.13 star, and the super-Earth EPIC 248782482 b (candidate, T$\rm_{eq} = 417^{+58}_{-24} \, K$, S$_{\rm p} = 5.0 \pm 1.0 \, S_{\oplus}$, R$_{\rm p} = 1.3  \pm 0.1 \, R_{\oplus}$) located in a moderately bright star (m$\rm_{kep}$ = 12.59), thus being a good target for radial velocity follow-up and also for atmospheric characterization.

Next we will discuss the main features of each system, such as their statistical validation results, their physical parameters, and other individual aspects like habitability.


\subsection{K2-316 (EPIC 249384674)}

This target (T$_{\rm eff}$ = 3436 K, R = 0.38 $R_\odot$, $k_{p}$ = 14.79) was observed in campaign 15 and shows two signals with periods 1.133 and 5.260 days. The speckle images (acquired with GeminiS) and contrast curves exclude companions with $\delta_{mag}<4$ at 0.2\arcsec. Also, analysis of Pan-STARRS1 archival images and Gaia DR2 discard the presence of possible nearby contaminant stars. In the analysis with \texttt{vespa} we obtained independent false positive probabilities $\rm FPP=1.3 \times 10^{-3}$ for candidate b and $\rm FPP=4.2 \times 10^{-4}$ for candidate c. Both candidates separately are statistically validated. Anyway, considering that they are part of a multi-planetary system, we computed multiplicity-corrected false positive probabilities of $\rm FPP_{2}=7.2 \times 10^{-5}$ for candidate b and $\rm FPP_{2}=2.3 \times 10^{-5}$ for candidate c.

The non-detection of blended objects in speckle images, analysis of Pan-STARRS1 images and Gaia DR2, the FPPs  $<1\%$  from \texttt{vespa} and the low probability of multiple false positive signals makes this target a validated planetary system with two super-Earths (R$ _{\rm b}$ = 1.3 $R_\oplus$ and R$_{\rm c}$ = 1.8 $R_\oplus$).\\

\subsection{EPIC 249391469}
\begin{figure*}
    \includegraphics[width=15cm]{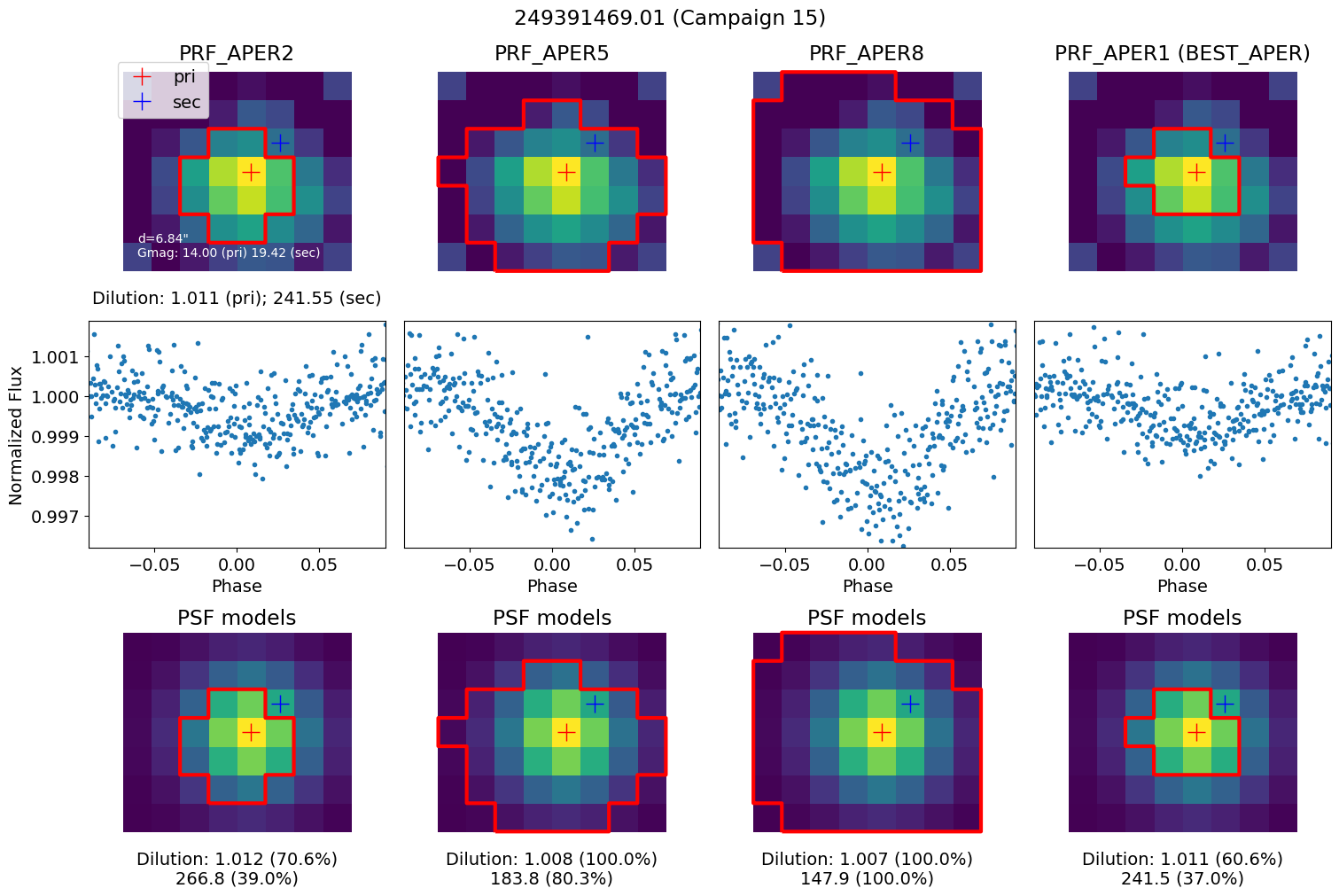}
    \centering
    \caption{Phase-folded light curves created using small and large apertures from the K2SFF photometric pipeline for EPIC 249391469. The red crosses corresponds to the primary star and the blue crosses to the faint nearby star. The central plots shows larger apertures (PRF\_APER5 and PRF\_APER8), which include the majority of the flux from the faint nearby star. For these apertures, the relative flux dimming is two times greater than in a situation in which the secondary source is left out of the aperture (PRF\_APER2 and PRF\_APER1). This result indicates that the faint nearby star (Gaia DR2 6239523158128298496) is the actual source of the signal. The transit depth implies a 14$\%$ dip, so the eclipsing binary scenario is the most likely, making this signal a false positive.}
    \label{fig:EPIC249391469}
\end{figure*}

This target (T$_{\rm eff}$ = 3503 K, R = 0.57 $R_\odot$, $k_{p}$ = 14.03, $mag\,  G=14.00$) was observed in campaign 15, showing a signal with a period of 1.075 days. Pan-STARRS1 image excludes stars with $\delta_{mag}<4.5$ at 2.5 arc seconds and shows a faint source ($mag\,  G=19.42$) with an angular separation of 5.85\arcsec and position angle PA = 172$^{\circ}$, which is within the optimal aperture of the K2SFF (see Figure \ref{fig:DSS2_aperture}). As discussed in Section \ref{candidates_caract}, the analysis of the phase-folded light curves created using small and large apertures from the K2SFF photometric pipeline shows a ~2x increase in the transit depth when using a large aperture that includes the majority of the flux from the faint nearby star (see Figure \ref{fig:EPIC249391469}). This indicates that the faint nearby star (Gaia DR2 6239523158128298496) is the actual source of the signal. The $9.5 \times 10^{-4}$ transit depth implies a 14$\%$ dip, so the eclipsing binary scenario is the most likely. As a result, this signal is considered as a false positive.

\subsection{K2-317 (EPIC 249557502)}

This target (T$_{\rm eff}$ = 3387 K, R = 0.38 $R_\odot$, $k_{p}$ = 16.40) was observed in campaign 15 and shows a transit-like signal with a period of 6.220 days. The speckle image (acquired with GeminiS) excludes companions with $\delta_{mag}<3$ at 0.1\arcsec. The analysis of close stellar sources in Pan-STARRS1 images and Gaia DR2 discard photometric contamination. The \texttt{vespa} analysis results in $\rm FPP=7.1 \times 10^{-5}$, and it is thus a validated mini-Neptune with R = 2.9 $R_\oplus$.

The optimistic Zsom model places the inner edge of the habitable zone in this system at $\rm d_{in}$ = 0.0546 $AU$  from the star (considering albedo = 0.8). Taking into account a derived semi-major axis a = 0.0542 $AU$, this candidate might be orbiting close to the inner edge of the habitable zone.

\subsection{K2-318 (EPIC 249826231)}

This target (T$_{\rm eff}$ = 3851 $K$, R = 0.55 $R_\odot$, $k_{p}$ = 14.47) was observed in campaign 15 and exhibits a signal with a period of 7.010 days. The speckle image (acquired with GeminiS) excludes companions with $\delta_{mag}<4$ at 0.2 \arcsec. Also the analysis of close stellar sources in Pan-STARRS-1 images and Gaia DR2 discard contamination. In the analysis with \texttt{vespa} we obtained $\rm FPP=2.2 \times 10^{-3}$, thus it is a validated planet. Our fit results in a super-Earth with R = 1.7 $R_\oplus$.\\

\subsection{EPIC 250001426}

This target was observed in campaign 15 and shows a signal with a period of 1.705 days. The speckle image (acquired with GeminiS) detected a secondary source at 0.23 arc seconds and $\Delta m \sim$ 0.4. Following the criteria discussed in Section \ref{candidates_caract}, this candidate is not validated. As the photometry of both stars is mixed, we can not derive stellar nor planetary parameters.

\subsection{EPIC 250099723}

This target (T$_{\rm eff}$ = 4176 K, R = 0.58 $R_\odot$, $k_{p}$ = 13.80) shows 5 transit events observed in campaign 15 with a period of 17.920 days. The speckle image (acquired with GeminiS) exclude companions  with $\delta_{mag}<4$ at 0.1\arcsec. The Pan-STARRS1 and Gaia DR2 analysis of close stellar sources discard contamination. However, the analysis with \texttt{vespa} results in $\rm FPP=5.6 \times 10^{-1}$, so it remains a low-quality candidate.The most likely false positive scenario is the BEB, as its contribution to the total computed FPP is $\rm FPP_{BEB}=5.5 \times 10^{-1}$. Additional high precision photometric follow-up is needed to improve the signal, for which the best fit is considerably flat-bottomed with the 5 transits observed by K2. 

The signal is modelled by a candidate super-Earth with R = 2.1 $R_\oplus$.The inner edge of the habitable zone in the optimistic Zsom  model is located at  $d_{in}$ = 0.1208 AU (adopting A = 0.8 albedo). Our estimated orbital radius for this candidate is a = 0.1417 AU, so it would be in or traversing the habitable zone of the star.\\

\subsection{K2-319 (EPIC 201663879) / EPIC 201663913}

In campaign 14, 3 transit events with a period of 26.718 days were detected in EPIC 201663913 (T$_{\rm eff}$ = 3874 K, R = 0.68 $R_\odot$, $k_{p}$ = 14.45). The image from Pan-STARRS1 shows a brighter star (K2-319, $k_{p}$ = 11.93) at an angular distance of 10\arcsec and PA = 102.4$^{\circ}$. Stellar parameters for K2-319 derived from \texttt{isochrones} are T$_{\rm eff}$ = 5440 K, R = 0.90 $R_\odot$, log $g$ = 4.5. The 26.718-day signal is also present at K2-319.

Suspecting contamination, we searched for transit signals in EPIC 201663913 with smaller apertures in its K2SFF corrected light curve, finding that the signal disappears and confirming that it comes from K2-319. Posterior light curve analysis of K2-319 showed, in addition to the initial signal, another signal with a period of 35.621 days and three transits, and also two more transit-like features that suggest the presence of more candidates.

The speckle image of K2-319 acquired at the 3.5 m WIYN telescope rules out the presence of blended targets with $\delta_{mag}<4$ at 0.5\arcsec. Also, analysis of close stars with Pan-STARRS1 images and Gaia DR2 discard possible contaminant sources (except EPIC 201663913, but as discussed, this M dwarf is not the source of the signals). We performed a statistical validation of this signal with \texttt{vespa}, obtaining $\rm FPP=1.2 \times 10^{-3}$ for candidate b, and $\rm FPP=5.0 \times 10^{-1}$ for candidate c. Considering that the candidates are in a multi-planetary system, the  multiplicity-corrected false positive probabilities are  $\rm FPP_{2}=3.0 \times 10^{-5}$ for candidate b and $\rm FPP_{2}=2.4 \times 10^{-2}$ for candidate c. Candidate b is considered to be statistically validated. Since the signal c originates from a multi-signal system, the most plausible scenario is a planetary origin. Even so, the analysis with \texttt{vespa} yields a FPP $>1\%$ so it remains in the candidate status, being the EB the main possible false positive scenario.

Assuming the stellar parameters indicated above for K2-319, the validated planet b is modeled as a mini-Neptune with R = 2.8 $R_\oplus$ and the candidate c as a super-Earth with R = 1.9 $R_\oplus$.

The optimistic Zsom model for habitable zones indicates for this system a inner edge situated at  $d_{in}$ = 0.3038 AU from the star (considering albedo = 0.8). Taking into account a derived semi-major axis a = 0.3501 AU for the candidate c, this might be in the habitable zone. 

With $k_{p}$ = 11.93, this system is a good target for radial velocity follow-up, in order to characterize the masses and detect the presence of more planets, and for their atmospheric characterization.

\subsection{K2-320 (EPIC 201796690)}
This target (T$_{\rm eff}$ = 3157 K, R = 0.30 $R_\odot$, $k_{p}$ = 15.80) was observed in campaign 14 and exhibits a transit-like signal with a period of 1.995 days. The Pan-STARRS1 image shows no blended sources, excluding stars with $\delta_{mag}<5.1$ at 2.5\arcsec. Also, analysis of Pan-STARRS1 images and Gaia DR2 discard possible nearby contaminant sources. Our fit points to a super-Earth candidate with R = 2.6 $R_\oplus$. The \texttt{vespa} analysis computed $\rm FPP=1.2 \times 10^{-3}$, and it is thus a statistically validated planet. 


\subsection{K2-321 (EPIC 248480671)}
This target (T$_{\rm eff}$ = 3855 K, R = 0.58 $R_\odot$, $k_{p}$ = 12.84) was observed in campaign 14 and shows a transit-like signal with a period of 2.298 days. The AO image acquired at Keck discard the presence of close stars. In particular, the observations showed no additional stellar companions to within a resolution $\sim 0.06\arcsec$ FWHM and out to 8\arcsec. Also, inspection of Pan-STARRS1 images and Gaia DR2 discard possible nearby contaminant sources. The analysis performed with \texttt{vespa} results in $\rm FPP=2.9 \times 10^{-5}$, making this a statistically validated planet, whose fit corresponds to a 2.0 $R_\oplus$ super-Earth.

\subsection{K2-322 (EPIC 248558190)}
This star (T$_{\rm eff}$ = 4141 K, R = 0.60 $R_\odot$, $k_{p}$ = 13.31) was observed in campaign 14 and shows a signal with an 8.205-day period. The speckle image (acquired at 3.5 m WIYN telescope) exclude blended objects with $\delta_{mag}<4$ at 0.2\arcsec and $\delta_{mag}<5$ at 0.5\arcsec. Analysis of Pan-STARRS1 images and Gaia DR2 shows a faint source (mag G = 20.22) at 12.6\arcsec, outside the optimal aperture (of radius $\sim$ 8\arcsec), but very close to the edge. Assuming depth = 0.8 ppt, we modelled the flux distribution, resulting in a partial flux of $\sim$ 12 $\%$ within the aperture, ruling out this faint star as the source of the signal. The analysis with \texttt{vespa} results in $\rm FPP=7.7 \times 10^{-5}$, thus making it a validated planet. The fit results in a super-Earth with 1.9 $R_\oplus$. Also a single transit event is detected in epoch 3105.28 BKJD showing a $\sim$ 2 ppt transit depth. A specific photometric or RV follow-up is necessary to confirm this signal, which could be associated with a $\sim$3$R_\oplus$ mini-Neptune (see Figure \ref{fig: lc_example}).

\subsection{K2-323 (EPIC 248616368)}
This star (T$_{\rm eff}$ = 3710 K, R = 0.55 $R_\odot$ \& $k_{p}$ = 14.13) was observed in campaign 14 and shows 3 transit events with a period of 24.930 days. The speckle image (acquired at the 3.5 m WIYN telescope) excludes close companions with $\delta_{mag}<3$ at 0.2\arcsec and $\delta_{mag}<4$ at 0.5\arcsec. The analysis of nearby stars with Pan-STARRS1 images and Gaia DR2 discard contaminant sources. The \texttt{vespa} analysis results in $\rm FPP=1.1 \times 10^{-5}$, thus making it a validated planet whose fit models a 2.1 $R_\oplus$ super-Earth.

This validated planet is moderately warm, with $\rm T_{eq}=318^{+24}_{-43} \, \,K$ and $\rm S_{p}=1.7 \pm 0.2 \, \, S_{\oplus}$.
The optimistic Zsom model for habitable zones indicates for this system an inner edge situated at  $d_{in}$ = 0.0922 AU from the star (considering albedo = 0.8). Taking into account a derived semi-major axis a = 0.1275 AU, this super-Earth might be in the habitable zone. 

\subsection{K2-324 (EPIC 248639308)}
This target (T$_{\rm eff}$ = 3752 K, R = 0.51 $R_\odot$, $k_{p}$ = 14.51) was observed in campaign 14 and exhibits a signal with a period of 3.262 days. The speckle image (acquired at the 3.5 m WIYN telescope) discards blended stars with $\delta_{mag}<3.5$ at 0.2\arcsec and $\delta_{mag}<5$ at 0.5\arcsec. The analysis of Pan-STARRS1 images and Gaia DR2 discard nearby contaminant stars. \texttt{vespa} statistical analysis concluded with $\rm FPP=3.9 \times 10^{-6}$, thus making it a validated planet whose fit corresponds to a super-Earth with 2.4 $R_\oplus$.

\subsection{EPIC 248775938}

This target (T$_{\rm eff}$ = 4663 K, R = 0.61 $R_\odot$, $k_{p}$ = 15.22) was observed in campaign 14 and
according to its \texttt{isochrones} fit, it has the highest temperature of the characterized stars in this work, with the exception of the early K-type star EPIC 201663879. It shows a transit-like signal with a period of 1.743 days. No close companions or blended stars are present in the images from Pan-STARRS1, excluding stars with $\delta_{mag}<4.7$ at 2.5\arcsec. Also, analysis of pan-STARRS1 images and Gaia DR2 shows no nearby sources, discarding contamination. Our fit points to a 5.1 $R_\oplus$ Neptune-like candidate, being the candidate with the largest radius presented in this work. The phase folded light curve is quite v-shaped, with the \texttt{vespa} analysis resulting in a computed $\rm FPP=2.0 \times 10^{-1}$. The most likely false positive scenario is the EB, as its contribution to the total computed FPP is $\rm FPP_{EB}=1.9 \times 10^{-1}$. 

\subsection{EPIC 248782482}
This star (T$_{\rm eff}$ = 3852 K, R = 0.50 $R_\odot$, $k_{p}$ = 12.59) was observed in campaign 14. It is the brightest of all the low-mass candidate hosts in our list (with the exception of the early K-type EPIC 201663879, $k_{p}$ = 11.93) and shows 5 transit events with a period of 16.222 days. Pan-STARRS1 image shows no evidence of close or blended sources, excluding stars with $\delta_{mag}<4.3$ at 2.5\arcsec. However, in Pan-STARRS1 images and Gaia DR2 we find a mag G = 19.77 star within the optimal aperture which we can't rule out as the source of the signal, so although \texttt{vespa} analysis computed $\rm FPP=1.3 \times 10^{-4}$, we can not validate this candidate. 

Our fit points to a 1.3 $R_\oplus$ super-Earth with $\rm T_{eq}=417^{+58}_{-24} \, \,K$ and $\rm S_{p}=5.0 \pm 0.1 \, \, S_{\oplus}$. The optimistic Zsom model for habitable zones indicates for this system an inner edge situated at  $d_{in}$ = 0.0910 AU from the star (considering albedo = 0.8). Taking into account a derived semi-major axis a = 0.0999 AU, this super-Earth candidate might be in the habitable zone. 

The moderately-high brightness of the star makes this candidate an interesting target for radial velocity follow up that could make it possible to confirm the planet, measure its mass, study the possible presence of more planets in the system and their atmospheric characterization.

\subsection{EPIC 246909566}
This target (T$_{\rm eff}$ = 3067 K, R = 0.24 $R_\odot$, $k_{p}$ = 17.23) was observed in campaign 13 and it is the coolest and faintest star in our candidate hosts list, showing a 1.925-day periodic signal. It is located in a moderately-crowded field with several nearby faint stars at angular distances > 6\arcsec. No blended sources are present in Pan-STARRS1 image, excluding stars with $\delta_{mag}<4.2$ at 2.5\arcsec. However, analysis of Pan-STARRS1 images and Gaia DR2 shows two stars within the optimal aperture (mag G = 13.81 at 12\arcsec and mag G = 20.51 at 4.8\arcsec) which we can not rule out as the sources of the signal. Furthermore, the \texttt{vespa} analysis computes $\rm FPP=5.2 \times 10^{-2}$. The most likely false positive scenario is the BEB with a contribution to the total FPP of $\rm FPP_{BEB}=4.8 \times 10^{-2}$. The fit suggests an Earth-size candidate with 1.1 $R_\oplus$. 

\subsection{EPIC 245944983}

This target was observed in campaign 12 and its light curve shows a signal with a period of 5.138 days. This is a binary system; according to Gaia DR2, both stars are situated at the same distance (140 pc) and exhibit almost identical proper motions ($\mu_{\alpha} \sim$ 109 mas, $\mu_{\delta} \sim$ 25 mas). The angular separation between the components is $\sim$ 4\arcsec.

The K2 detector has an image scale of 3.98\arcsec per pixel, so the photometry of both sources is blended in the light curve. This signal needs a specific photometric follow-up with a resolution of $\sim 1\arcsec$ to distinguish in which component the signal originates and to model the candidate parameters properly.

\subsection{K2-325 (EPIC 246074965)}

This star (T$_{\rm eff}$ = 3287 K, R = 0.30 $R_\odot$, $k_{p}$ = 16.28) was observed in campaign 12 and shows a transit-like signal with a period of 6.930 days. The image from Pan-STARRS1 shows no evidence of close or blended companions, excluding stars with $\delta_{mag}<4.9$ at 2.5\arcsec. Also the Pan-STARRS1 images and Gaia DR2 analysis discard nearby contaminant stars. Our fit models a super-Earth with R = 2.2 $R_\oplus$ and \texttt{vespa} analysis computed a $\rm FPP=6.3 \times 10^{-5}$, thus making it a statistically validated planet.

The optimistic Zsom model for habitable zones indicates for this system an inner edge situated at  $d_{in}$ = 0.0411 AU from the star (considering albedo = 0.8). Taking into account a derived semi-major axis a = 0.0419 AU, this super-Earth might be in the habitable zone.

\subsection{EPIC 246163416}

This target was observed in campaign 12 and exhibits a signal with a period of 0.877 days. The speckle image (acquired with NESSI instrument at WIYN-3.5m) detected a secondary source at 0.66\arcsec and $\Delta m$ = 1.39, so we can not distinguish which star is the source of the signal, nor derive the stellar or planetary parameters. 

\subsection{EPIC 246313886}

This target (T$_{\rm eff}$ = 3949 K, R = 0.50 $R_\odot$, $k_{p}$ = 14.48) was observed in campaign 12 and shows a periodic signal with 1.827-day period. Pan-STARRS1 image shows no close or blended stars, excluding stars with $\delta_{mag}<4.9$ at 2.5\arcsec. Also analysis of Pan-STARRS1 images and Gaia DR2, discards contamination from nearby stars. We fit the signal as an Earth-size candidate with 1.0 $R_\oplus$ and the \texttt{vespa} analysis resulted in $\rm FPP=6.0 \times 10^{-3}$.However the computed signal-to-noise ratio does not meet the condition SNR > 10, so we consider this signal as a candidate.

\subsection{EPIC 246331347}

This target (T$_{\rm eff}$ = 3408 K, R = 0.52 $R_\odot$, $k_{p}$ = 15.46) was observed in campaign 12 and shows a signal with a period of 1.082 days. The Pan-STARRS1 image does not show close or blended stars, excluding stars with $\delta_{mag}<5.9$ at 2.5\arcsec. Analysis of Pan-STARRS1 images and Gaia DR2 discard contamination from nearby stars. However, the \texttt{vespa} analysis computed $\rm FPP=2.0 \times 10^{-2}$ with all the contribution to the FPP from the BEB scenario. Our fit points to a super-Earth candidate with 2.2 $R_\oplus$. 

\subsection{EPIC 246331418}

This target (T$_{\rm eff}$ = 4408 K, R = 0.43 $R_\odot$, $k_{p}$ = 14.47) was observed in campaign 12 and shows two transit-like signals with periods 3.350 and 9.320 days. However, the Pan-STARRS1 image shows a faint star at an angular distance of 3\arcsec (PA = 125.8$^{\circ}$), smaller than the pixel scale of K2 and well within the optimal aperture. According to Gaia DR2, both stars lie at the same distance ($\pi=6.34  \pm  0.04\arcsec$ and $\pi=5.54  \pm  1.20\arcsec$ for the primary and secondary respectively) and present the same proper motions ($\mu_{\alpha}=-19.89 \pm 0.08$ $mas \cdot yr^{-1}$, $\mu_{\delta}=-53.64\pm 0.06$ $mas \cdot yr^{-1}$ for the primary and $\mu_{\alpha}=-20.25 \pm 2.51$ $mas \cdot yr^{-1}$, $\mu_{\delta}=-53.89\pm 2.61$ $mas \cdot yr^{-1}$ for the secondary), making it a binary system.
 
With the distance and the G magnitude from Gaia DR2, we estimated an absolute magnitude for the secondary of $M_{G} \sim$ 14.3. With this $M_{G}$, we estimated from the \citet{pecaut2013intrinsic} tabulated parameters a M5V spectral type for the secondary, with $\rm T \sim 3030\, \,K$, $\rm R \sim 0.20\, \, R_{\odot}$ and $\rm M \sim 0.15\, \, M_{\odot}$.

A planetary origin for the signals, both in the same star, is the most likely scenario as the other combinations involving transiting planets and eclipsing binaries have extremely low probabilities. However, it is not possible to distinguish which of both stars is the source of the signals, so they are considered as candidates.
Even so, taking into account the $6.3 \times 10^{-4}$ and $1.0 \times 10^{-3}$ mag transit depths for candidates b and c, if both signals were originated in the secondary, it would imply a $9\%$ and $16\%$ decreases in the secondary flux for candidates b and c respectively. As discussed above, adopting $\rm R \sim 0.20\, \,R_{\odot}$ for the star, the radii of the candidates would be estimated at $\rm \sim 0.6 \, \, R_{J}$ (b) and $\rm \sim 0.8 \, \, R_{J}$ (c). Finding two gas giants on a low mass star is extremely unlikely as only a few Jupiter-mass exoplanets have been found orbiting low mass stars \citep{1998A&A...338L..67D,1998ApJ...505L.147M,2019Sci...365.1441M}. All these considerations make the primary as the most likely host for the candidates.

Assuming the primary as the host star, our fit results in candidates with radii 1.0 and 1.4 $R_\oplus$. The computed false positive probabilities from \texttt{vespa} are $\rm FPP=9.5 \times 10^{-4}$ (b) and $\rm FPP=2.2 \times 10^{-3}$ (c), and the multiplicity-corrected false positive probabilities are $\rm FPP_{2}=3.4 \times 10^{-5}$ (b) and $\rm FPP_{2}=7.9 \times 10^{-5}$ (c), anyway these signals are not validated, as discussed before.

\subsection{K2-326 (EPIC 246472939)}

This target (T$_{\rm eff}$ = 3924 K, R = 0.69 $R_\odot$, $k_{p}$ = 14.87) was observed in campaign 12 and shows a candidate signal with a period of 1.256 days. Pan-STARRS1 image shows no close or blended stars, excluding stars with $\delta_{mag}<5.1$ at 2.5\arcsec. Also analysis of Pan-STARRS1 images and Gaia DR2 discard nearby contaminant stars. We fit this signal as a 2.3 $R_\oplus$ super-Earth and the \texttt{vespa} analysis computed $\rm FPP=1.0 \times 10^{-3}$, thus it is a validated planet.\\


\section{Conclusions}
\label{conclusions}

Our analysis of the photometry of 20038 cool stars from campaigns 12, 13, 14 and 15 of the K2 mission resulted in a catalogue of {\validated} new statistically validated planets, {\candidates} planetary candidates and 1 false positive, distributed among {\totalstars} stars.

We computed the stellar parameters for {\starswithparams} stars from their colours and accurate distances from Gaia. We derived the candidate parameters for all the systems (except for the binary EPIC 245944983, EPIC 250001426 and EPIC 246163416, both with close stars detected in speckle images, and the false positive EPIC 249391469) and also presented their phase-folded light curves along with their best fit.

Six of these signals belong to three multi-planetary systems, with two signals in each system (EPIC 249384674, EPIC 201663879 and EPIC 246331418).

Both validated planets and candidates from this work cover a wide range of orbital periods, between 0.877 d and 35.621 d, but they mostly have short periods, with a median value of 3.350 d. 

{\validatedbright} validated planets and {\candidatesbright} candidates are located in moderately bright stars ($\rm m_{kep}< 13$), making these systems good targets for radial velocity follow-up and atmospheric characterization. {\validatedhabitable} validated planets and {\candidateshabitable} candidates have estimated orbital radius within the limits of the habitable zone according to the optimistic Zsom model. 
Of special interest are the warm super-Earth K2-323 b (validated, T$_{\rm eq} = 318^{+24}_{-43} \, K$, S$_{\rm p} = 1.7\pm 0.2 \, S_{\oplus}$, R$_{\rm p} = 2.1\pm 0.1 \, R_{\oplus} $ ), located in a m$\rm_{kep}$ = 14.13 star, and the super-Earth EPIC 248782482 b (candidate, T$_{\rm eq} = 417^{+58}_{-24} \, K$, S$_{\rm p} = 5.0 \pm 1.0 \, S_{\oplus}$, R$_{\rm p} = 1.28  \pm 0.1 \, R_{\oplus}$ ) located in a moderately bright star (m$\rm_{kep}$ = 12.59), thus being a good targets for radial velocity follow-up and also for atmospheric characterization with transmission techniques with high resolution
spectrographs such as ESPRESSO at VLT \citep{2014AN....335....8P,2018haex.bookE.157G}.

Future radial velocity and photometric follow-up are necessary in order to confirm the candidates presented in this work. Meanwhile, detailed analysis of the light curves acquired by missions such as the Kepler Space Telescope, with the aim of detecting undiscovered candidate signals, will continue to provide valuable information for our understanding of planetary systems.

\section*{Acknowledgements}


ACG, EDA, SLSG, CGG, FGR and JCJ would like to acknowledge Spanish ministry project
AYA2017-89121-Pystems. LB acknowledges financial support from the PGC 2018 project PGC2018-101948-B-I00 (MICINN, FEDER). 


This paper includes data collected by the Kepler mission and obtained from the MAST data archive at the Space Telescope Science Institute (STScI). Funding for the Kepler mission is provided by the NASA Science Mission Directorate. STScI is operated by the Association of Universities for Research in Astronomy, Inc., under NASA contract NAS 5–26555.


The Pan-STARRS1 Surveys (PS1) and the PS1 public science archive have been made possible through contributions by the Institute for Astronomy, the University of Hawaii, the Pan-STARRS Project Office, the Max-Planck Society and its participating institutes, the Max Planck Institute for Astronomy, Heidelberg and the Max Planck Institute for Extraterrestrial Physics, Garching, The Johns Hopkins University, Durham University, the University of Edinburgh, the Queen's University Belfast, the Harvard-Smithsonian Center for Astrophysics, the Las Cumbres Observatory Global Telescope Network Incorporated, the National Central University of Taiwan, the Space Telescope Science Institute, the National Aeronautics and Space Administration under Grant No. NNX08AR22G issued through the Planetary Science Division of the NASA Science Mission Directorate, the National Science Foundation Grant No. AST–1238877, the University of Maryland, Eotvos Lorand University (ELTE), the Los Alamos National Laboratory, and the Gordon and Betty Moore Foundation.


This research has made use of the Exoplanet Follow-up Observation Program website, which is operated by the California Institute of Technology, under contract with the National Aeronautics and Space Administration under the Exoplanet Exploration Program.


This research has made use of the NASA Exoplanet Archive, which is operated by the California Institute of Technology, under contract with the National Aeronautics and Space Administration under the Exoplanet Exploration Program.


This work has made use of data from the European Space Agency (ESA) mission
{\it Gaia} (\url{https://www.cosmos.esa.int/gaia}), processed by the {\it Gaia}
Data Processing and Analysis Consortium (DPAC,
\url{https://www.cosmos.esa.int/web/gaia/dpac/consortium}). Funding for the DPAC
has been provided by national institutions, in particular the institutions
participating in the {\it Gaia} Multilateral Agreement.

The Digitized Sky Surveys were produced at the Space Telescope Science Institute under U.S. Government grant NAG W-2166. The images of these surveys are based on photographic data obtained using the Oschin Schmidt Telescope on Palomar Mountain and the UK Schmidt Telescope. The plates were processed into the present compressed digital form with the permission of these institutions.


Some of the observations in the paper made use of the NN-EXPLORE Exoplanet and Stellar Speckle Imager (NESSI). NESSI was funded by the NASA Exoplanet Exploration Program and the NASA Ames Research Center. NESSI was built at the Ames Research Center by Steve B. Howell, Nic Scott, Elliott P. Horch, and Emmett Quigley.


We thank Joaquin Gonz\'alez-Nuevo for useful discussions.


\bibliographystyle{mnras}
\bibliography{refs} 


\appendix

\section{Appendix}
\label{Appendix}

\begin{landscape}
\begin{table}
\renewcommand{\arraystretch}{1.6}
\begin{tabular}{lcccccccccc}
\hline
EPIC ID & Campaign & Kp [$mag$] & g [$mag$] & r [$mag$] & i [$mag$] & B [$mag$] & V [$mag$] & J [$mag$] & H [$mag$] & K [$mag$] \\
\hline
249384674 (1) & 15 & 14.794 & $17.272 \pm 0.002$ & $16.044 \pm 0.007$ & $14.798 \pm 0.004$ & --- & --- & $12.717 \pm 0.023$ & $12.144 \pm 0.026$ & $11.911 \pm 0.023$ \\
249391469     & 15 & 14.025 & $15.765 \pm 0.070$ & $14.365 \pm 0.010$ & $13.279 \pm 0.030$ & $16.515 \pm 0.081$ & $14.952 \pm 0.045$ & $11.423 \pm 0.024$ & $10.756 \pm 0.021$ & $10.554 \pm 0.024$ \\
249557502 (1) & 15 & 16.400 & $18.463 \pm 0.006$ & $17.164 \pm 0.003$ & $15.920 \pm 0.003$ & --- & --- & $13.928 \pm 0.035$ & $13.293 \pm 0.030$ & $12.990 \pm 0.036$ \\
249826231     & 15 & 14.466 & $16.086 \pm 0.030$ & $14.652 \pm 0.020$ & $13.772 \pm 0.050$ & $16.667 \pm 0.060$ & $15.269 \pm 0.049$ & $12.042 \pm 0.024$ & $11.341 \pm 0.022$ & $11.152 \pm 0.023$ \\
250001426 (2)     & 15 & 13.681 & $15.662 \pm 0.010$ & $14.303 \pm 0.010$ & $12.832 \pm 0.040$ & $16.361 \pm 0.140$ & $14.826 \pm 0.010$ & $10.658 \pm 0.024$ & $10.114 \pm 0.022$ & $9.829 \pm 0.021$ \\
250099723     & 15 & 13.803 & $15.182 \pm 0.040$ & $13.839 \pm 0.020$ & $13.212 \pm 0.070$ & $15.822 \pm 0.060$ & $14.400 \pm 0.020$ & $11.657 \pm 0.027$ & $10.977 \pm 0.024$ & $10.797 \pm 0.021$ \\
201663879 (3) & 14 & 11.932 & $12.601 \pm 0.050$ & $11.924 \pm 0.080$ & $11.645 \pm 0.020$ & $13.000 \pm 0.110$ & $12.172 \pm 0.070$ & $10.705 \pm 0.026$ & $10.313 \pm 0.023$ & $10.223 \pm 0.025$ \\
201663913 (3) & 14 & 14.451 & $16.482 \pm 0.004$ & $15.126 \pm 0.004$ & $14.049 \pm 0.004$ & --- & --- & $12.193 \pm 0.027$ & $11.543 \pm 0.029$ & $11.333 \pm 0.025$ \\
201796690 (4) & 14 & 15.798 & $18.565 \pm 0.008$ & $17.082 \pm 0.005$ & $15.626 \pm 0.004$ & --- & --- & $13.411 \pm 0.032$ & $12.823 \pm 0.024$ & $12.508 \pm 0.023$ \\
248480671     & 14 & 12.840 & $14.457 \pm 0.010$ & $13.118 \pm 0.010$ & $12.147 \pm 0.050$ & $15.127 \pm 0.040$ & $13.677 \pm 0.050$ & $10.462 \pm 0.027$ & $9.796 \pm 0.024$ & $9.567 \pm 0.021$ \\
248558190     & 14 & 13.307 & $14.715 \pm 0.060$ & $13.371 \pm 0.030$ & $12.703 \pm 0.040$ & $15.317 \pm 0.060$ & $13.941 \pm 0.020$ & $11.194 \pm 0.023$ & $10.569 \pm 0.023$ & $10.413 \pm 0.021$ \\
248616368     & 14 & 14.132 & $15.722 \pm 0.110$ & $14.387 \pm 0.080$ & $13.451 \pm 0.080$ & $16.418 \pm 0.060$ & $14.932 \pm 0.110$ & $11.648 \pm 0.023$ & $11.002 \pm 0.024$ & $10.789 \pm 0.021$ \\
248639308     & 14 & 14.509 & $16.142 \pm 0.070$ & $14.793 \pm 0.030$ & $13.809 \pm 0.060$ & $16.783 \pm 0.040$ & $15.399 \pm 0.020$ & $12.092 \pm 0.026$ & $11.457 \pm 0.022$ & $11.237 \pm 0.026$ \\
248775938      & 14 & 15.223 & $16.564 \pm 0.020$ & $15.314 \pm 0.060$ & $14.649 \pm 0.070$ & $17.244 \pm 0.106$ & $15.879 \pm 0.030$ & $13.087 \pm 0.026$ & $12.394 \pm 0.023$ & $12.202 \pm 0.026$ \\
248782482     & 14 & 12.585 & $14.088 \pm 0.060$ & $12.745 \pm 0.030$ & $11.941 \pm 0.050$ & $14.783 \pm 0.040$ & $13.358 \pm 0.040$ & $10.319 \pm 0.021$ & $9.702 \pm 0.022$ & $9.488 \pm 0.023$ \\
246909566     & 13 & 17.231 & $19.506 \pm 0.012$ & $17.910 \pm 0.007$ & $16.256 \pm 0.005$ & --- & --- & $13.802 \pm 0.029$ & $13.210 \pm 0.040$ & $12.980 \pm 0.024$ \\
245944983 (5)     & 12 & 12.738 & $14.154 \pm 0.050$ & $12.752 \pm 0.040$ & $12.131 \pm 0.030$ & $14.725 \pm 0.050$ & $13.332 \pm 0.020$ & $11.277 \pm 0.061$ & $10.618 \pm 0.074$ & $10.478 \pm 0.063$ \\
246074965     & 12 & 16.278 & $18.311 \pm 0.007$ & $16.900 \pm 0.005$ & $15.407 \pm 0.004$ & --- & --- & $13.177 \pm 0.026$ & $12.595 \pm 0.030$ & $12.352 \pm 0.028$ \\
246163416 (2)   & 12 & 13.482 & $15.113 \pm 0.090$ & $13.725 \pm 0.020$ & $12.783 \pm 0.030$ & $15.812 \pm 0.050$ & $14.285 \pm 0.040$ & $10.995 \pm 0.026$ & $10.415 \pm 0.026$ & $10.162 \pm 0.023$ \\
246313886     & 12 & 14.481 & $16.042 \pm 0.051$ & $14.523 \pm 0.050$ & $13.813 \pm 0.069$ & $16.763 \pm 0.130$ & $15.163 \pm 0.048$ & $12.312 \pm 0.031$ & $11.625 \pm 0.023$ & $11.427 \pm 0.023$ \\
246331347 (4) & 12 & 15.459 & $17.575 \pm 0.005$ & $16.129 \pm 0.004$ & $15.051 \pm 0.005$ & --- & --- & $13.173 \pm 0.026$ & $12.525 \pm 0.022$ & $12.304 \pm 0.023$ \\
246331418 (4) & 12 & 14.472 & $16.530 \pm 0.004$ & $15.096 \pm 0.005$ & $14.636 \pm 0.002$ & --- & --- & $12.381 \pm 0.024$ & $11.780 \pm 0.024$ & $11.571 \pm 0.021$ \\
246472939     & 12 & 14.867 & $16.409 \pm 0.020$ & $14.781 \pm 0.040$ & $14.206 \pm 0.030$ & $16.926 \pm 0.060$ & $15.342 \pm 0.070$ & $12.977 \pm 0.024$ & $12.231 \pm 0.026$ & $12.091 \pm 0.026$ \\
\hline
\multicolumn{3}{l}{(1) \textit{gri} photometry extracted from Pan-STARRS DR1.} \\
\multicolumn{6}{l}{(2) Close star detected in speckle / AO images.}\\
\multicolumn{6}{l}{(3) Signals present at EPIC 201663913 are contamination from the early K-type star EPIC 201663879.} \\
\multicolumn{6}{l}{(4) \textit{gri} photometry extracted from SDSS DR15.} \\
\multicolumn{6}{l}{(5) Binary star.}\\
\end{tabular}
\caption{Photometry of candidate host stars.}
\label{tab:stars-photo}
\end{table}

\end{landscape}

\begin{landscape}
\begin{table}
\renewcommand{\arraystretch}{1.6}
\begin{tabular}{lccllccc}
\hline
    EPIC ID & Campaign & Name & Distance [$pc$] & Teff [$K$] & Mass [$M_\odot$] & Radius [$R_\odot$] & logg \\
    \hline
    249384674       & 15 & K2-316 & $112.84 \pm 0.96$ & $3436 \pm 3$ & $0.408 \pm 0.005$ & $0.379 \pm 0.003$ & $4.891 \pm 0.002$ \\ 
    249391469       & 15 & --& $101.56 \pm 0.40$ & $3503 \pm 63$ & $0.337 \pm 0.079$ & $0.567 \pm 0.018$ & $4.447 \pm 0.130$ \\
    249557502       & 15 & K2-317 & $177.08 \pm 3.34$ & $3387 \pm 5$ & $0.403 \pm 0.006$ & $0.379 \pm 0.006$ & $4.885 \pm 0.011$ \\
    249826231       & 15 & K2-318 & $148.49 \pm 0.99$ & $3851 \pm 83$ & $0.561 \pm 0.050$ & $0.552 \pm 0.017$ & $4.701 \pm 0.070$ \\
    250001426 (1)     & 15 & --& $------$ & $-----$ & $------$ & $------$ & $------$ \\
    250099723       & 15 & --& $143.29 \pm 0.58$ & $4176 \pm 78$ & $0.603 \pm 0.017$ & $0.576 \pm 0.007$ & $4.696 \pm 0.012$ \\
    201663879 (2)   & 14 & K2-319 & $210.56 \pm 1.37$ & $5440 \pm 115$ & $0.954 \pm 0.055$ & $0.898 \pm 0.013$ & $4.512 \pm 0.055$ \\
    201663913 (2,3) & 14 & --& $215.94 \pm 2.19$ & $3874 \pm 72$ & $0.649 \pm 0.016$ & $0.678 \pm 0.020$ & $4.588 \pm 0.026$ \\
    201796690       & 14 & K2-320 & $109.07 \pm 1.04$ & $3157 \pm 7$ & $0.117 \pm 0.011$ & $0.300 \pm 0.003$ & $4.550 \pm 0.036$ \\ 
    248480671       & 14 & K2-321 & $77.86 \pm 0.32$ & $3855 \pm 65$ & $0.599 \pm 0.027$ & $0.583 \pm 0.016$ & $4.683 \pm 0.042$ \\
    248558190       & 14 & K2-322 & $123.86 \pm 0.31$ & $4141 \pm 96$ & $0.630 \pm 0.033$ & $0.602 \pm 0.013$ & $4.677 \pm 0.041$ \\
    248616368       & 14 & K2-323 & $118.72 \pm 0.55$ & $3710 \pm 118$ & $0.478 \pm 0.089$ & $0.549 \pm 0.024$ & $4.630 \pm 0.127$ \\
    248639308       & 14 & K2-324& $137.18 \pm 1.60$ & $3752 \pm 57$ & $0.517 \pm 0.041$ & $0.505 \pm 0.012$ & $4.743 \pm 0.057$ \\ 
    248775938       & 14 & --& $305.58 \pm 6.75$ & $4663 \pm 144$ & $0.625 \pm 0.023$ & $0.608 \pm 0.014$ & $4.665 \pm 0.019$ \\
    248782482       & 14 & --& $62.16 \pm 0.16$ & $3852 \pm 52$ & $0.518 \pm 0.029$ & $0.503 \pm 0.008$ & $4.749 \pm 0.037$ \\
    246909566       & 13 & --& $98.51 \pm 1.43$ & $3067 \pm 5$ & $0.121 \pm 0.007$ & $0.236 \pm 0.004$ & $4.772 \pm 0.027$ \\
    245944983 (4)   & 12 & --& $------$ & $------$ & $------$ & $------$ & $------$ \\
    246074965 (3)   & 12 & K2-325 & $111.98 \pm 1.02$ & $3287 \pm 67$ & $0.274 \pm 0.008$ & $0.298 \pm 0.009$ & $4.840 \pm 0.026$ \\
    246163416 (1)      & 12 & --& $------$ & $-----$ & $------$ & $------$ & $------$
 \\
    246313886       & 12 & --& $156.46 \pm 0.95$ & $3949 \pm 54$ & $0.523 \pm 0.015$ & $0.503 \pm 0.005$ & $4.753 \pm 0.012$ \\
    246331347       & 12 & --& $190.96 \pm 2.29$ & $3408 \pm 17$ & $0.245 \pm 0.027$ & $0.516 \pm 0.009$ & $4.399 \pm 0.062$ \\
    246331418    & 12 & --& $157.71 \pm 1.06$ & $4408 \pm 5$ & $0.276 \pm 0.006$ & $0.427 \pm 0.004$ & $4.617 \pm 0.013$ \\
    246472939       & 12 & K2-326 & $284.58 \pm 3.37$ & $3924 \pm 51$ & $0.507 \pm 0.051 $ & $0.688 \pm 0.014$ & $4.462 \pm 0.059$ \\
    \hline 
    \multicolumn{6}{l}{(1) Close blended source detected in speckle / AO images. Unreliable stellar parameters.}  \\
    \multicolumn{6}{l}{(2) Signals present at EPIC 201663913 are contamination from the early K-type star EPIC 201663879.}\\
    \multicolumn{6}{l}{(3) Stellar parameters from Dressing (2019).} \\
    \multicolumn{6}{l}{(4) Binary star. Unreliable stellar parameters.} \\
    \end{tabular}
    \caption{Candidate hosts stars parameters.}
    \label{tab:starparams}
\end{table}
\end{landscape}

\begin{landscape}
\begin{table}
\large
\small
\renewcommand{\arraystretch}{1.6}
\begin{tabular}{cccccccccccc}
    \hline
    Candidate & C & Name & $\rm P_{orb}$ [$d$] & a [$AU$] & $\rm R_{p}$ [$R_{\oplus}$] & $\rm T_{eq}$ [$K$] & Epoch (BKJD) [$d$] & i [$deg$] &  b & Transit Depth [$mag$] & SNR \\
    \hline
    249384674 b     & 15 & K2-316 b & 1.133 $\pm$ 0.001 &  $0.0147^{+0.0010}_{-0.0031}$ & $1.33^{+0.10}_{-0.09}$ & $841^{+105}_{-54}$ & 3158.605 &  $87.02^{+2.19}_{-3.98}$ & $0.43^{+0.32}_{-0.30}$ & $0.00104 \pm 0.00002$ & 15.5\\
    249384674 c     & 15 & K2-316 c & 5.260 $\pm$ 0.007 &  $0.0582^{+0.0081}_{-0.0122}$ & $1.83^{+0.13}_{-0.12}$ & $423^{+53}_{-26}$ & 3161.123 &  $89.31^{+0.49}_{-1.06}$ & $0.40^{+0.34}_{-0.28}$ & $0.00270 \pm 0.00006$ & 14.8 \\
    249391469 b (1)   & 15 & --& 1.075 $\pm$ 0.003 & $--$   & $--$ & $--$ & 3158.860 & $--$ & $--$ & $0.00095 \pm 0.00001$ & 28.3 \\
    249557502 b     & 15 & K2-317 b & 6.220 $\pm$ 0.007 & $0.0542^{+0.0065}_{-0.0103}$  & $2.93^{+0.19}_{-0.17}$ & $432^{+48}_{-24}$ & 3161.132 & $89.22^{+0.54}_{-0.93}$ & $0.43^{+0.29}_{-0.28}$ & $0.00129 \pm 0.00014$ & 13.6\\
    249826231 b     & 15 & K2-318 b & 7.010 $\pm$ 0.007 & $0.0911^{+0.0140}_{-0.0185}$  & 1.66 $\pm$ 0.13 & $456^{+55}_{-32}$ & 3164.194 & $89.38^{+0.44}_{-0.76}$ & $0.40^{+0.29}_{-0.27}$ & $0.00092 \pm 0.00002$ & 11.7 \\
    250001426 b (2)    & 15 & -- & $1.705^{+0.004}_{-0.003}$  & $--$  & $--$  & $--$ & 3159.286 & $--$ & $--$ & $0.00056 \pm 0.00001$ & 17.1 \\
    250099723 b     & 15 & --& $17.920^{+0.013}_{-0.014}$  & $0.1417^{+0.0158}_{-0.0285}$  & 2.07 $\pm$ 0.09 & $407^{+48}_{-23}$ & 3170.786 & $89.57^{+0.31}_{-0.50}$ & $0.40^{+0.29}_{-0.28}$ & $0.00097 \pm 0.00004$ & 11.8\\
    201663879 b (3) & 14 & K2-319 b & $26.680^{+0.002}_{-0.003}$  & $0.1516^{+0.0115}_{-0.0258}$  & $2.79^{+0.11}_{-0.10}$ & $641^{+61}_{-31}$ & 3082.592 & $89.45^{+0.40}_{-0.66}$ & $0.35^{+0.28}_{-0.24}$ & $0.00090 \pm 0.00001$ & 19.1 \\
    201663879 c (3) & 14 & --&  $35.621^{+0.003}_{-0.002}$  & $0.3501^{+0.0449}_{-0.0648}$  & $1.89^{+0.14}_{-0.13}$ & $420^{+47}_{-27}$ & 3078.974 & $89.77^{+0.17}_{-0.31}$ & $0.35^{+0.31}_{-0.26}$ & $0.00045 \pm 0.00002$ & 9.4\\
    201796690 b & 14 & K2-320 b & 1.995 $\pm$ 0.001  & $0.0142^{+0.0026}_{-0.0030}$  & $2.62^{+0.23}_{-0.15}$ & $698^{+87}_{-57}$ & 3075.602 & $87.84^{+1.41}_{-1.97}$ & $0.49^{+0.25}_{-0.30}$ & $0.00606 \pm 0.00006$ & 17.2\\
    248480671 b     & 14 & K2-321 b & 2.298 $\pm$ 0.001  & $0.0410^{+0.0035}_{-0.0048}$  & $1.95^{+0.10}_{-0.12}$ & $701^{+45}_{-30}$ & 3076.170 & $88.94^{+0.73}_{-1.12}$ & $0.28^{+0.24}_{-0.19}$ & $0.00120 \pm 0.00001$ & 27.9 \\
    248558190 b  & 14 & K2-322 b & 8.205 $\pm$ 0.003  & $0.0594^{+0.0174}_{-0.0221}$  & $1.92^{+0.29}_{-0.18}$ & $635^{+165}_{-78}$ & 3081.022 & $88.15^{+1.11}_{-2.01}$ & $0.68^{+0.21}_{-0.33}$ & $0.00077 \pm 0.00002$ & 19.8 \\
    248616368 b     & 14 & K2-323 b & 24.930 $\pm$ 0.006  & $0.1275^{+0.0055}_{-0.0072}$  & $2.10^{+0.09}_{-0.11}$ & $318^{+24}_{-43}$ & 3081.251 & $89.88^{+1.08}_{-1.29}$ & $0.14^{+0.11}_{-0.09}$ & $0.00286 \pm 0.00011$ & 19.3\\
    248639308 b & 14 & K2-324 b & 3.262 $\pm$ 0.001  & $0.0331^{+0.0037}_{-0.0066}$  & $2.43^{+0.15}_{-0.13}$ & $707^{+83}_{-38}$ & 3076.146 & $88.33^{+1.06}_{-1.84}$ & $0.42^{+0.28}_{-0.25}$ & $0.00216 \pm 0.00002$ & 24.2\\
    248775938 b     & 14 & --& 1.754 $\pm$ 0.001  & $0.0365^{+0.0038}_{-0.0053}$  & $5.06^{+0.49}_{-0.44}$ & $873^{+75}_{-54}$ & 3074.851 & $88.26^{+1.17}_{-1.73}$ & $0.39^{+0.27}_{-0.26}$ & $0.00568 \pm 0.00003$ & 30.5 \\
    248782482 b     & 14 & --&  $16.222^{+0.012}_{-0.014}$  & $0.0999^{+0.0124}_{-0.0229}$  & $1.28^{+0.11}_{-0.09}$ & $417^{+58}_{-24}$ & 3080.007 & $89.49^{+0.37}_{-0.72}$ & $0.39^{+0.33}_{-0.27}$ & $0.00056 \pm 0.00002$ & 9.2\\
    246909566 b   & 13 & --&  $1.925^{+0.003}_{-0.004}$  & $0.0251^{+0.0048}_{-0.0047}$  & $1.07^{+0.09}_{-0.08}$ & $452^{+48}_{-38}$ & 2988.974 & $89.13^{+0.61}_{-1.02}$ & $0.36^{+0.28}_{-0.24}$ & $0.00216 \pm 0.00002$ & 10.8 \\
    245944983 b (4) & 12 & -- & 5.138 $\pm$ 0.002 & $--$  & $--$ & $--$ & 2910.267 & $--$ & $--$ & $0.00073 \pm 0.00001$ & 19.6\\
    246074965 b & 12 & K2-325 b & $6.930^{+0.004}_{-0.003}$  & $0.0419^{+0.0046}_{-0.0058}$  & $2.20^{+0.14}_{-0.15}$ & $423^{+33}_{-21}$ & 2910.818 & $89.37^{+0.40}_{-0.68}$ & $0.34^{+0.26}_{-0.22}$ & $0.00732 \pm 0.00011$ & 15.8\\
    246163416 b (2)  & 12 & --& $0.877^{+0.003}_{-0.003}$  & $--$  & $--$ & $--$ & 2905.852 & $--$ & $--$ & $0.00050 \pm 0.00001$ & 6.2\\
    246313886 b     & 12 & --& 1.827 $\pm$ 0.002  & $0.0187^{+0.0038}_{-0.0043}$  & $0.99^{+0.11}_{-0.09}$ & $986^{+138}_{-88}$ & 2906.474 & $87.19^{+2.07}_{-3.88}$ & $0.43^{+0.32}_{-0.31}$ & $0.00035 \pm 0.00002$ & 10.0\\
    246331347 b     & 12 & --& 1.082 $\pm$ 0.001  & $0.0162^{+0.0027}_{-0.0034}$  & $2.22^{+0.21}_{-0.18}$ & $928^{+119}_{-68}$ & 2906.474 & $86.19^{+2.59}_{-5.24}$ & $0.46^{+0.32}_{-0.30}$ & $0.00162 \pm 0.00003$ & 15.8\\
    246331418 b     & 12 & --& 3.350 $\pm$ 0.002  & $0.0266^{+0.0064}_{-0.0060}$  & 0.97 $\pm$ 0.11 & $851^{+117}_{-86}$ & 2908.005 & $88.21^{+1.30}_{-2.36}$ & $0.44^{+0.31}_{-0.31}$ & $0.00063 \pm 0.00003$ & 6.5 \\
    246331418 c     & 12 & --&  9.320 $\pm$ 0.001  & $0.0724^{+0.0123}_{-0.0178}$  & $1.43^{+0.13}_{-0.14}$ & $516^{+78}_{-39}$ & 2911.661 & $89.30^{+0.49}_{-0.92}$ & $0.44^{+0.32}_{-0.29}$ & $0.00109 \pm 0.00005$ & 8.7 \\
    246472939 b     & 12 & K2-326 b&  $1.256^{+0.001}_{-0.002}$  & $0.0198^{+0.0036}_{-0.0041}$  & $2.28^{+0.18}_{-0.18}$ & $1114^{+138}_{-90}$ & 2906.045 & $85.63^{+2.87}_{-4.24}$ & $0.48^{+0.26}_{-0.29}$ & $0.00100 \pm 0.00003$ & 11.4\\
    \hline

\multicolumn{7}{l}{(1) False positive. The signal comes from Gaia DR2 6239523158128298496.}\\
\multicolumn{7}{l}{(2) Close blended source detected in speckle / AO images. Unreliable stellar and therefore planetary parameters.}\\
\multicolumn{7}{l}{(3) Signal from an early K-type star.}\\
\multicolumn{7}{l}{(4) Candidate signal from a binary star.}\\

\end{tabular}
\caption{Parameters of planetary candidates presented in this work.}
\label{tab:planet_params}
\end{table}
\end{landscape}

\begin{figure*}
\centering
\begin{subfigure}{8cm}
\includegraphics[width=8.0cm]{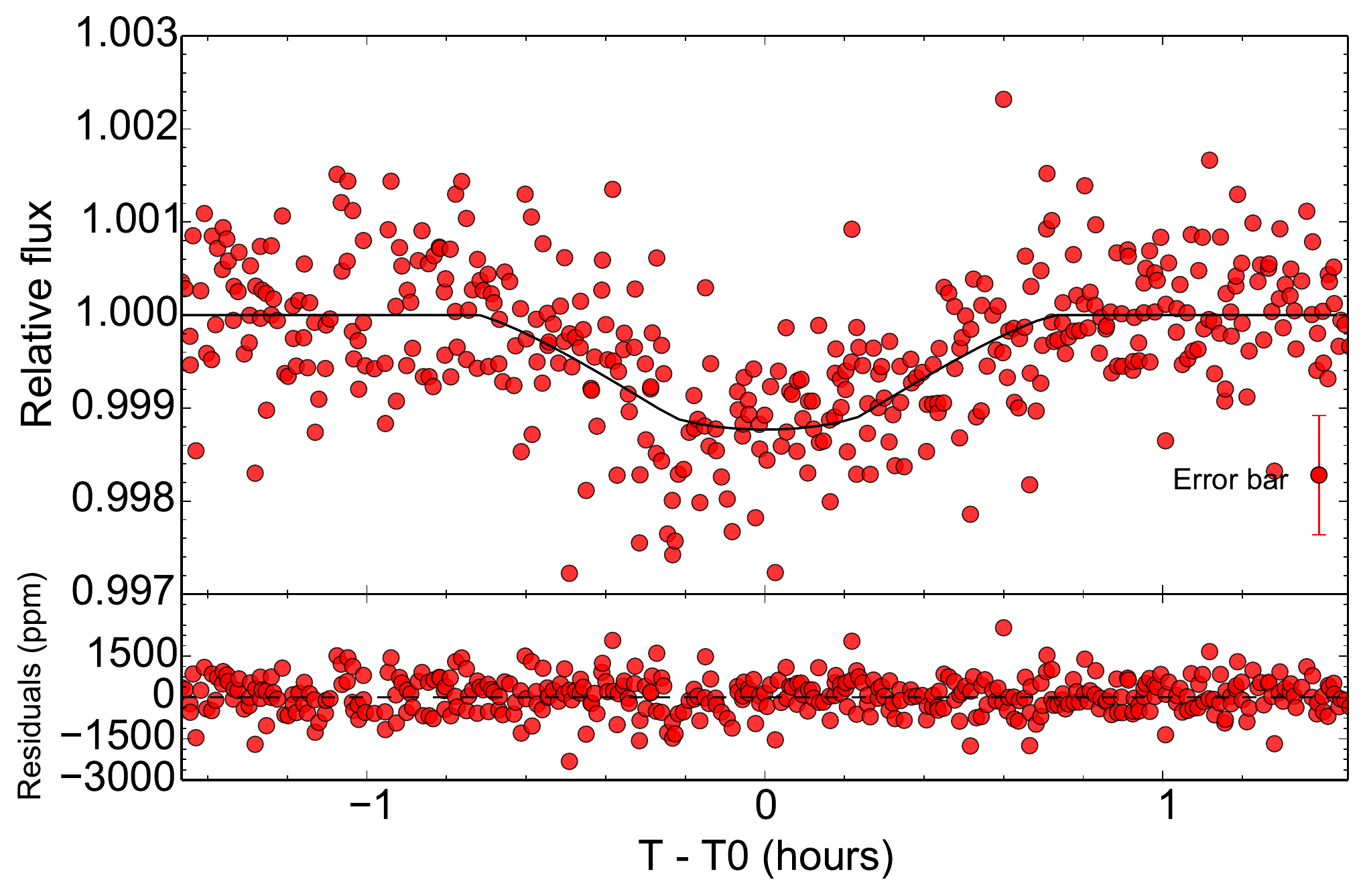}
\caption*{\quad \quad \, \,  K2-316 b (EPIC 249384674 b)}
\end{subfigure}
\begin{subfigure}{8cm}
\includegraphics[width=8.0cm]{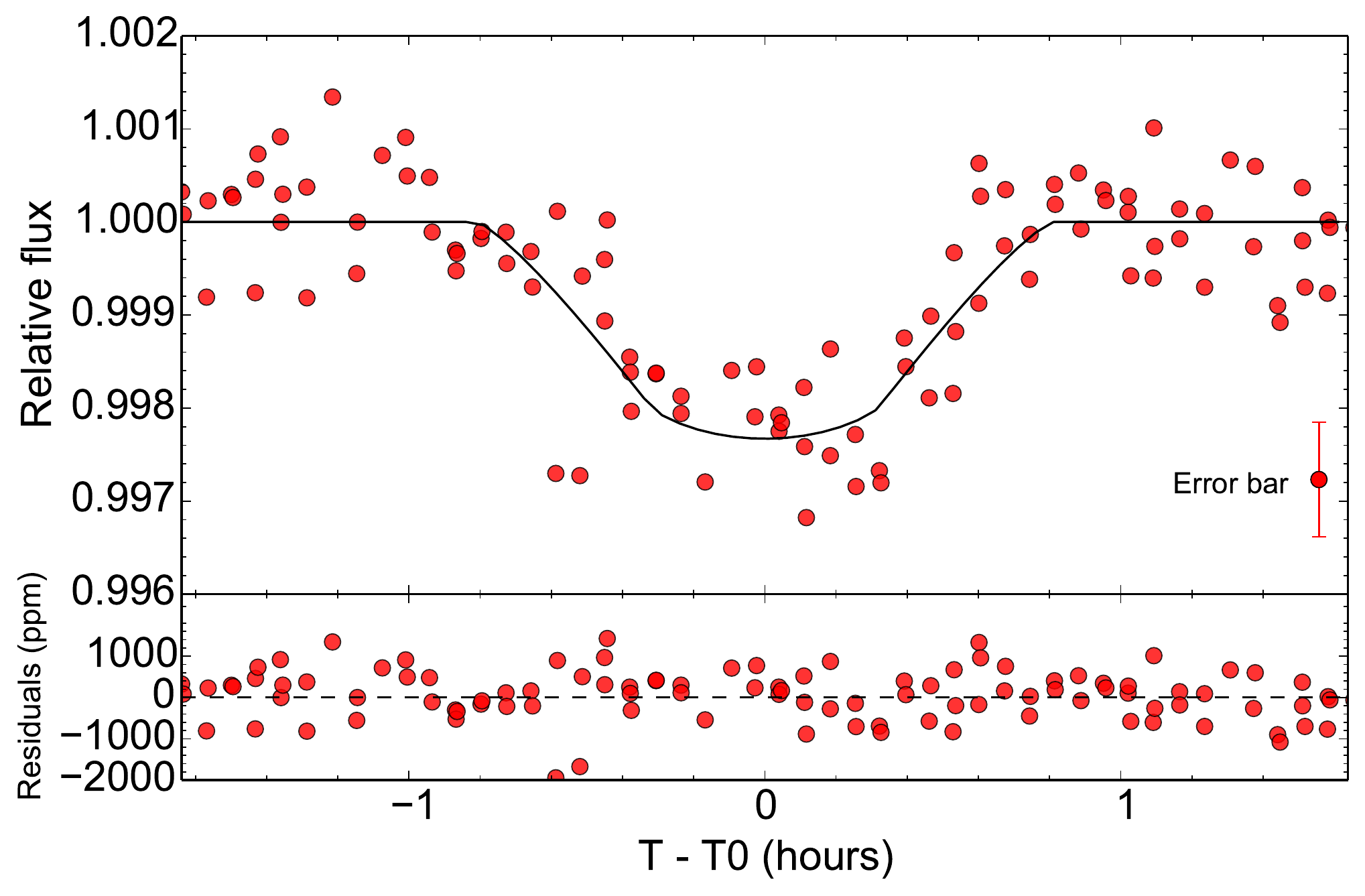}
\caption*{\quad \quad \, \, K2-316 c (EPIC 249384674 c)}
\end{subfigure}%

\begin{subfigure}{8cm}
\includegraphics[width=8.0cm]{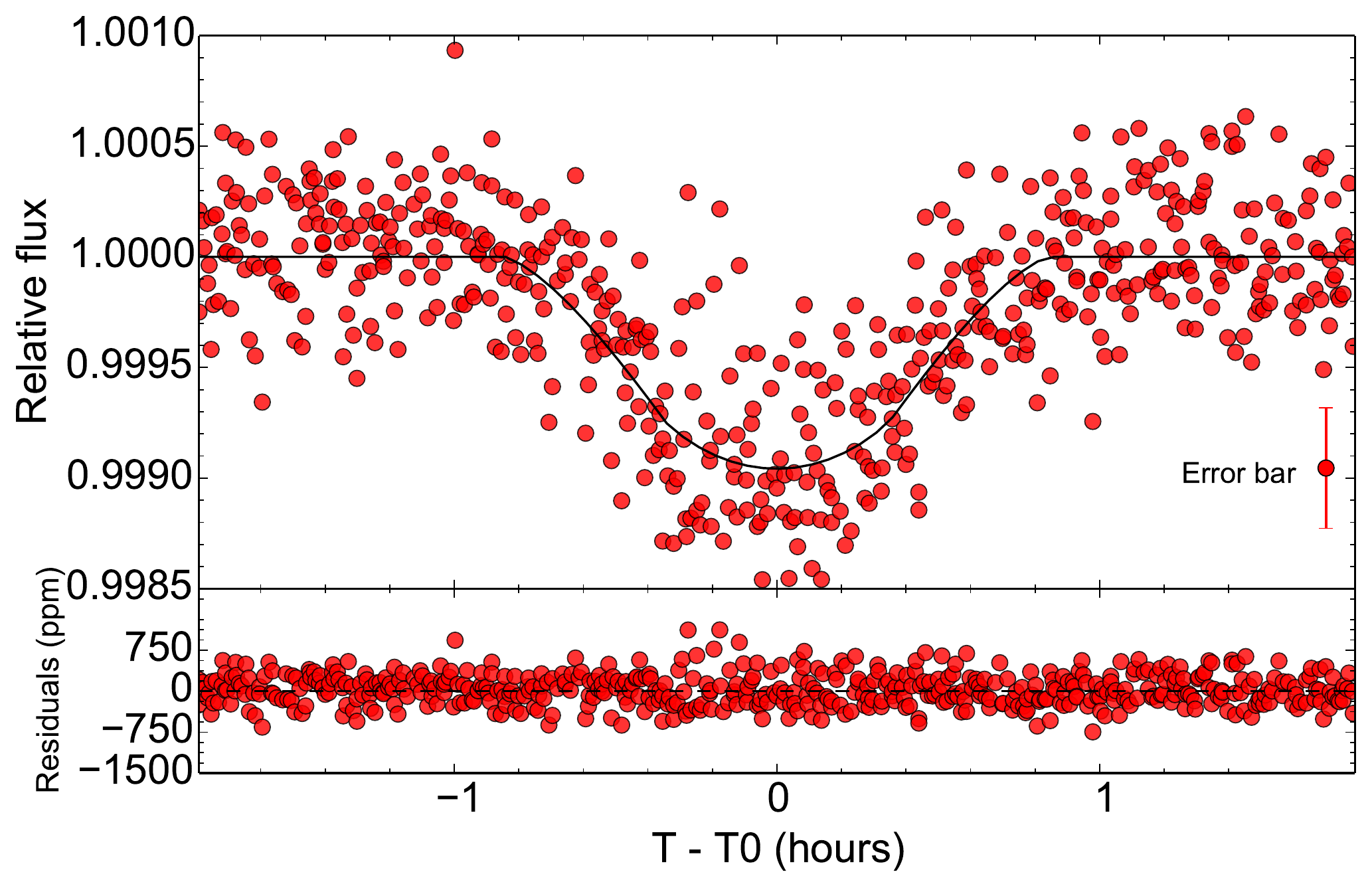}
\caption*{\quad \quad \, \, \, EPIC 249391469 b}
\end{subfigure}
\begin{subfigure}{8cm}
\includegraphics[width=8.0cm]{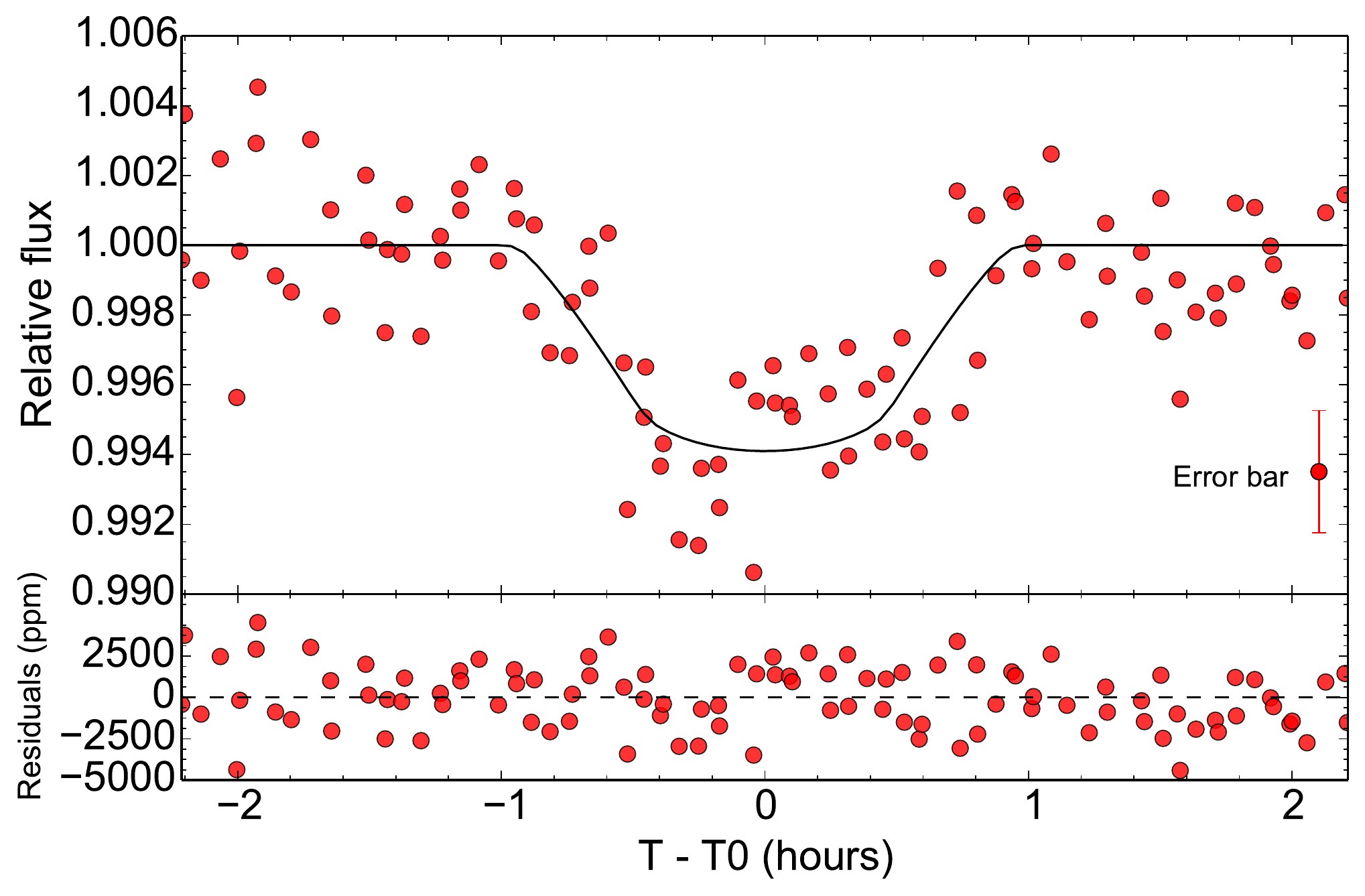}
\caption*{\quad \quad \, \, K2-317 b (EPIC 249557502 b)}
\end{subfigure}

\begin{subfigure}{8cm}
\includegraphics[width=8.0cm]{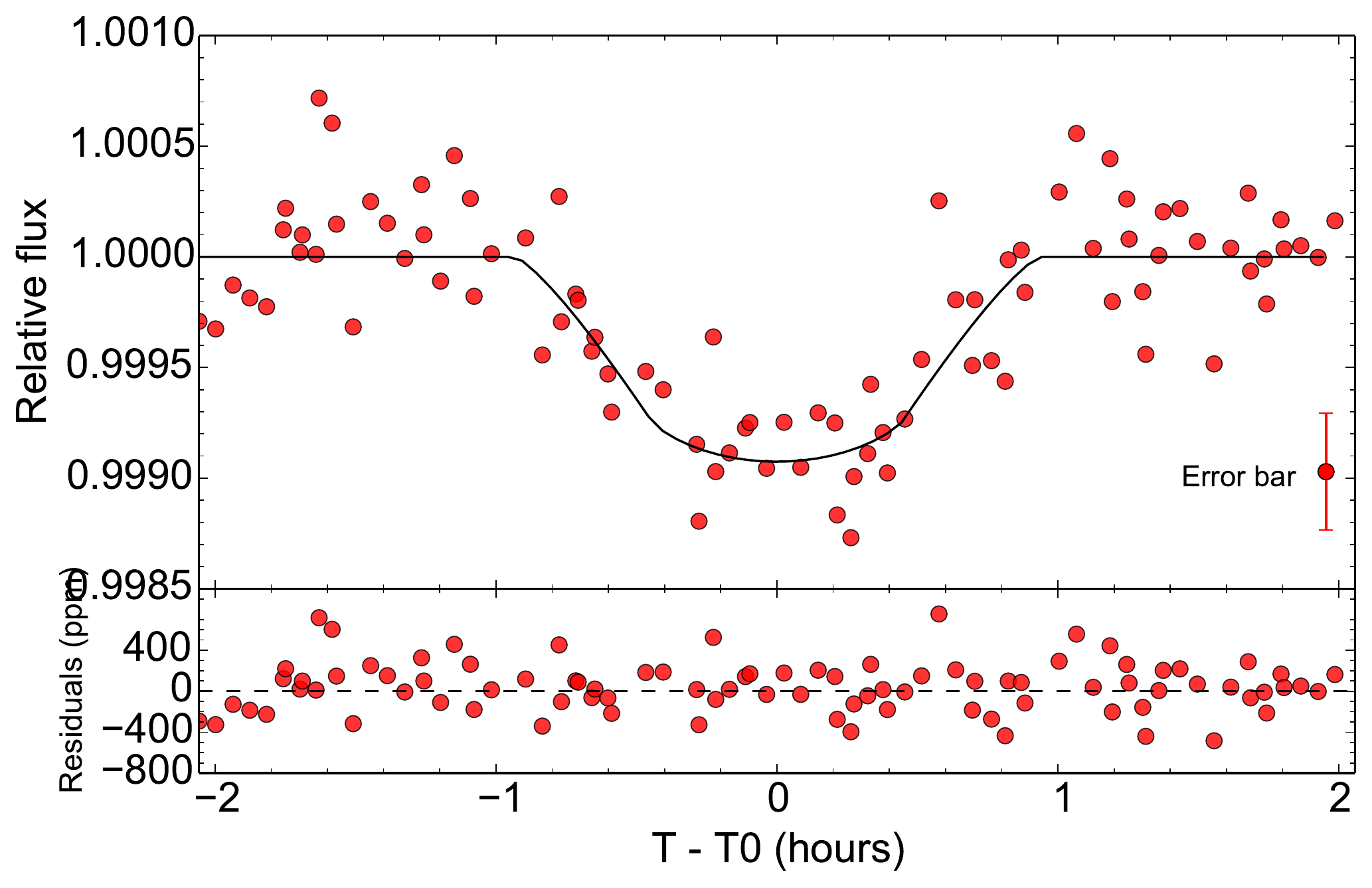}
\caption*{\quad \quad \, \, K2-318 b (EPIC 249826231b)}
\end{subfigure}%
\begin{subfigure}{8cm}
\includegraphics[width=8.0cm]{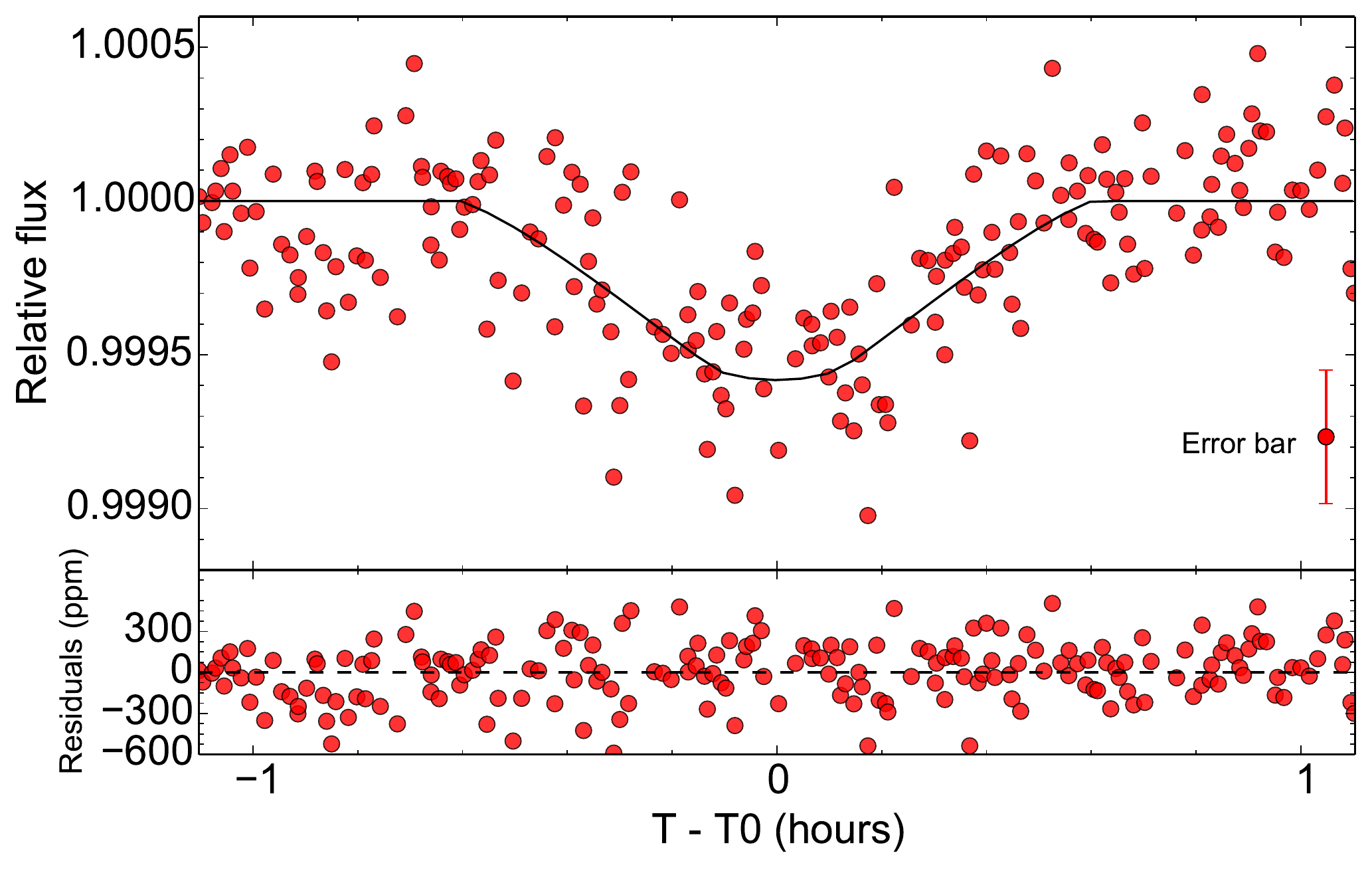}
\caption*{\quad \quad \, \, \, \, EPIC 250001426 b}
\end{subfigure}%

\begin{subfigure}{8cm}
\includegraphics[width=8.0cm]{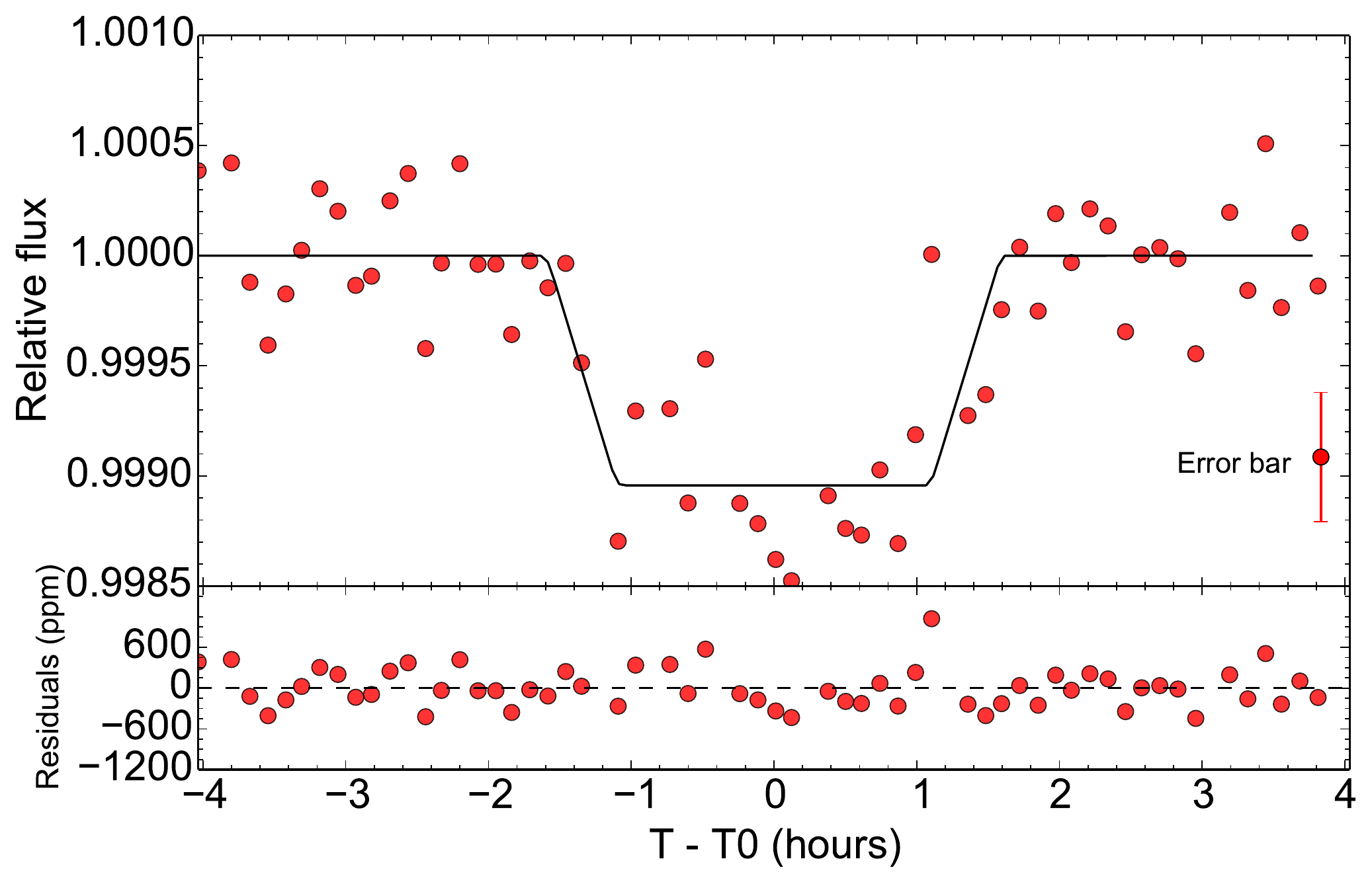}
\caption*{\quad \quad \, \, \,\, EPIC 250099723 b}
\end{subfigure}
\begin{subfigure}{8cm}
\includegraphics[width=8.0cm]{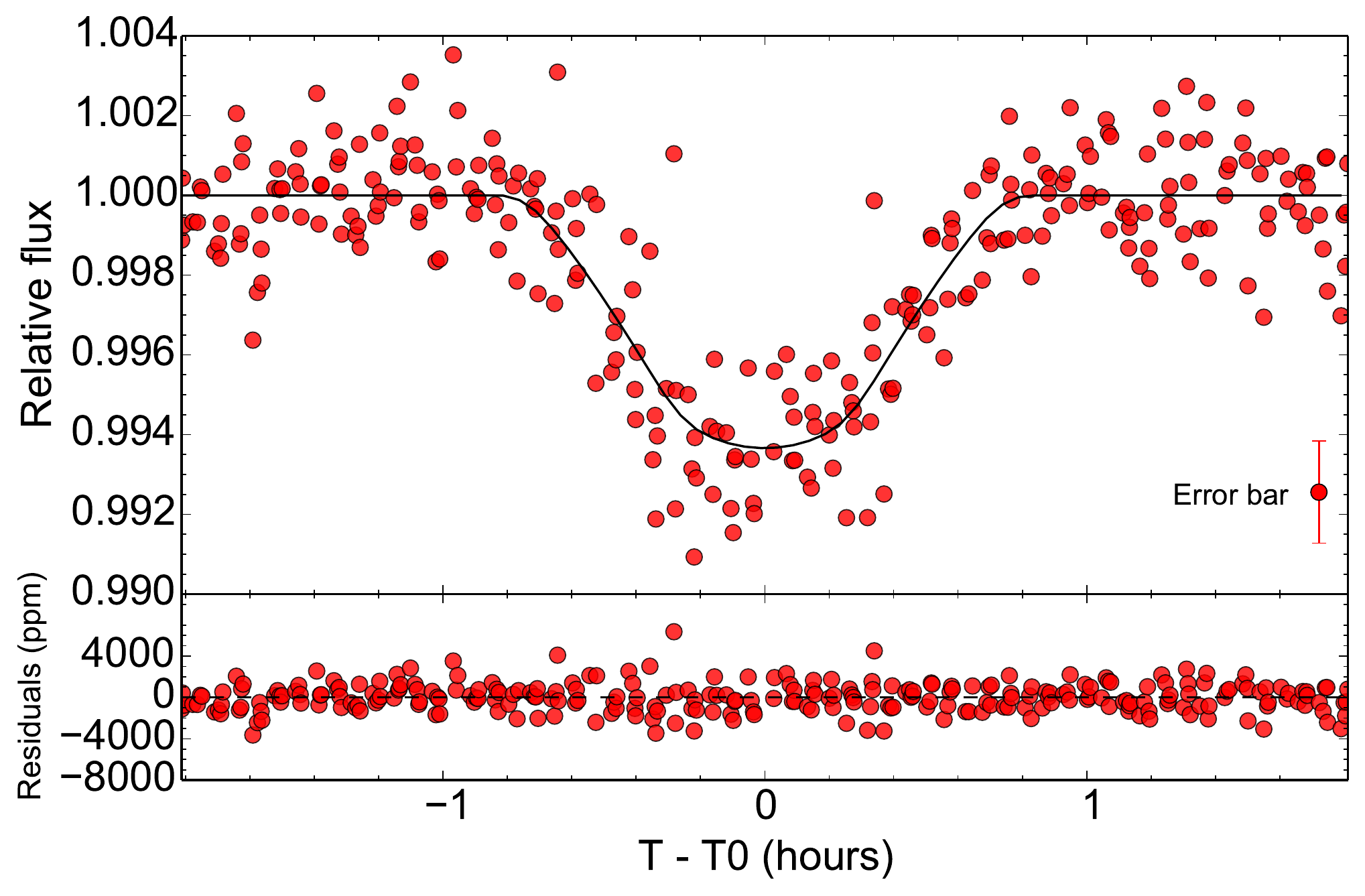}
\caption*{\quad \quad \, \, K2-320 b (EPIC 201796690 b, C14)}
\end{subfigure}

\caption{Phase-folded transits and best fit for planetary candidates of campaign 15 and 14 (K2-320 b).}
 
\label{fig:phase_folded1}
\end{figure*}


\begin{figure*}
\centering

\begin{subfigure}{8cm}
\includegraphics[width=8.0cm]{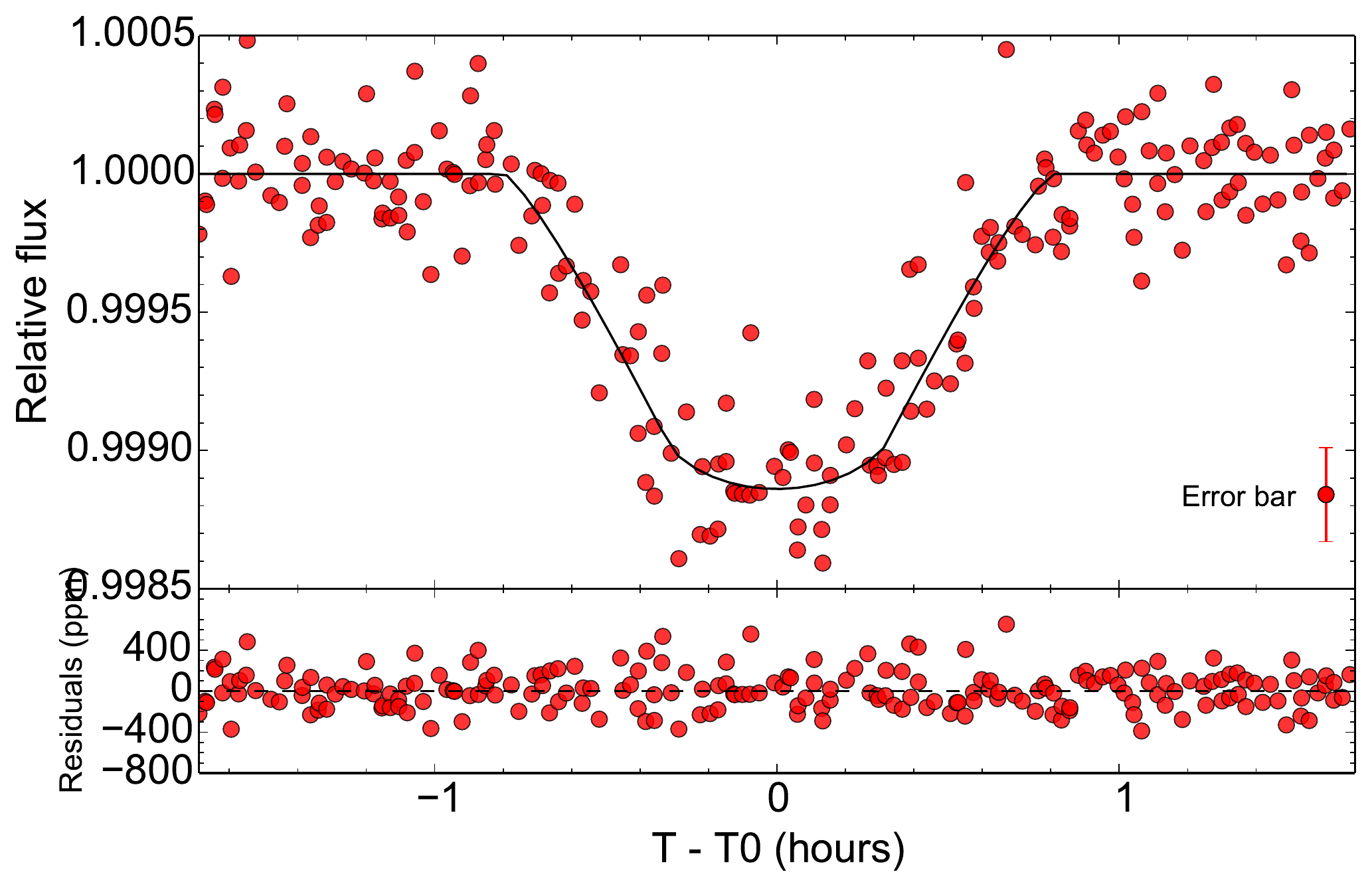}
\caption*{\quad \quad \, \, K2-321 b (EPIC 248480671 b)}
\end{subfigure}%
\begin{subfigure}{8cm}
\includegraphics[width=8.0cm]{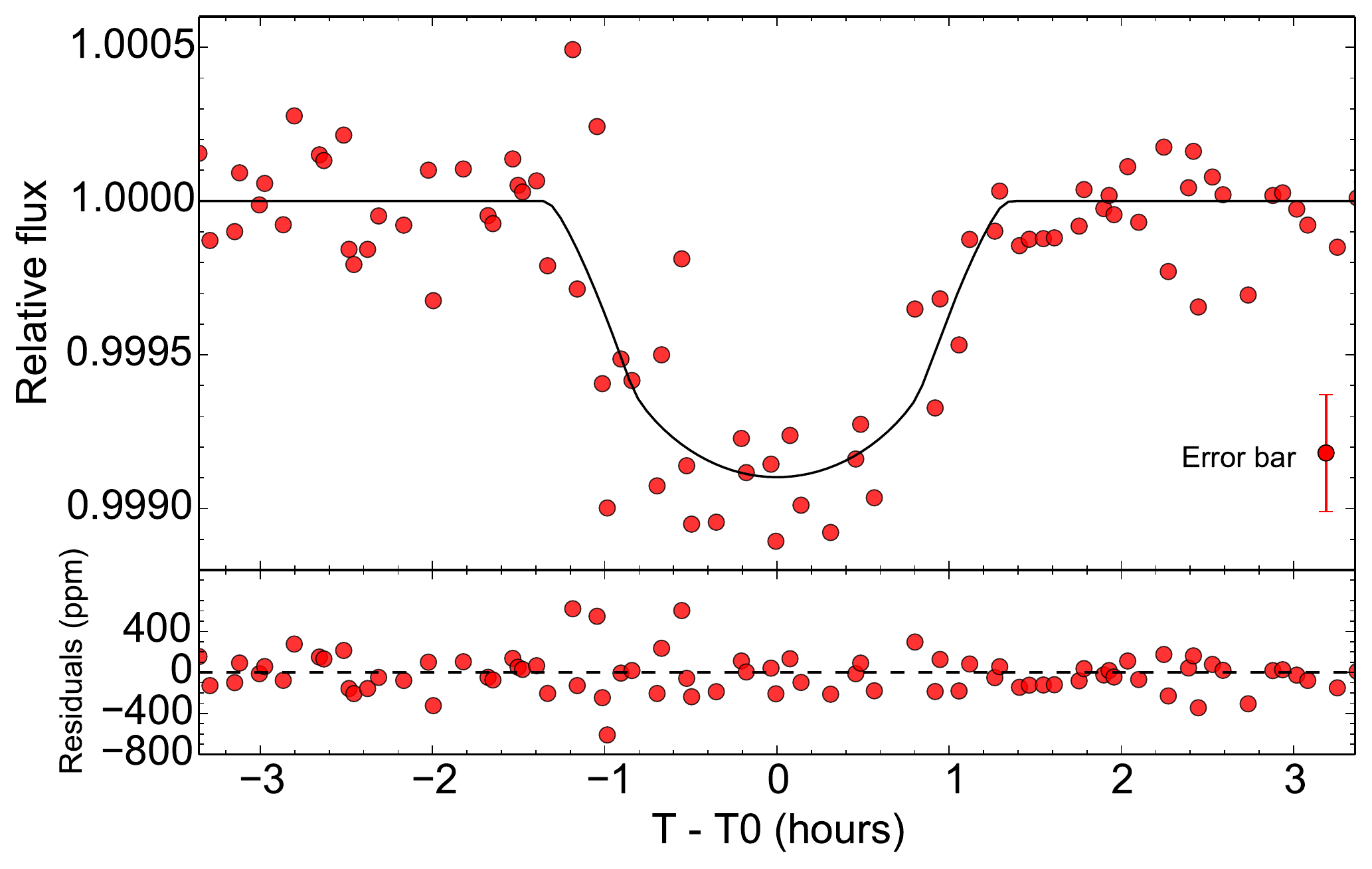}
\caption*{\quad \quad \, \, K2-322 b (EPIC 248558190 b)}
\end{subfigure}

\begin{subfigure}{8.1cm}
\includegraphics[width=8.1cm]{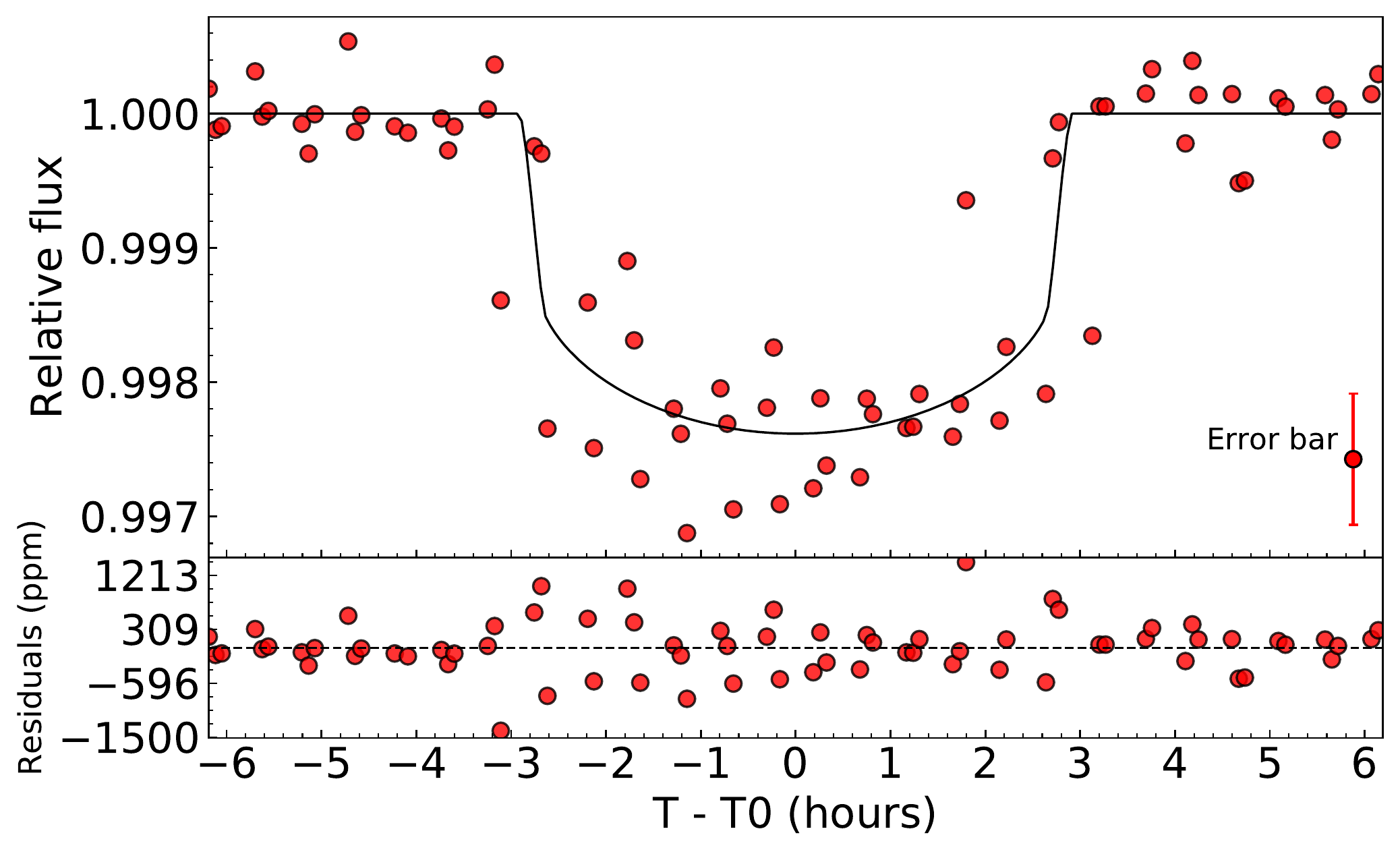}
\caption*{\quad \quad \, \, K2-323 b (EPIC 248616368 b)}
\end{subfigure}%
\begin{subfigure}{8cm}
\includegraphics[width=7.9cm]{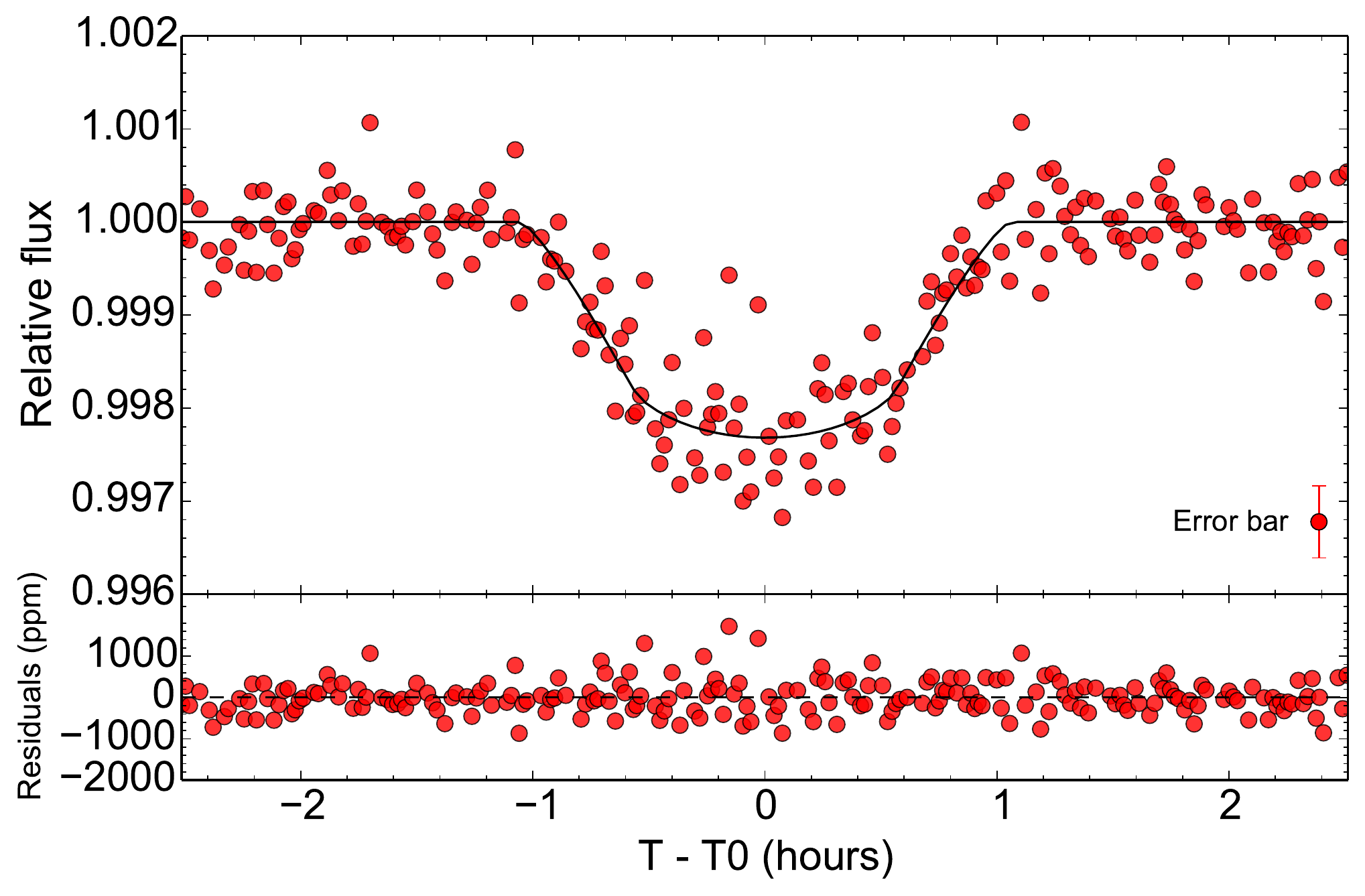}
\caption*{\quad \quad \, \, K2-324 b (EPIC 248639308 b)}
\end{subfigure}

\begin{subfigure}{8cm}
\includegraphics[width=8.0cm]{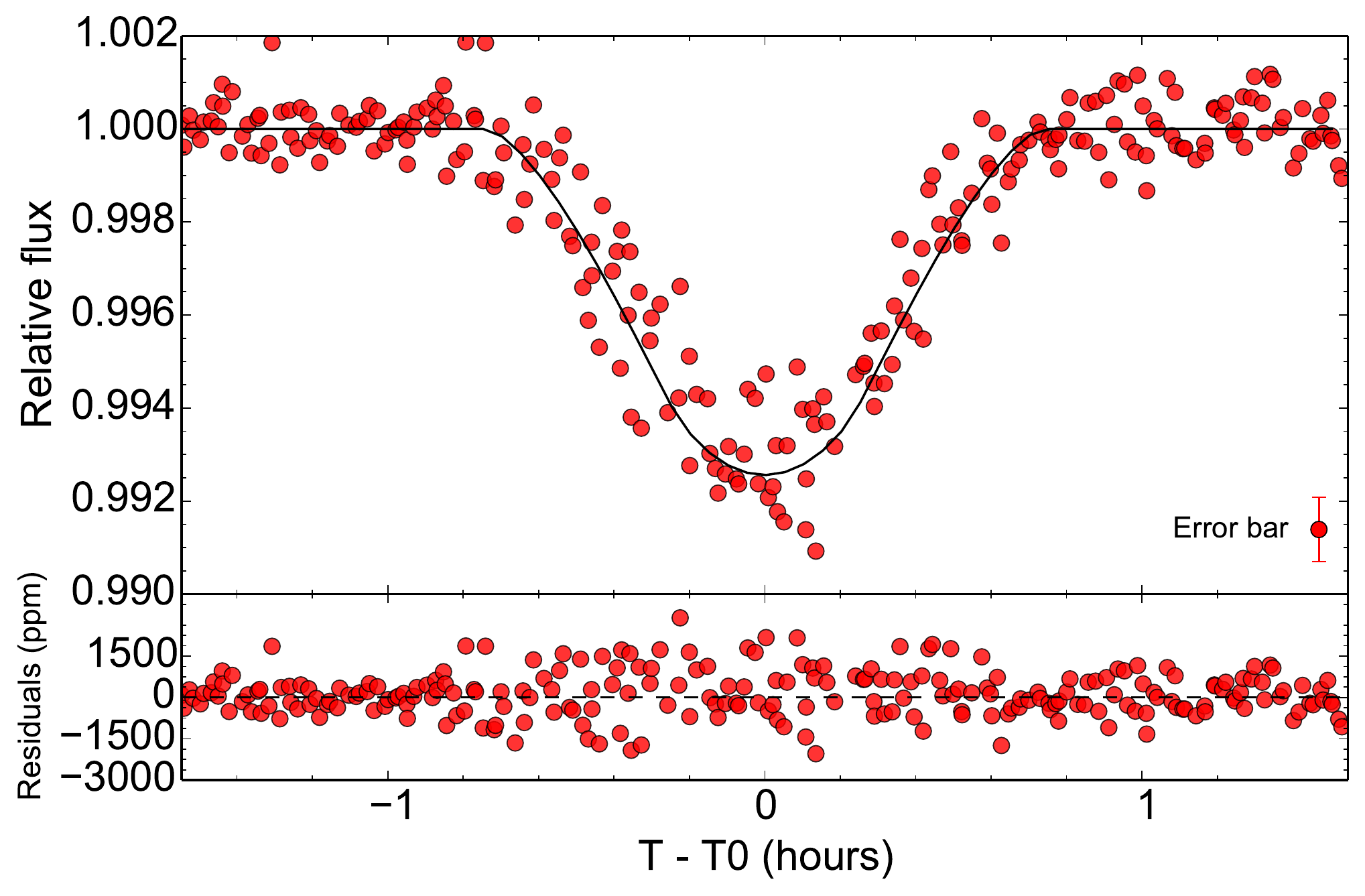}
\caption*{\quad \quad \, \, \, \, EPIC 248775938 b}
\end{subfigure}%
\begin{subfigure}{8cm}
\includegraphics[width=8.0cm]{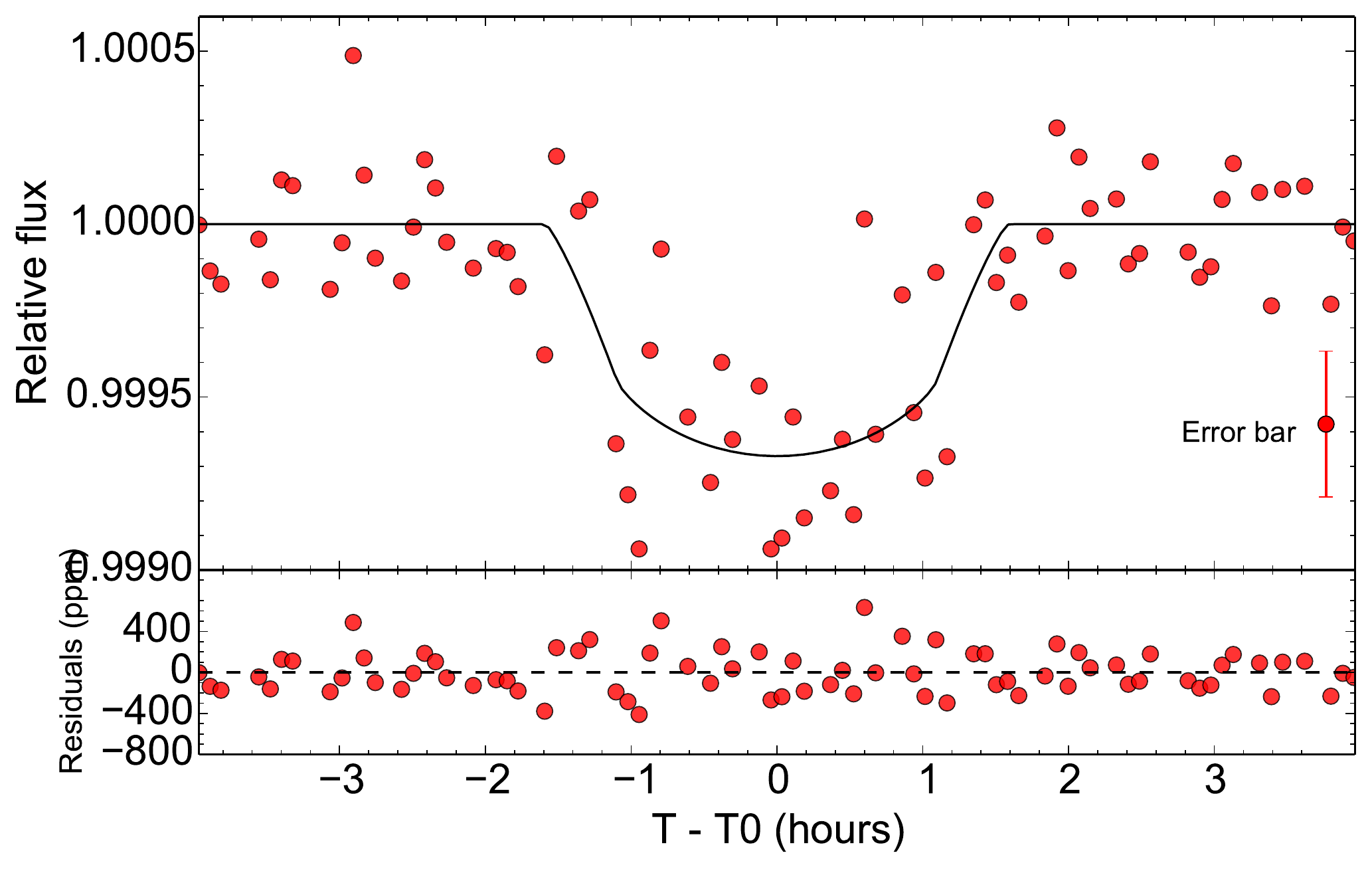}
\caption*{\quad \quad \, \, \, \, EPIC 248782482 b}
\end{subfigure}

\begin{subfigure}{8cm}
\includegraphics[width=8.0cm]{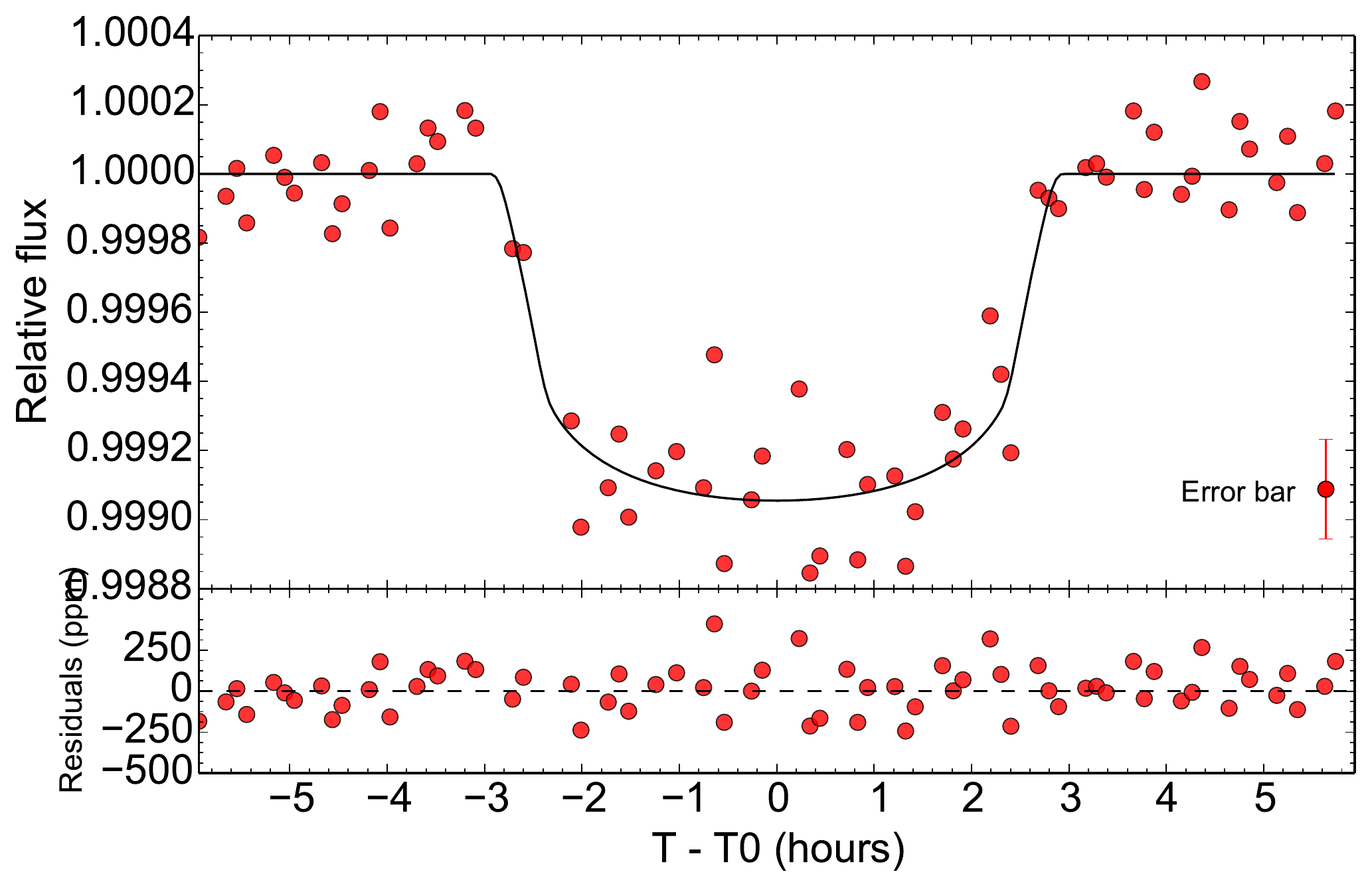}
\caption*{\quad \quad \, \, K2-319 b (EPIC 201663879 b)}
\end{subfigure}%
\begin{subfigure}{8cm}
\centering\includegraphics[width=8.0cm]{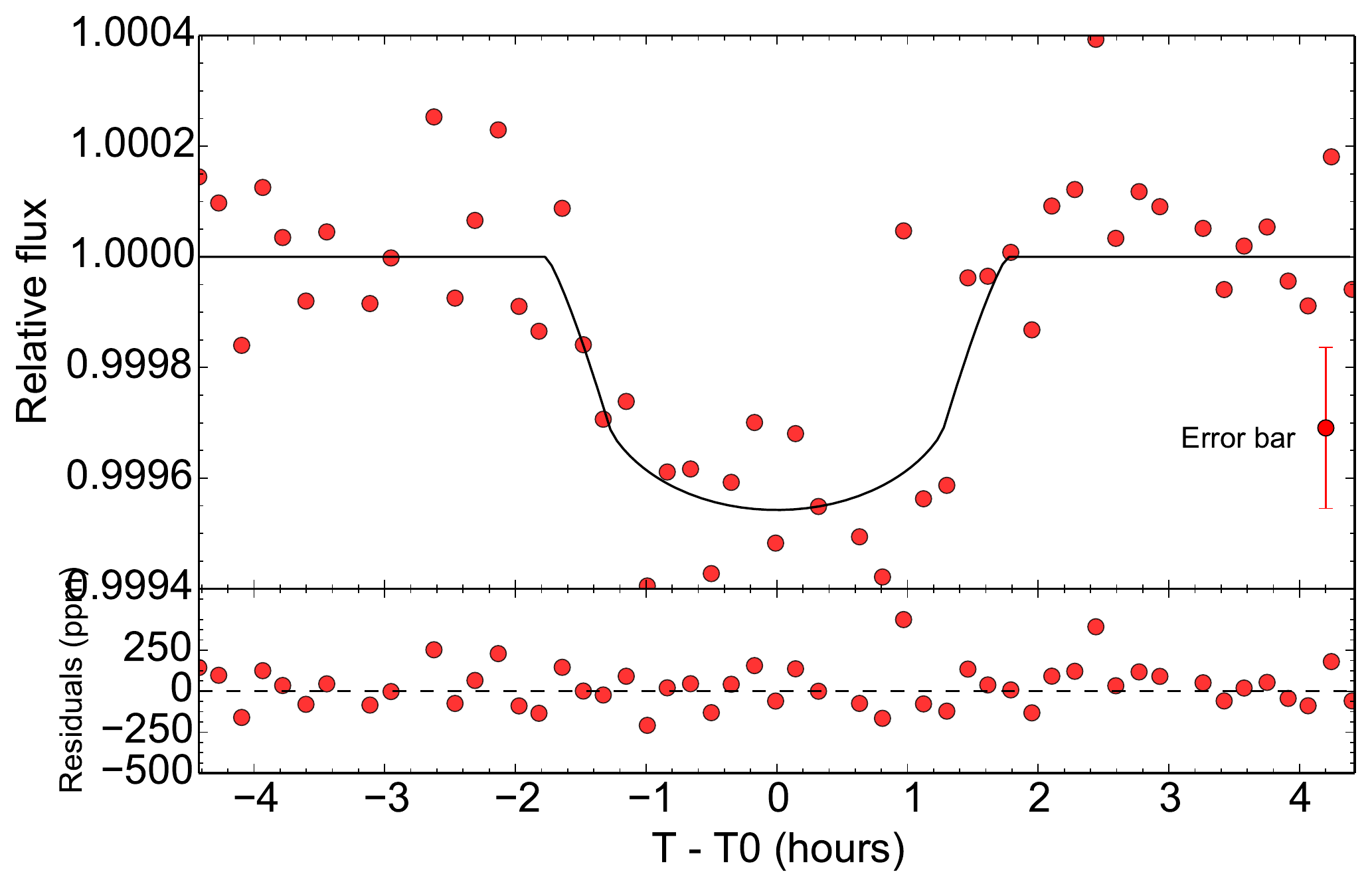}
\caption*{\quad \quad \, \, \,\,\, EPIC 201663879 c}
\end{subfigure}

\caption{Phase-folded transits and best fit for planetary candidates of campaign 14.}
\label{fig:phase_folded2}
\end{figure*}


\begin{figure*}
\centering
\begin{subfigure}{8cm}
\centering\includegraphics[width=6.5cm]{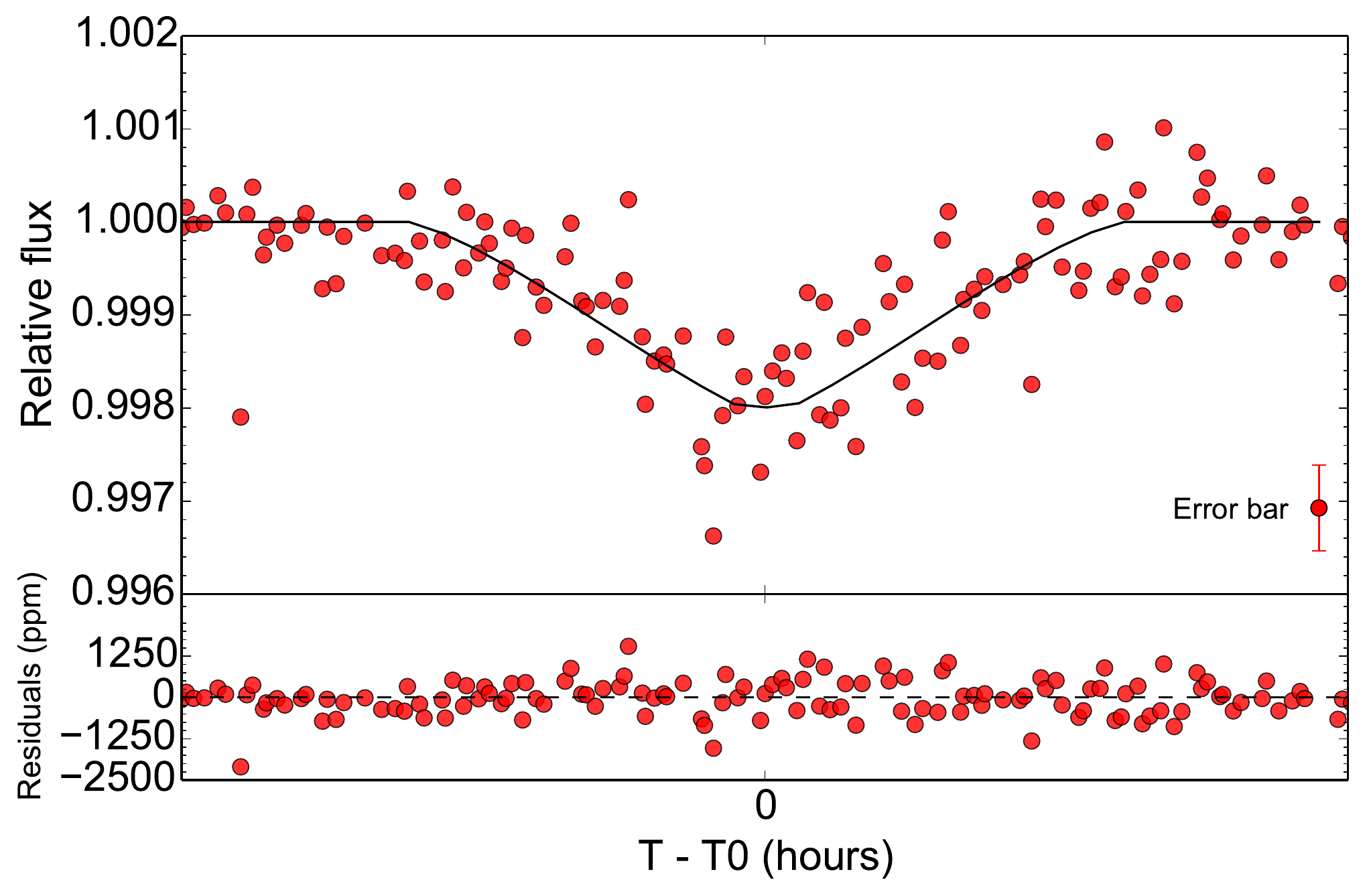}
\caption*{\quad \quad \, \, EPIC 246909566 b (C13)}
\end{subfigure}%
\begin{subfigure}{8cm}
\centering\includegraphics[width=6.5cm]{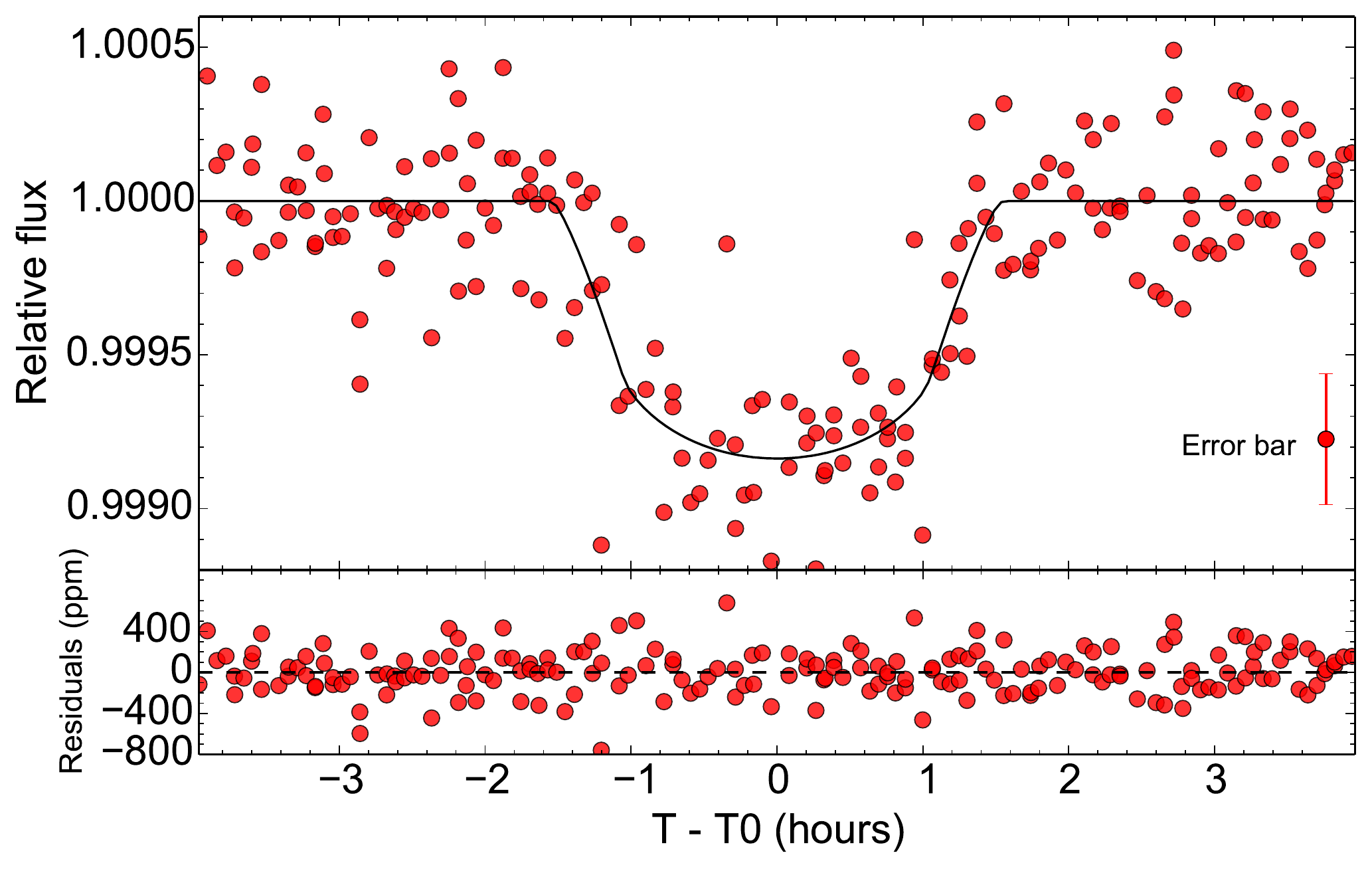}
\caption*{\quad \quad \, \, EPIC 2485944983 b}
\end{subfigure}%

\begin{subfigure}{8cm}
\centering\includegraphics[width=6.5cm]{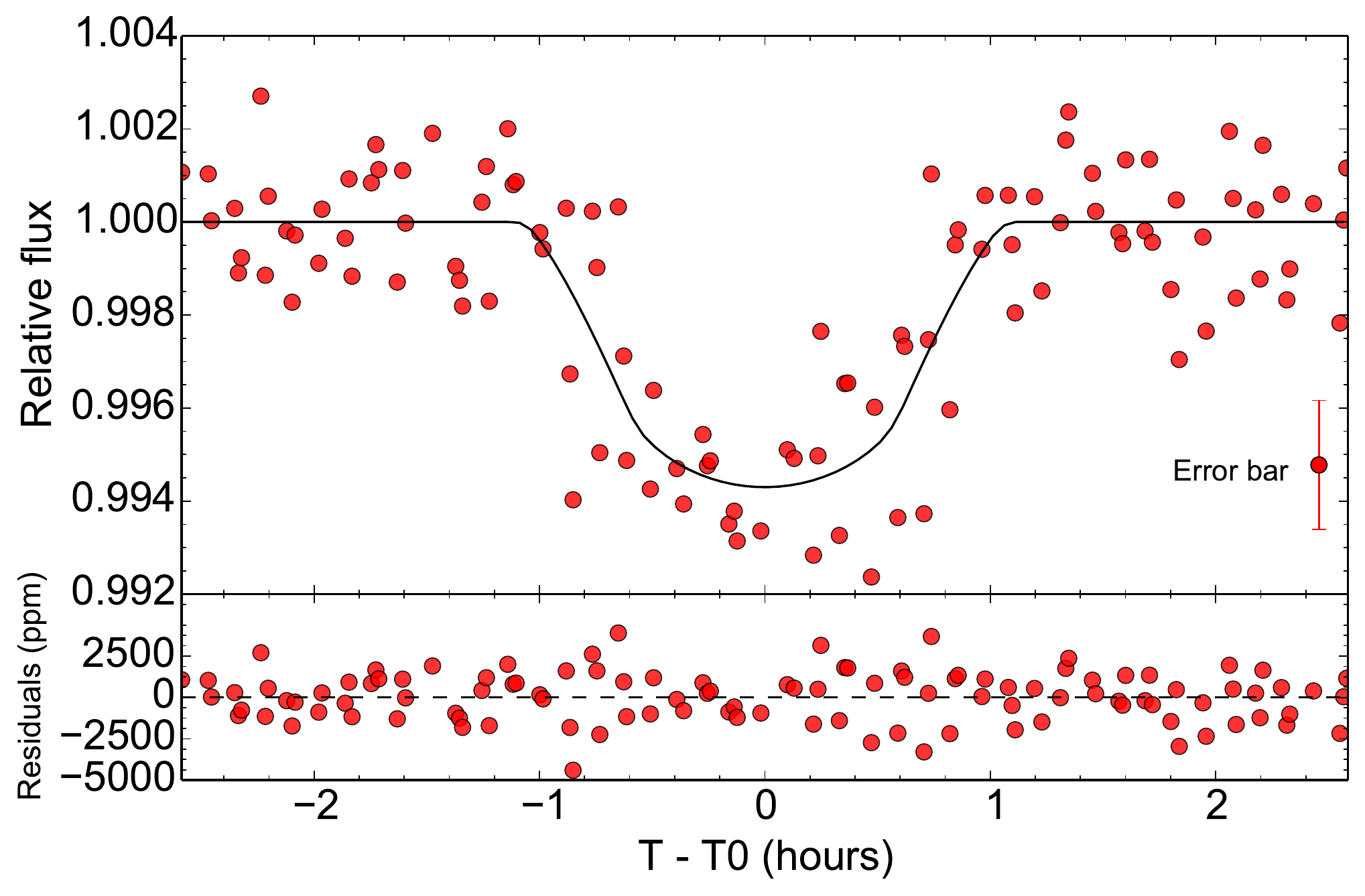}
\caption*{\quad \quad \, \, K2-325 b (EPIC 246074965 b)}
\end{subfigure}
\begin{subfigure}{8cm}
\centering\includegraphics[width=6.5cm]{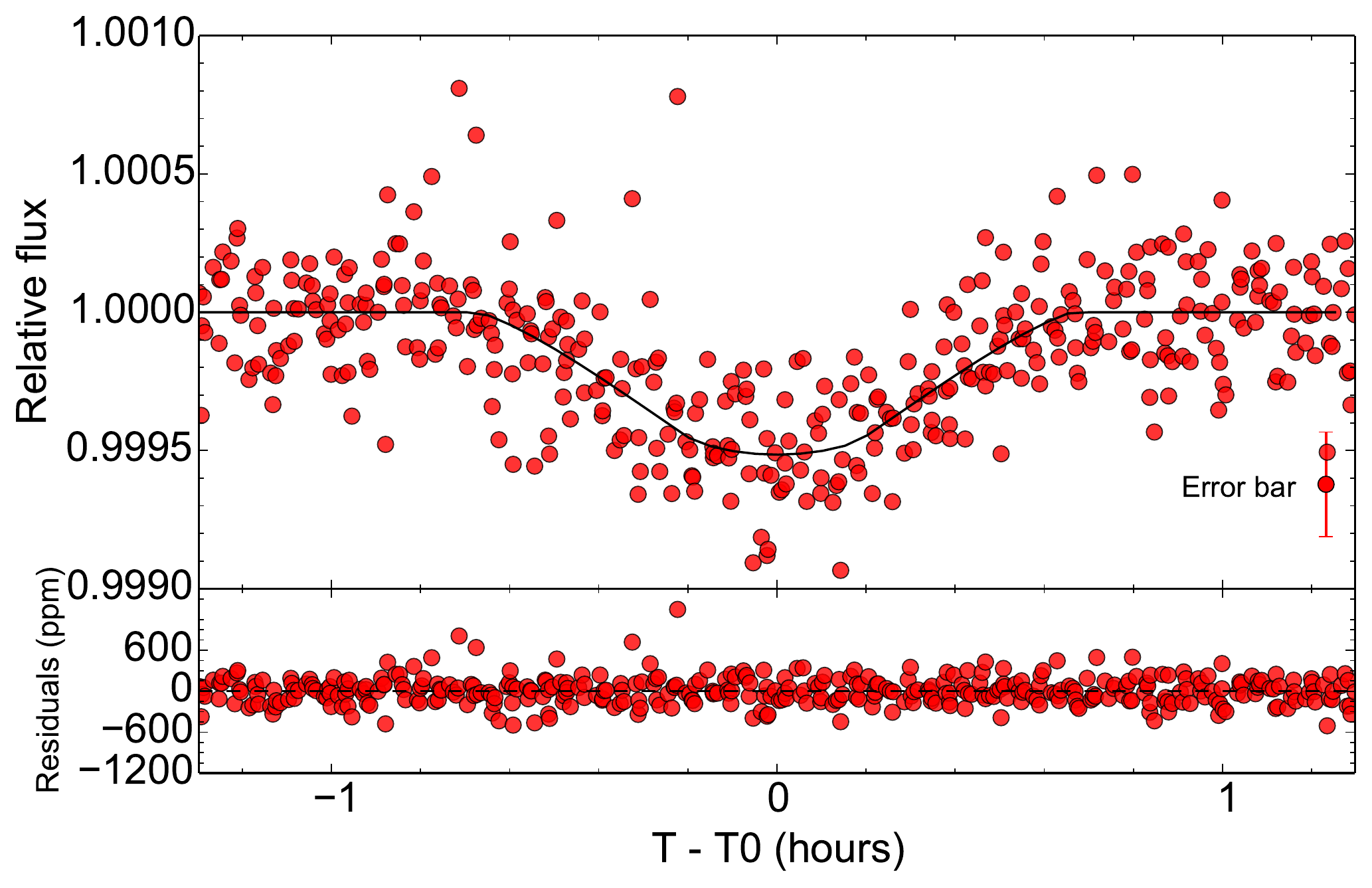}
\caption*{\quad \quad \, \, \,\, EPIC 246163416 b}
\end{subfigure}

\begin{subfigure}{8cm}
\centering\includegraphics[width=6.5cm]{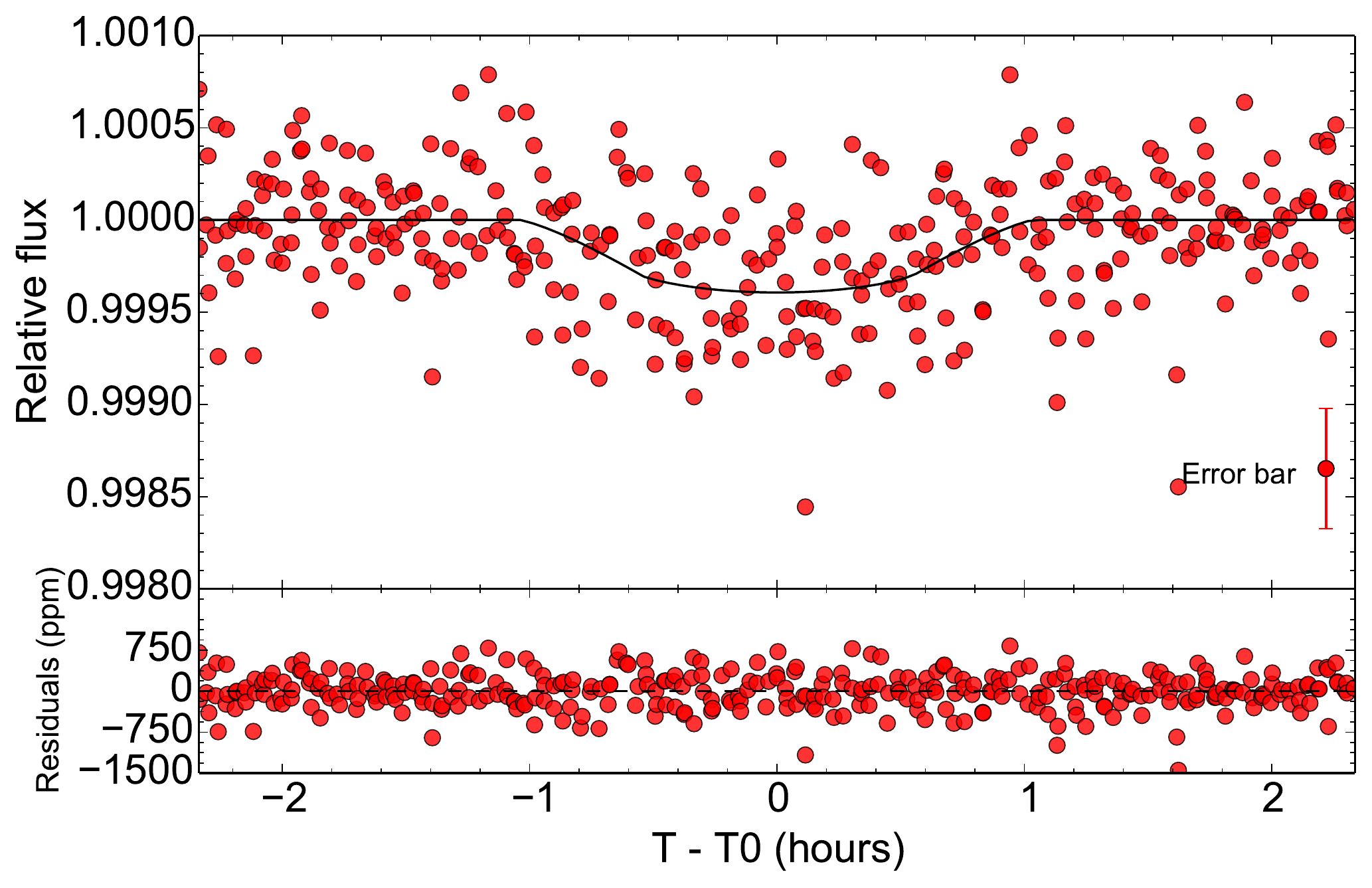}
\caption*{\quad \quad \, \, \,\, EPIC 246313886 b}
\end{subfigure}
\begin{subfigure}{8cm}
\centering\includegraphics[width=6.5cm]{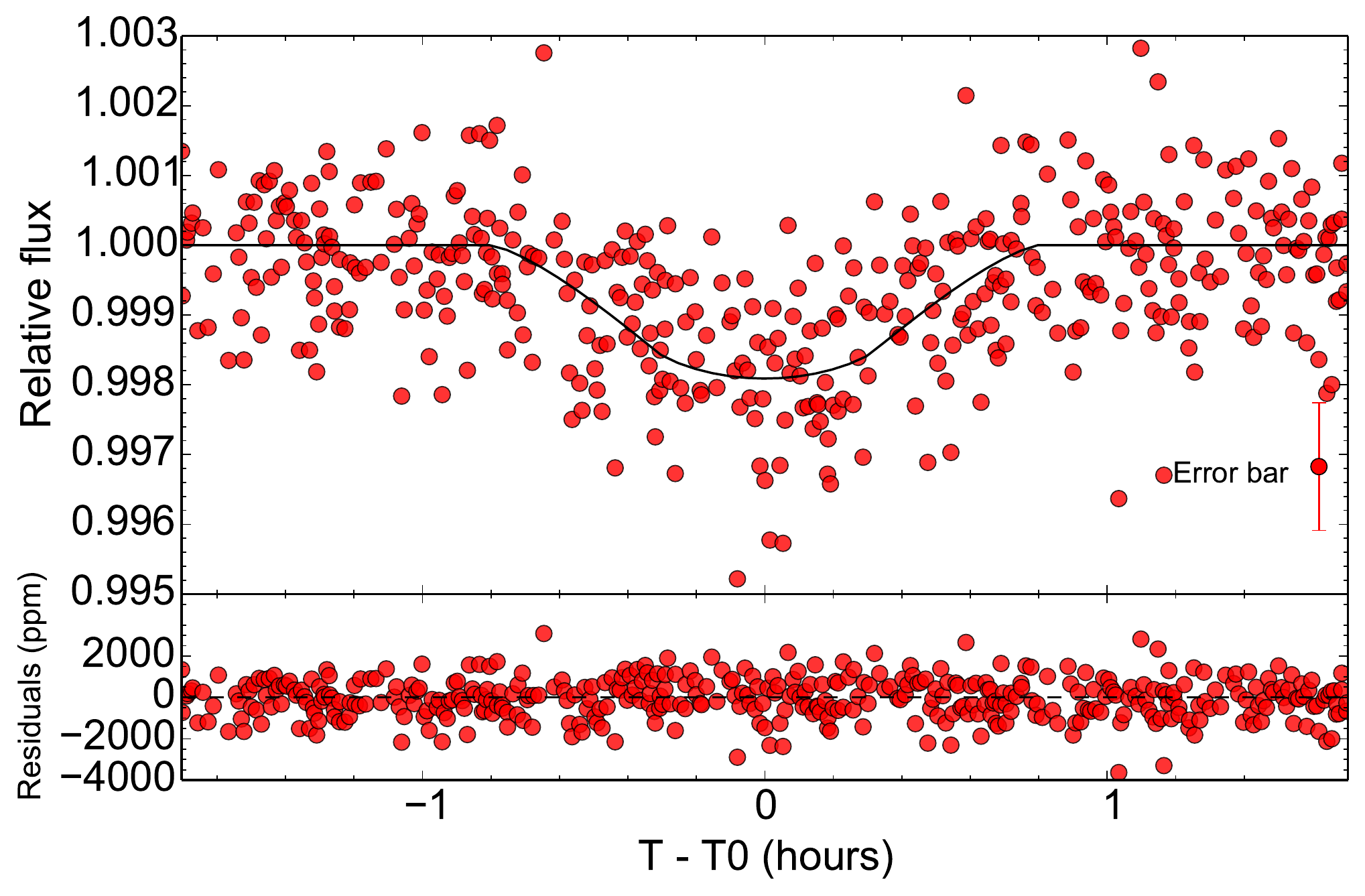}
\caption*{\quad \quad \, \, \,\, EPIC 246331347 b}
\end{subfigure}%

\begin{subfigure}{8cm}
\centering\includegraphics[width=6.5cm]{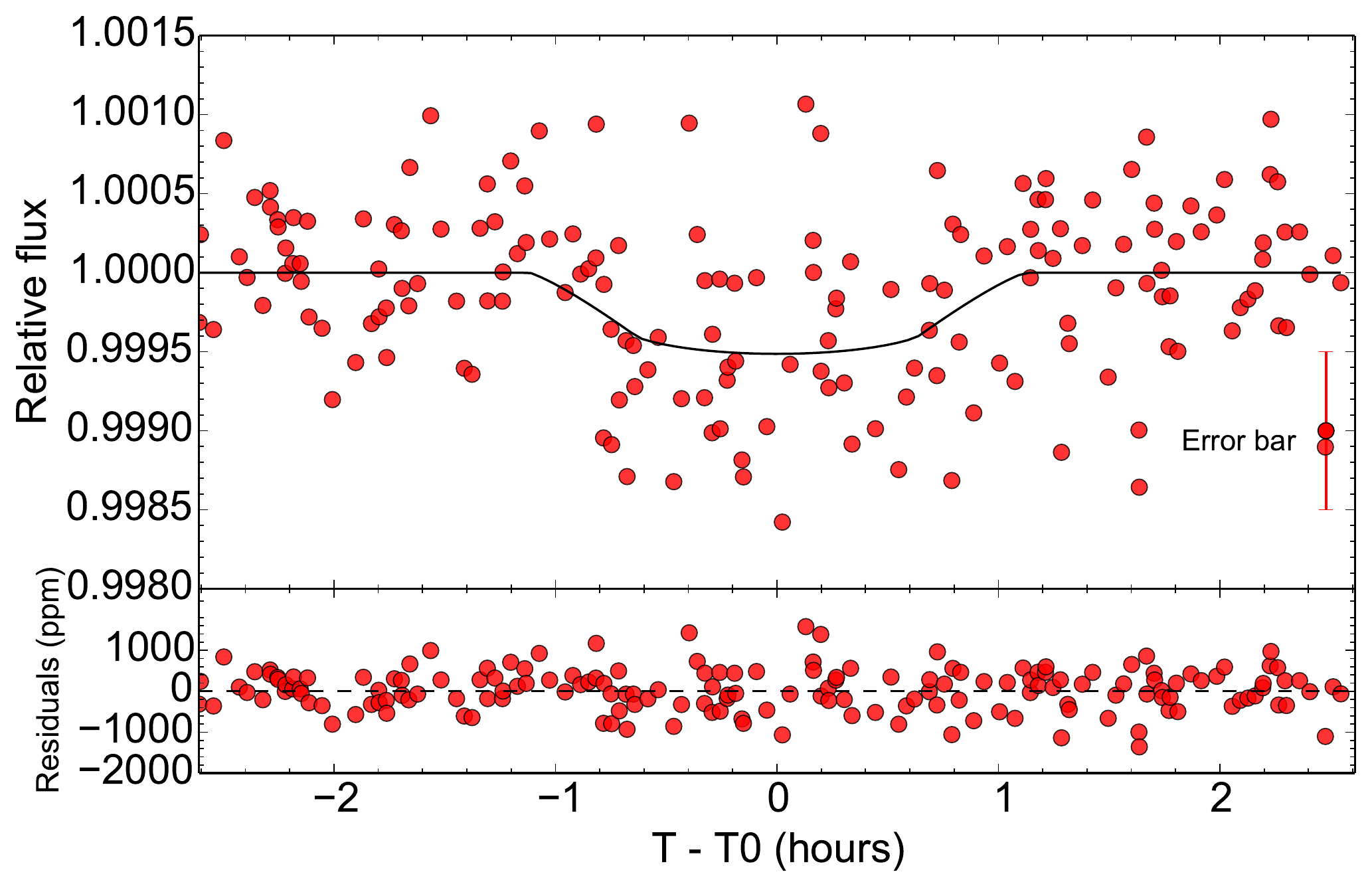}
\caption*{\quad \quad \, \, \,\, EPIC 246331418 b}
\end{subfigure}
\begin{subfigure}{8cm}
\centering\includegraphics[width=6.5cm]{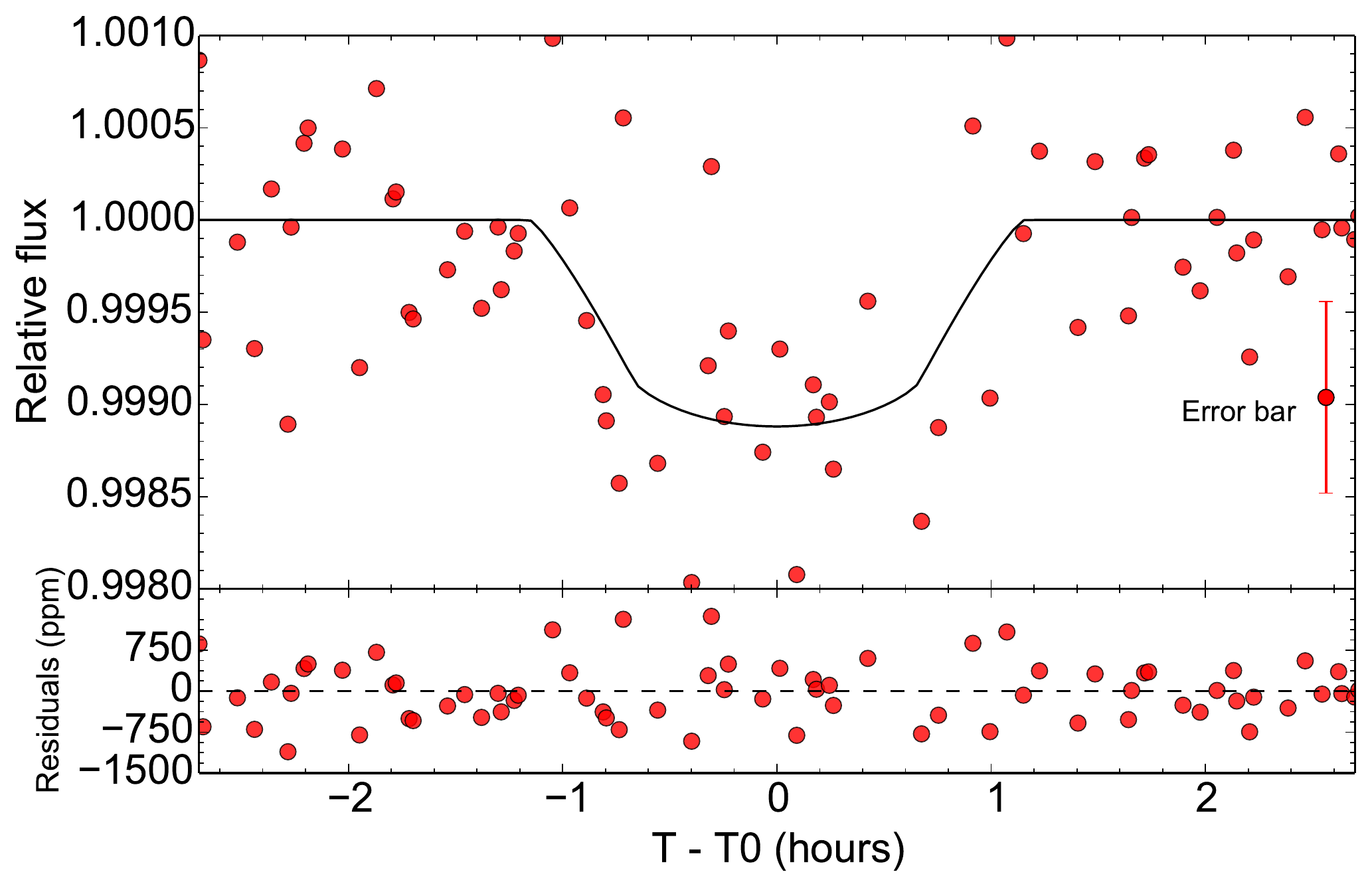}
\caption*{\quad \quad \, \, \,\, EPIC 246331418 c}
\end{subfigure}%

\begin{subfigure}{8cm}
\centering\includegraphics[width=6.5cm]{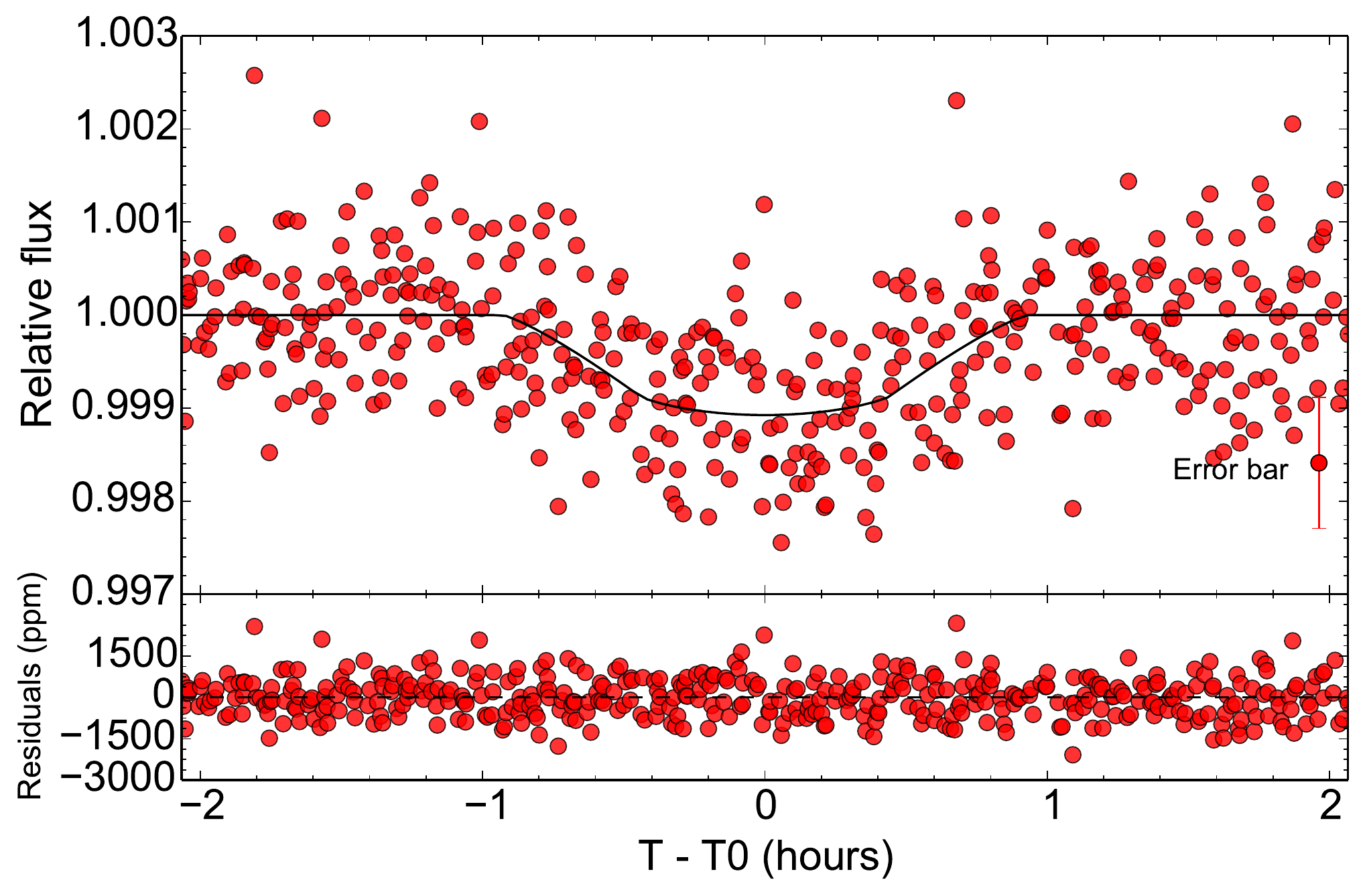}
\caption*{\quad \quad \, \, K2-326 b (EPIC 246472939 b)}
\end{subfigure}%

\caption{Phase-folded transits for candidates of campaign 12 and 13 (EPIC 246909566b).}
\label{fig:phase_folded3}
\end{figure*}

\begin{figure*}
\centering

\begin{subfigure}{9cm}
\centering\includegraphics[width=9cm]{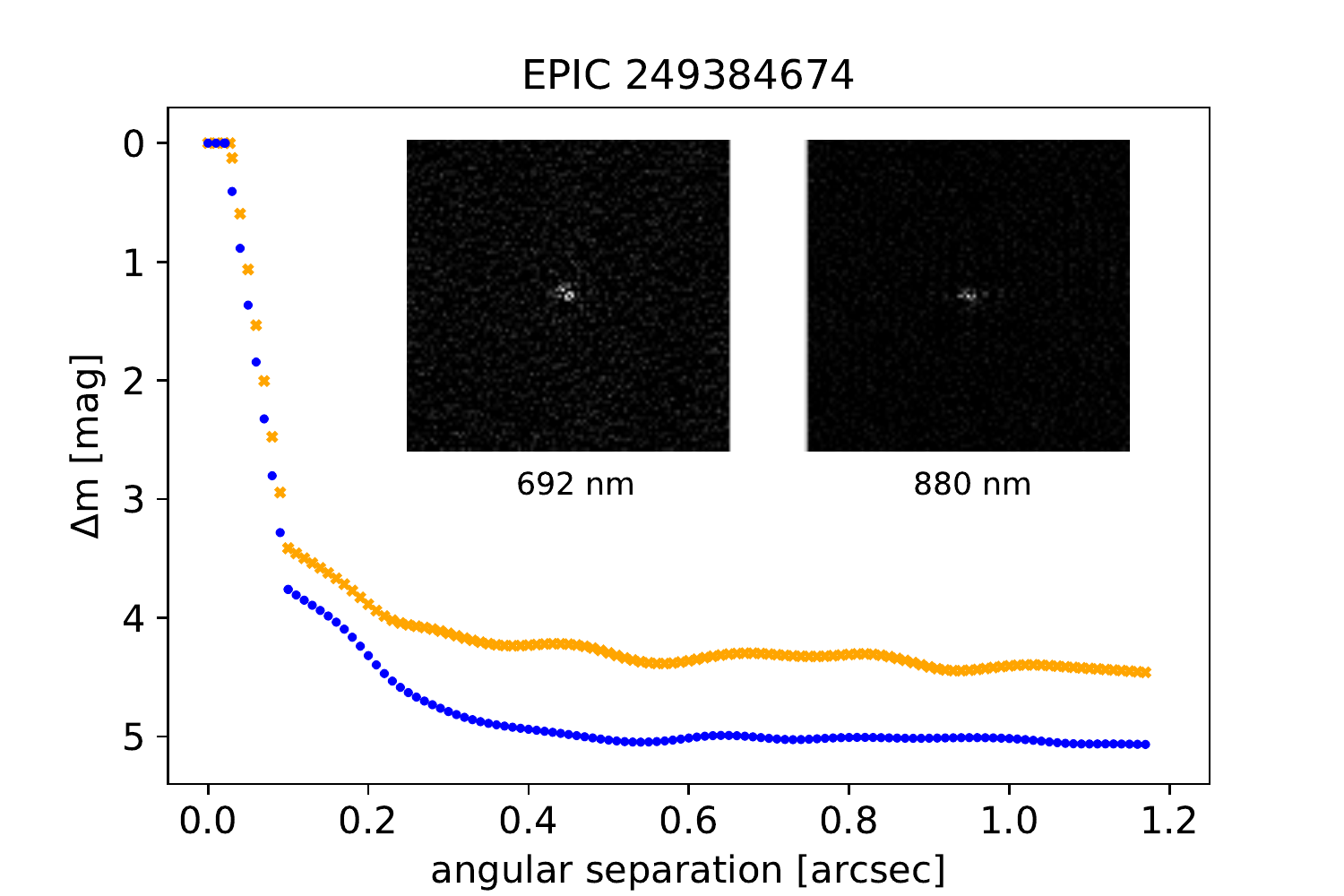}
\end{subfigure}%
\begin{subfigure}{9cm}
\centering\includegraphics[width=9cm]{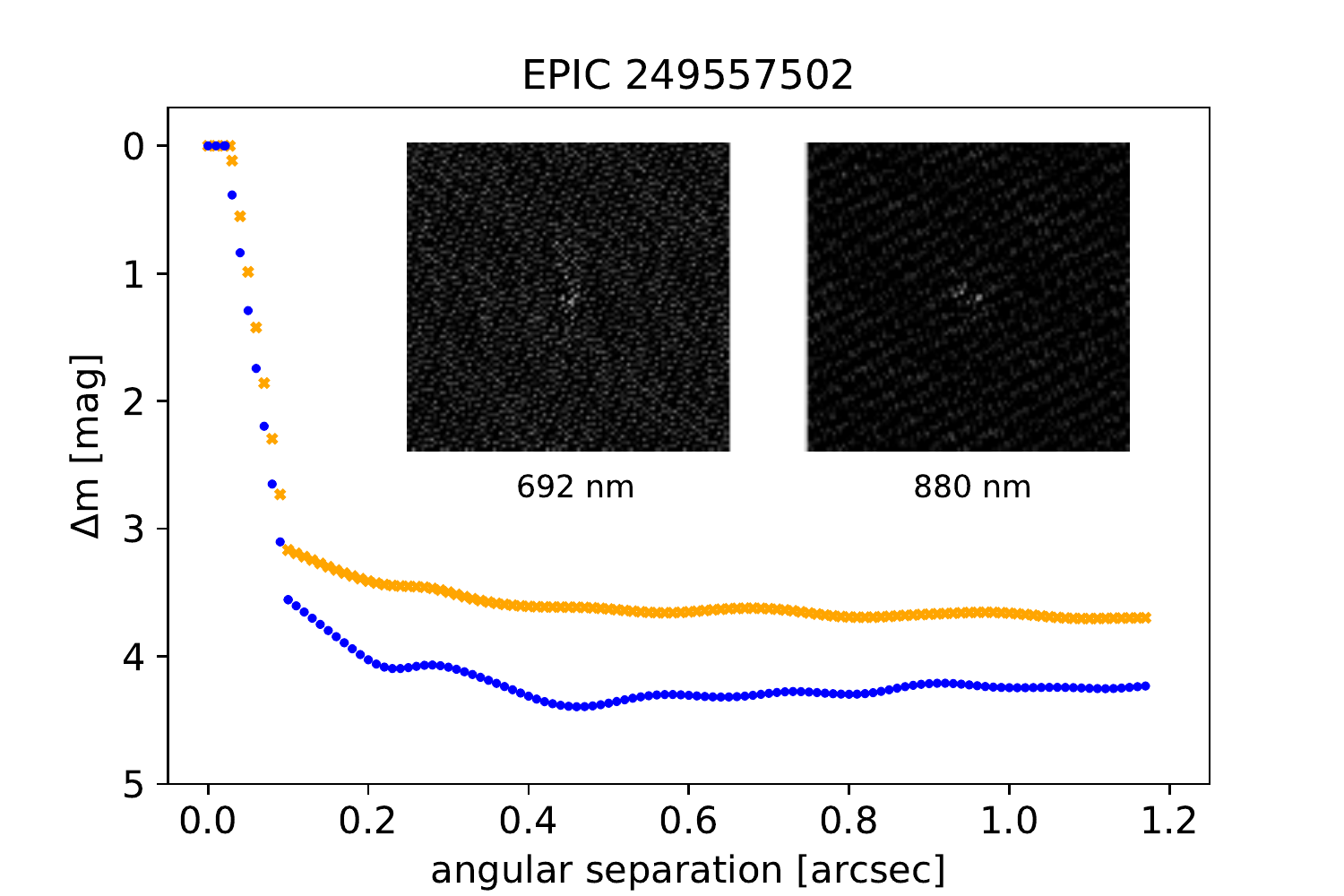}
\end{subfigure}

\begin{subfigure}{9cm}
\centering\includegraphics[width=9cm]{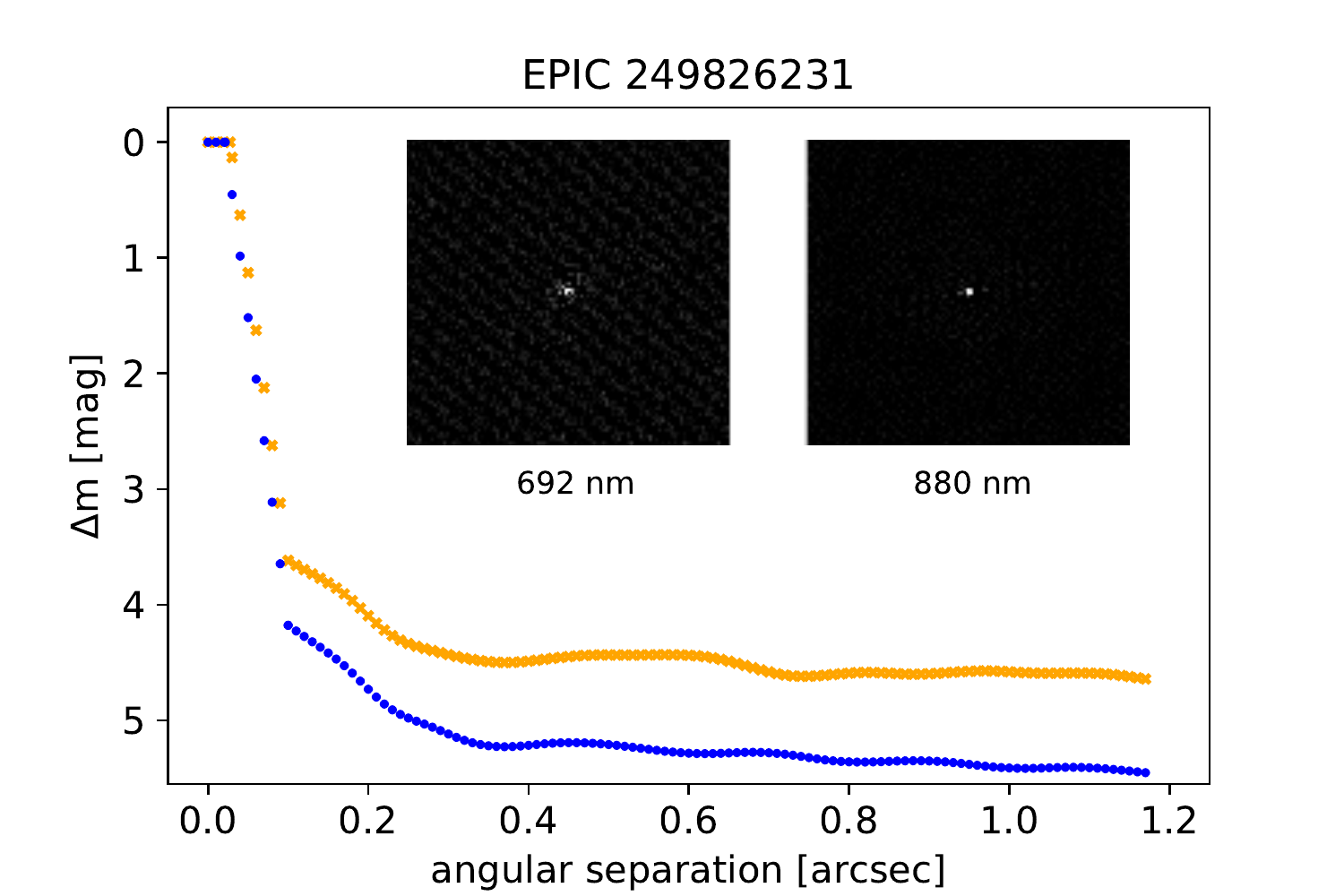}
\end{subfigure}%
\begin{subfigure}{9cm}
\centering\includegraphics[width=9cm]{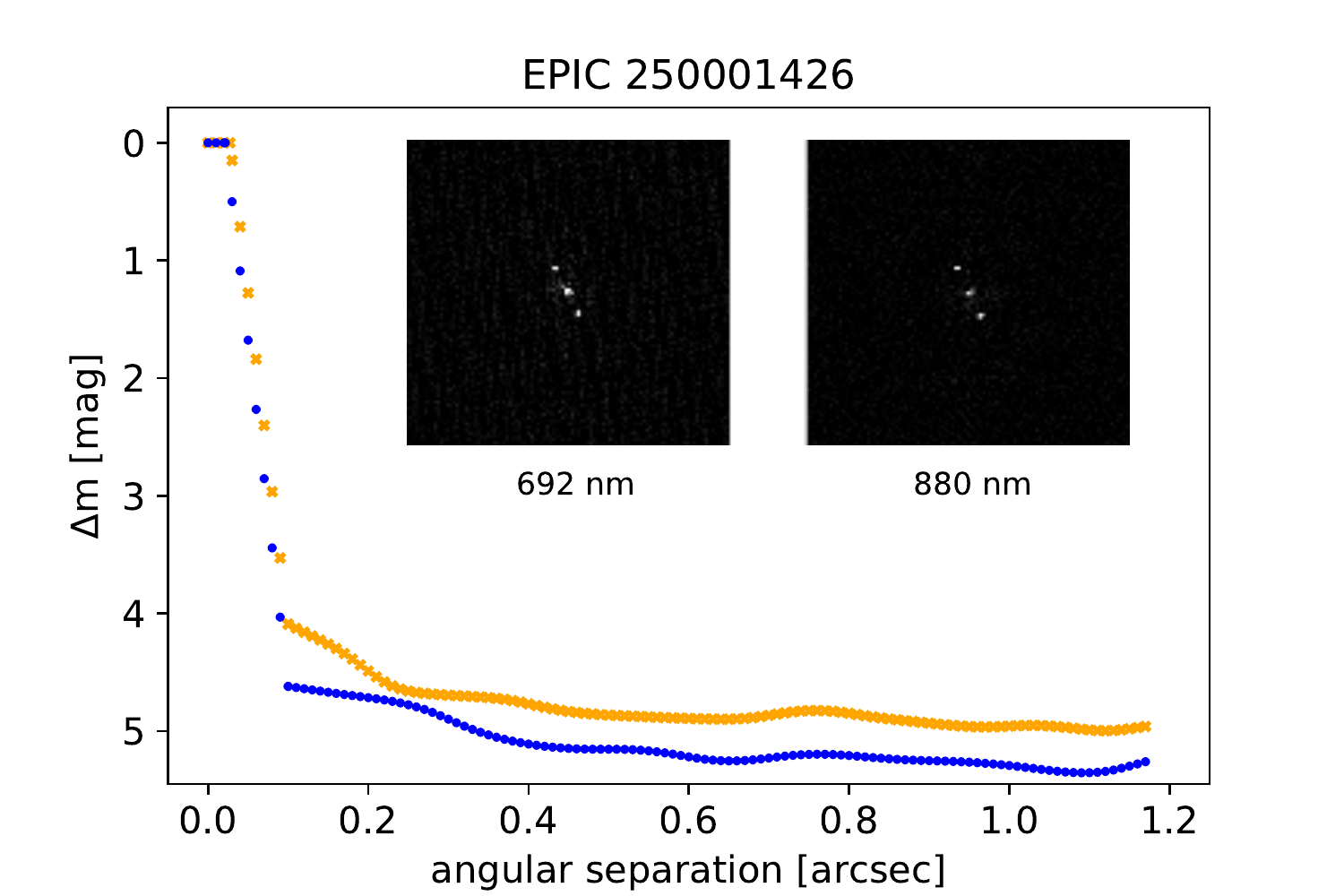}
\end{subfigure}%

\begin{subfigure}{9cm}
\centering\includegraphics[width=9cm]{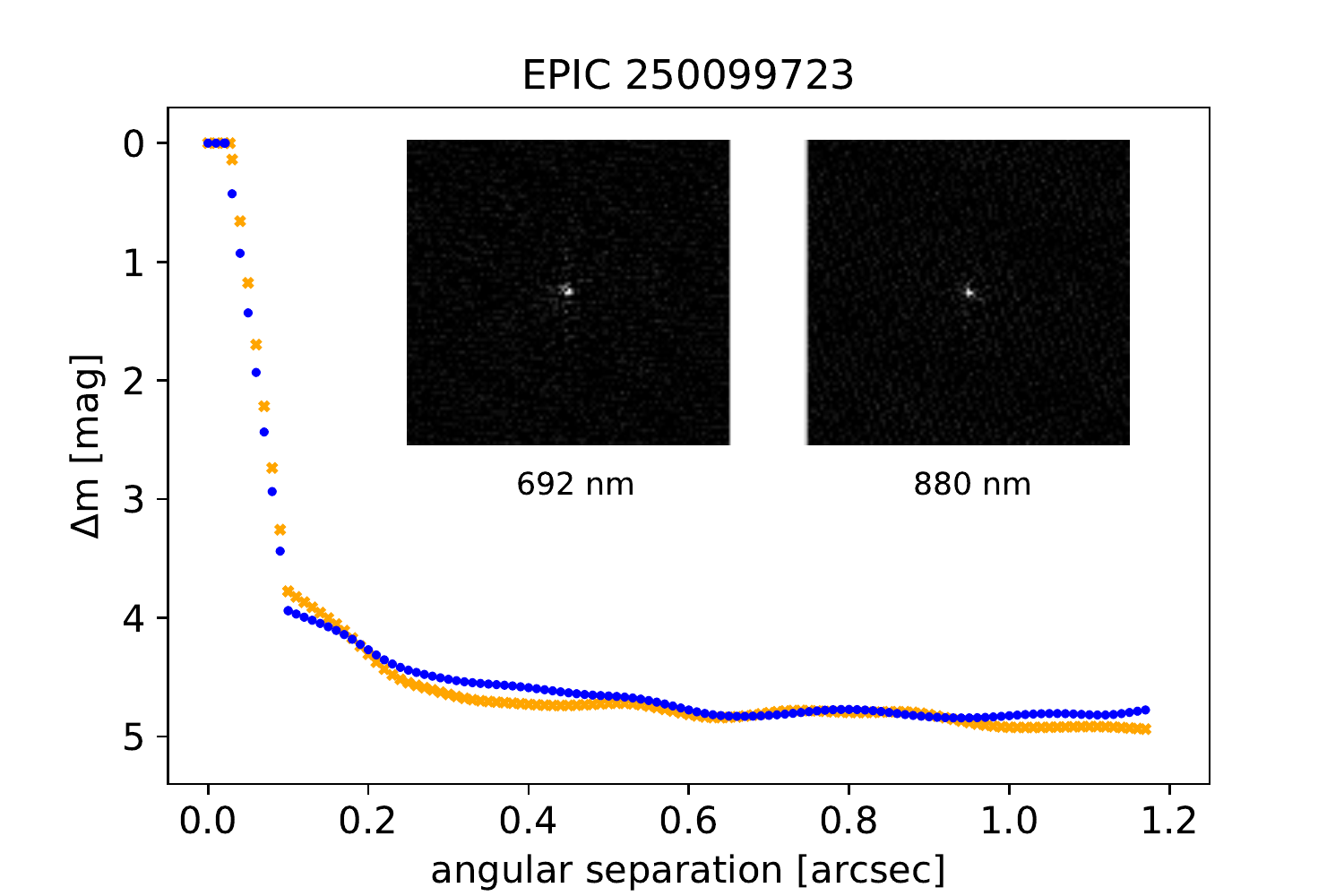}
\end{subfigure}%
\begin{subfigure}{9cm}
\centering\includegraphics[width=9cm]{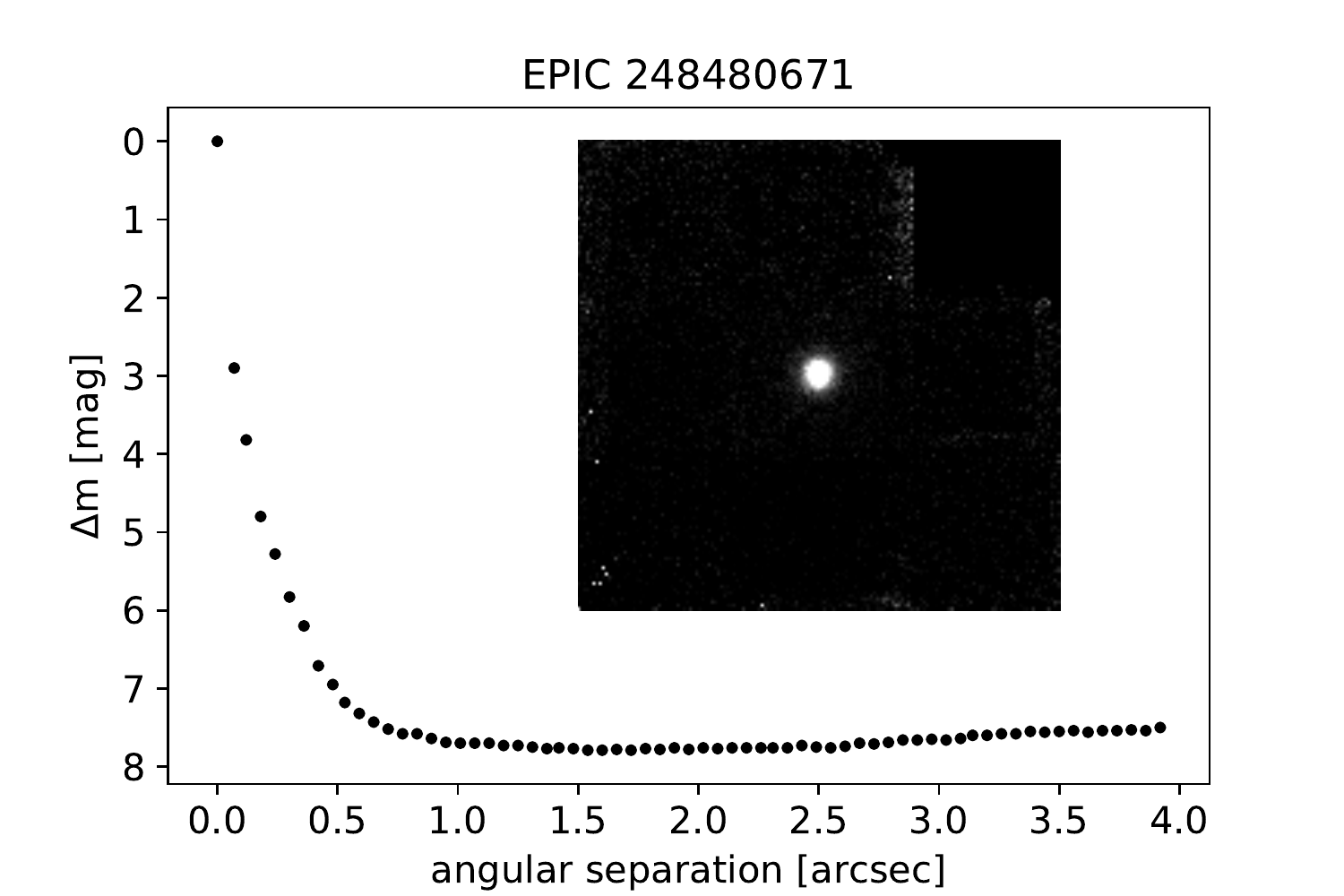}
\end{subfigure}%
\caption{Speckle images and background plots for 5 stars from C15 and an AO image for EPIC 248480671 (C14). The 5 images from C15 stars have been acquired at GeminiS-8m telescope at 880 nm (orange crosses) and 692 nm (blue dots) in a field of 3 x 3 arc seconds. The AO image for EPIC 248480671, was acquired with NIRC2 at Keck-10m telescope (field 8 x 8 arc seconds).}
 \label{fig:AO_1}
\end{figure*}

\begin{figure*}
\centering

\begin{subfigure}{9cm}
\centering\includegraphics[width=9cm]{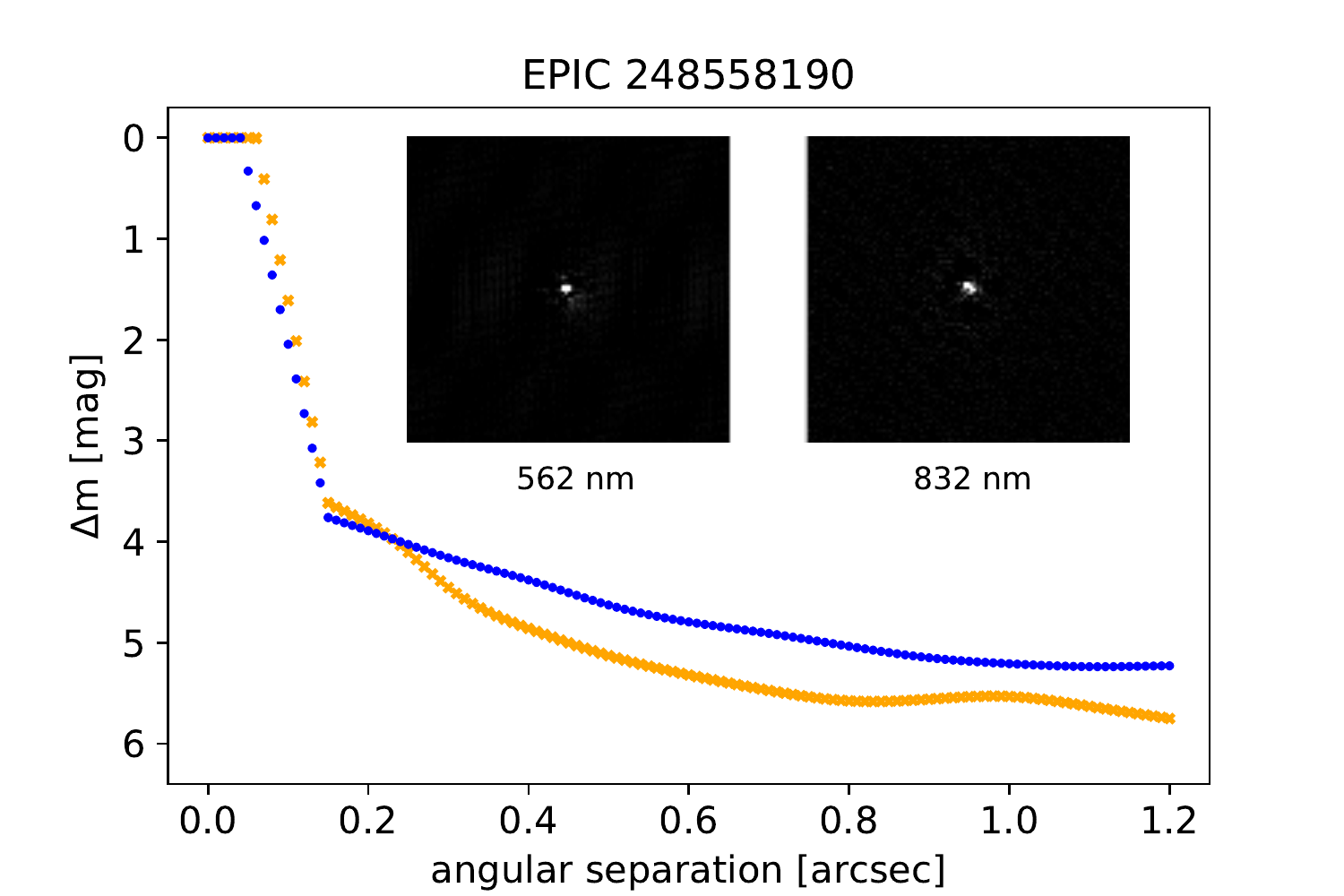}
\end{subfigure}%
\begin{subfigure}{9cm}
\centering\includegraphics[width=9cm]{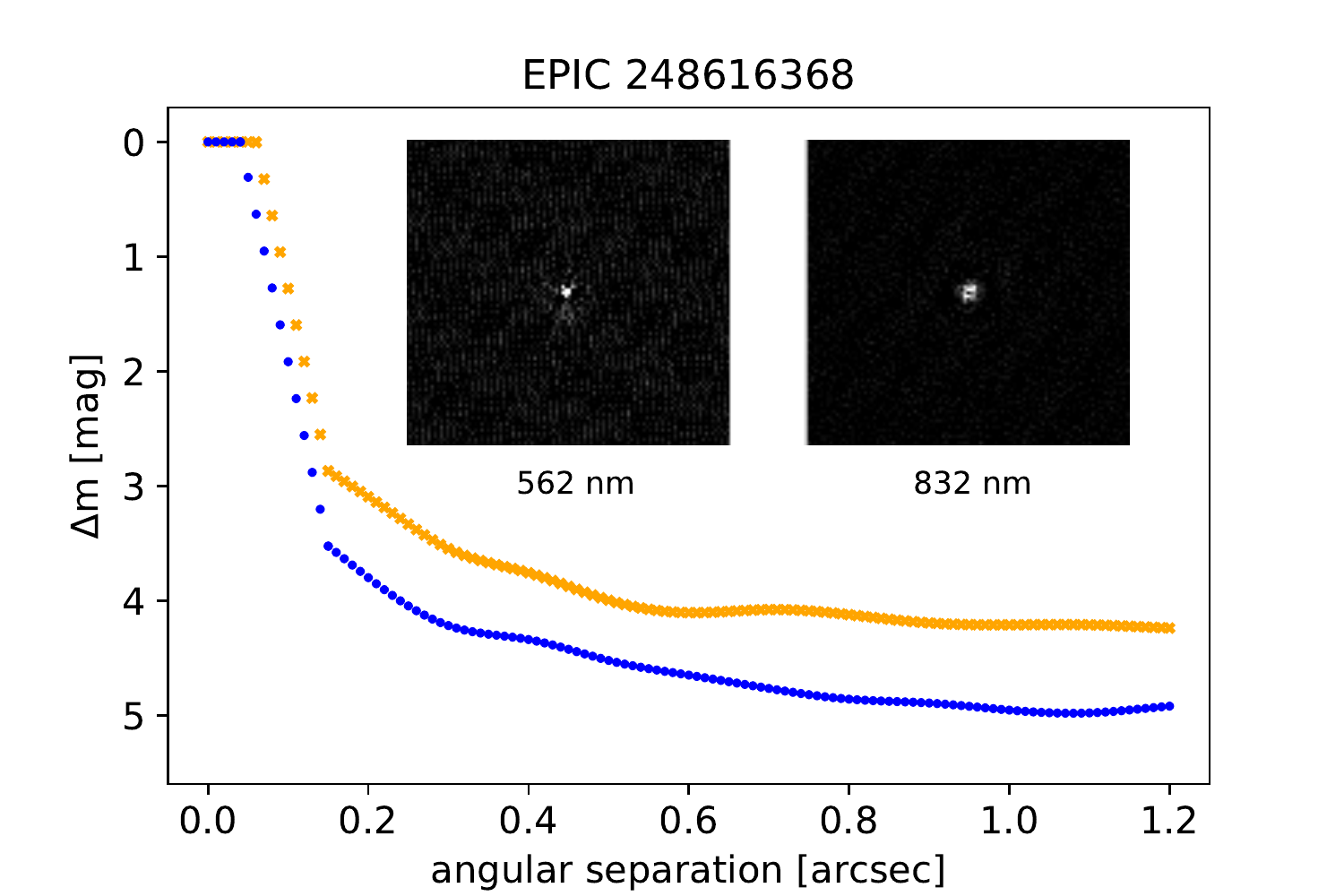}
\end{subfigure}%

\begin{subfigure}{9cm}
\centering\includegraphics[width=9cm]{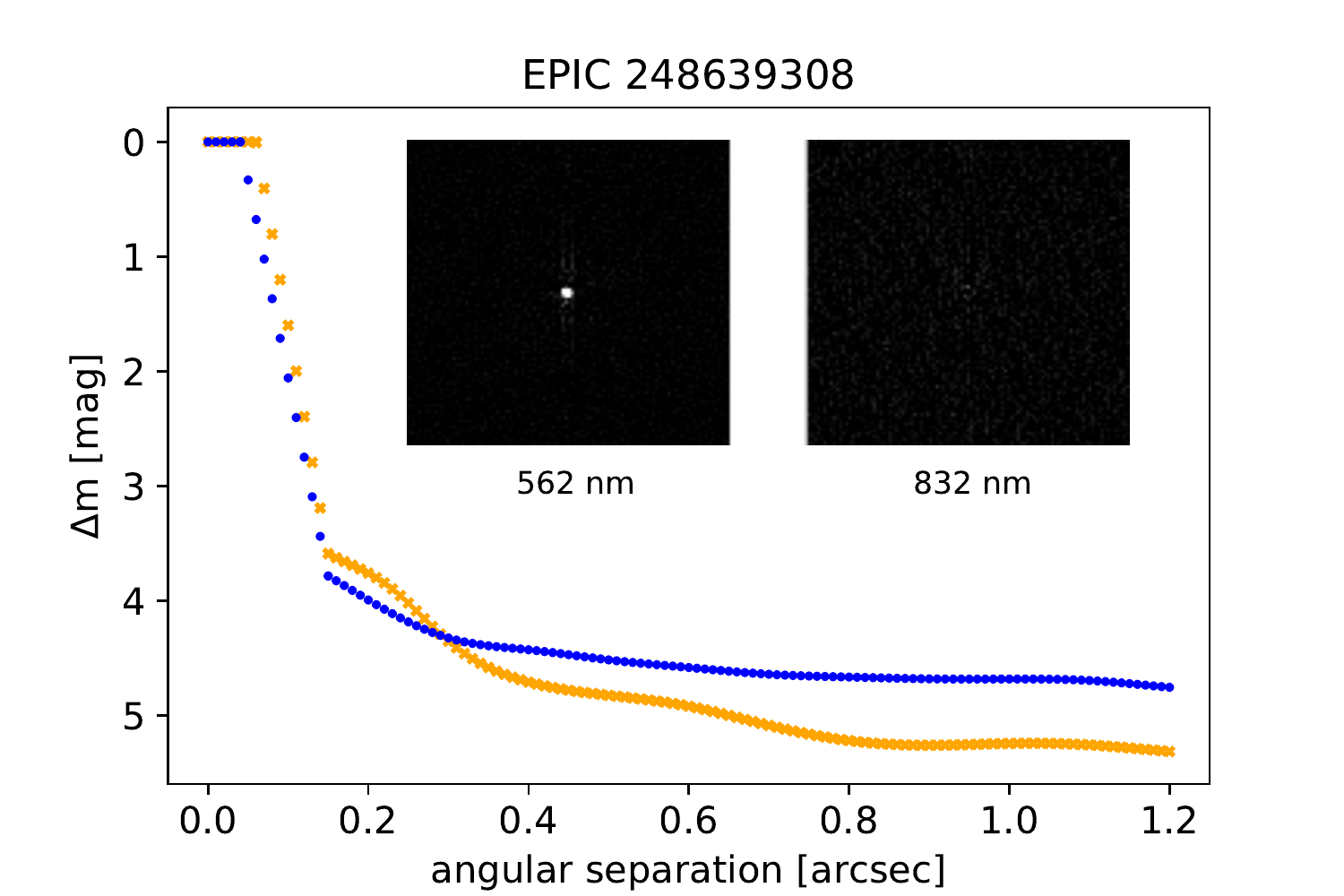}
\end{subfigure}%
\begin{subfigure}{9cm}
\centering\includegraphics[width=9cm]{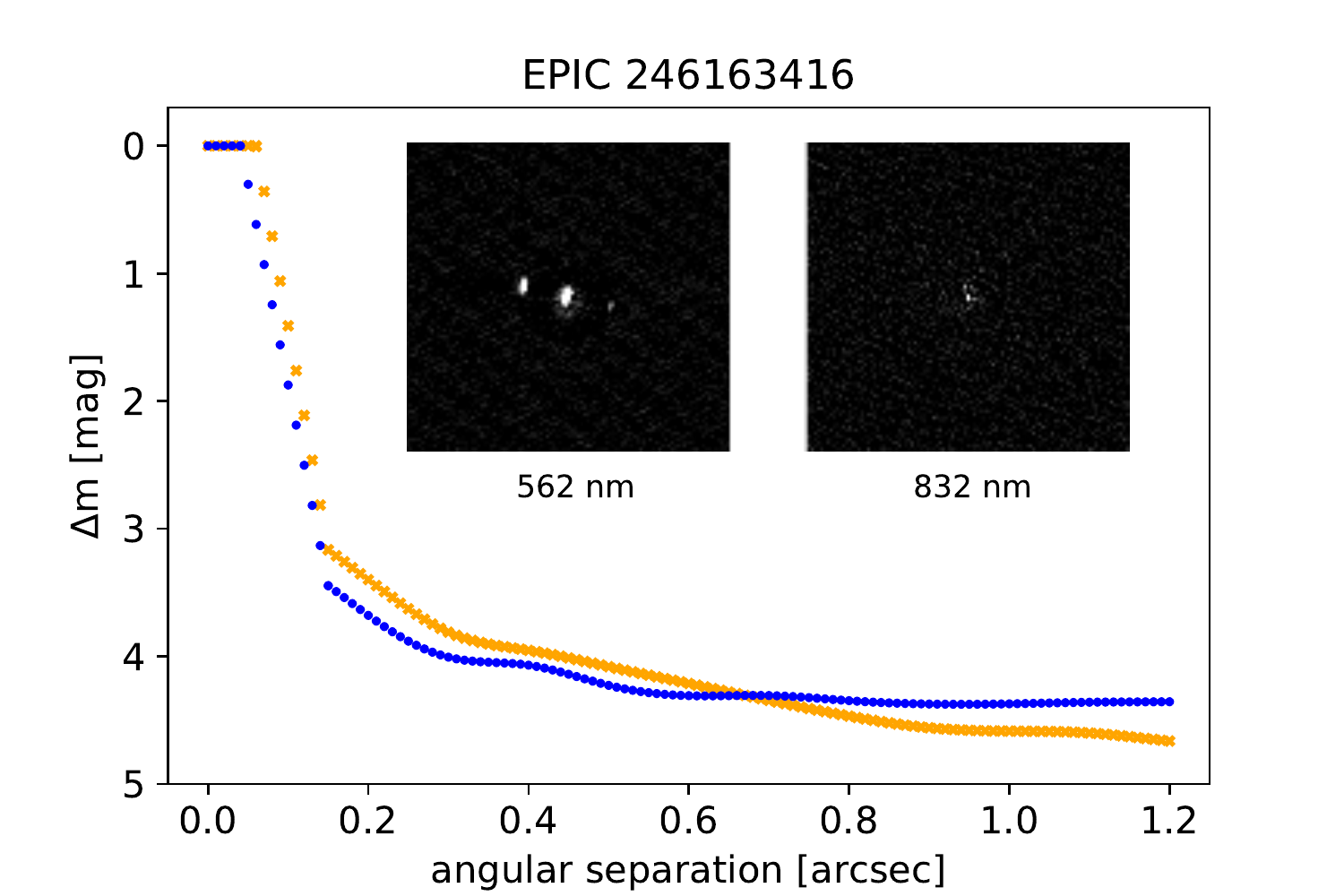}
\end{subfigure}%

\begin{subfigure}{9cm}
\centering\includegraphics[width=9cm]{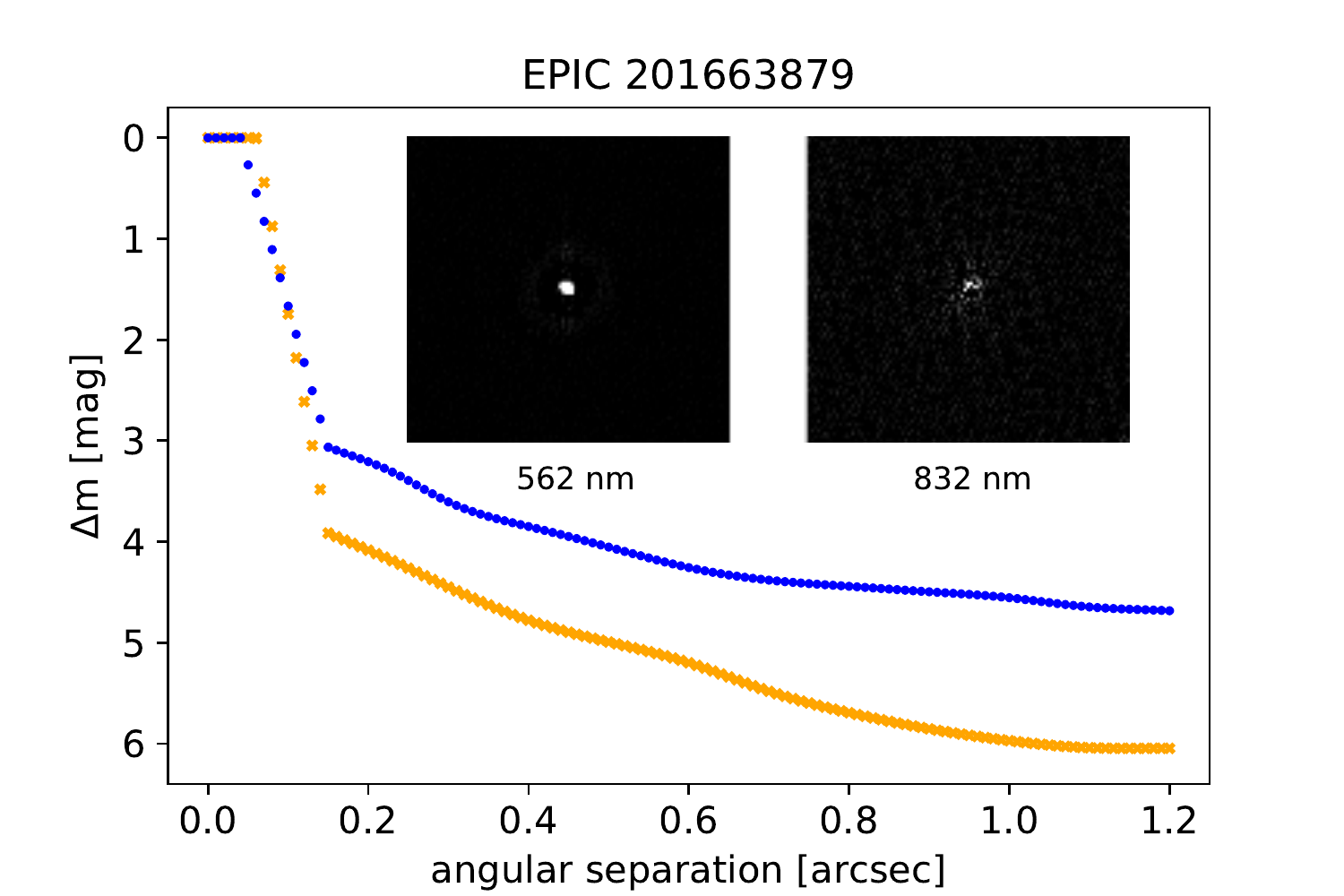}
\end{subfigure}

\caption{Speckle images and background plot for 4 stars from C14 and 1 from C12 (EPIC 246163416) . All images were acquired with NESSI instrument at WIYN-3.5m telescope at 832 nm (orange crosses) and 562 nm (blue dots). The field of the images is 4.65 x 4.65 arc seconds .}

\label{fig:AO_2}
 \end{figure*}

\begin{figure*}
\centering

\begin{subfigure}{6cm}
\centering\includegraphics[width=5cm]{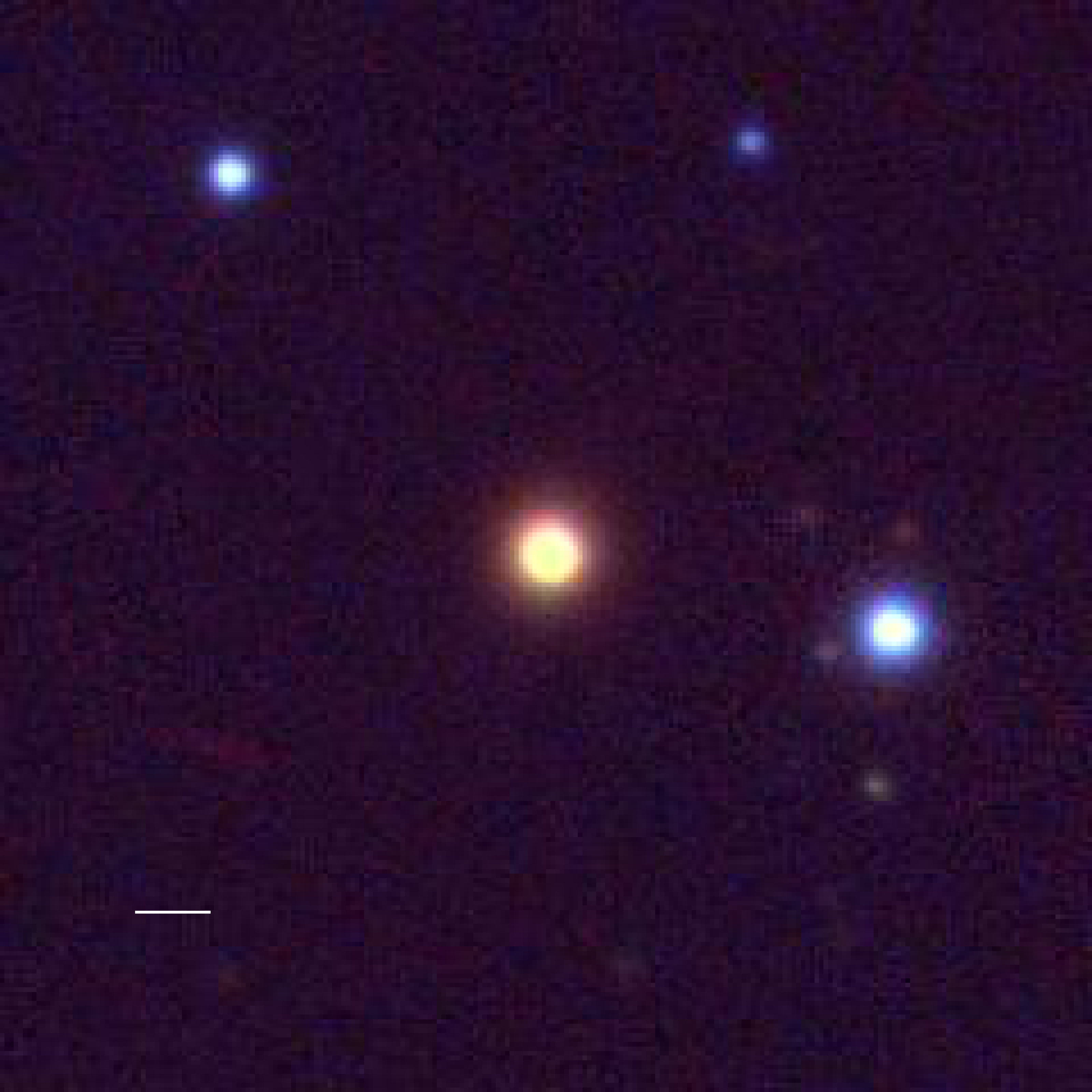}
\caption*{K2-316 (EPIC 249384674)}
\end{subfigure}%
\begin{subfigure}{6cm}
\centering\includegraphics[width=5cm]{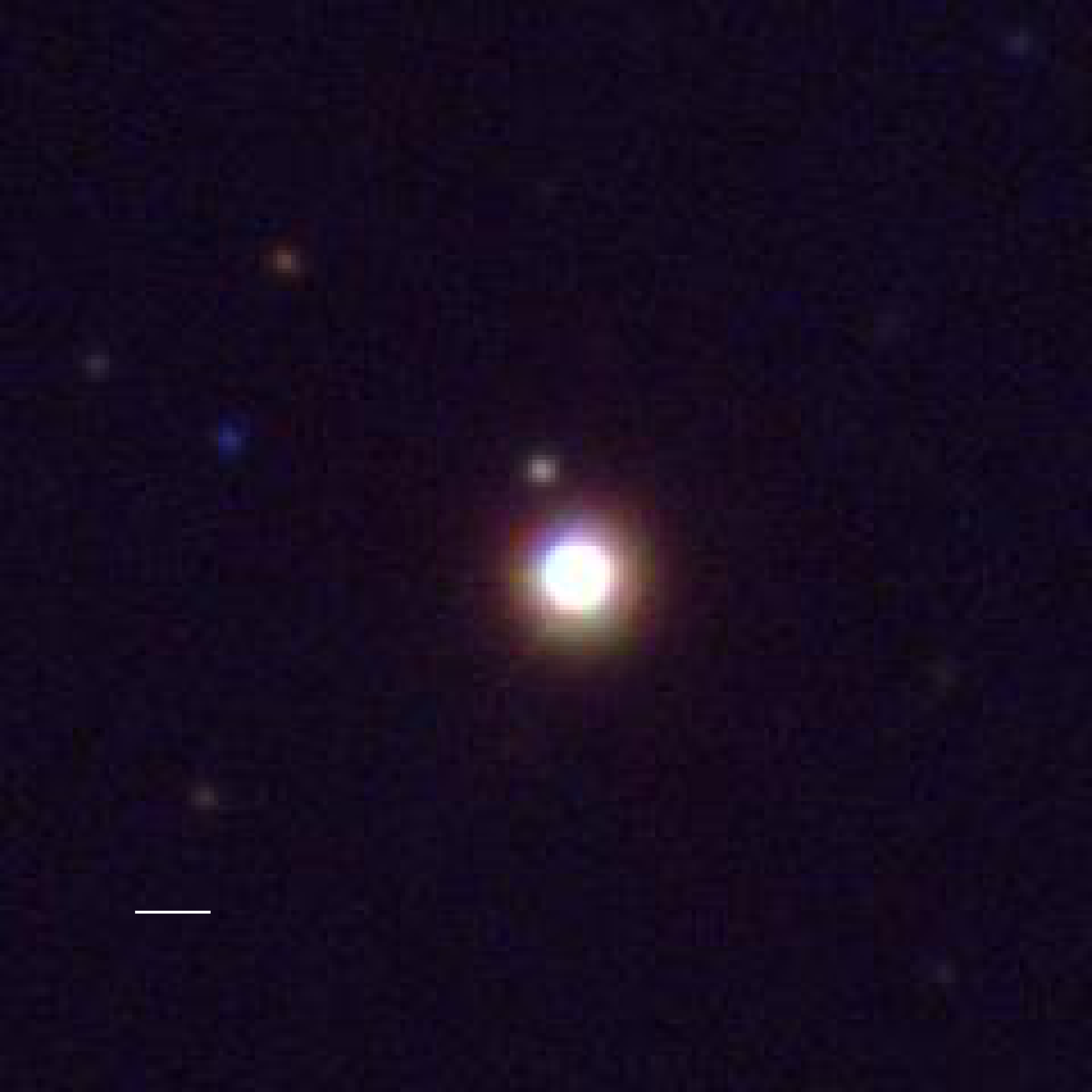}
\caption*{EPIC 249391469}
\end{subfigure}%
\begin{subfigure}{6cm}
\centering\includegraphics[width=5cm]{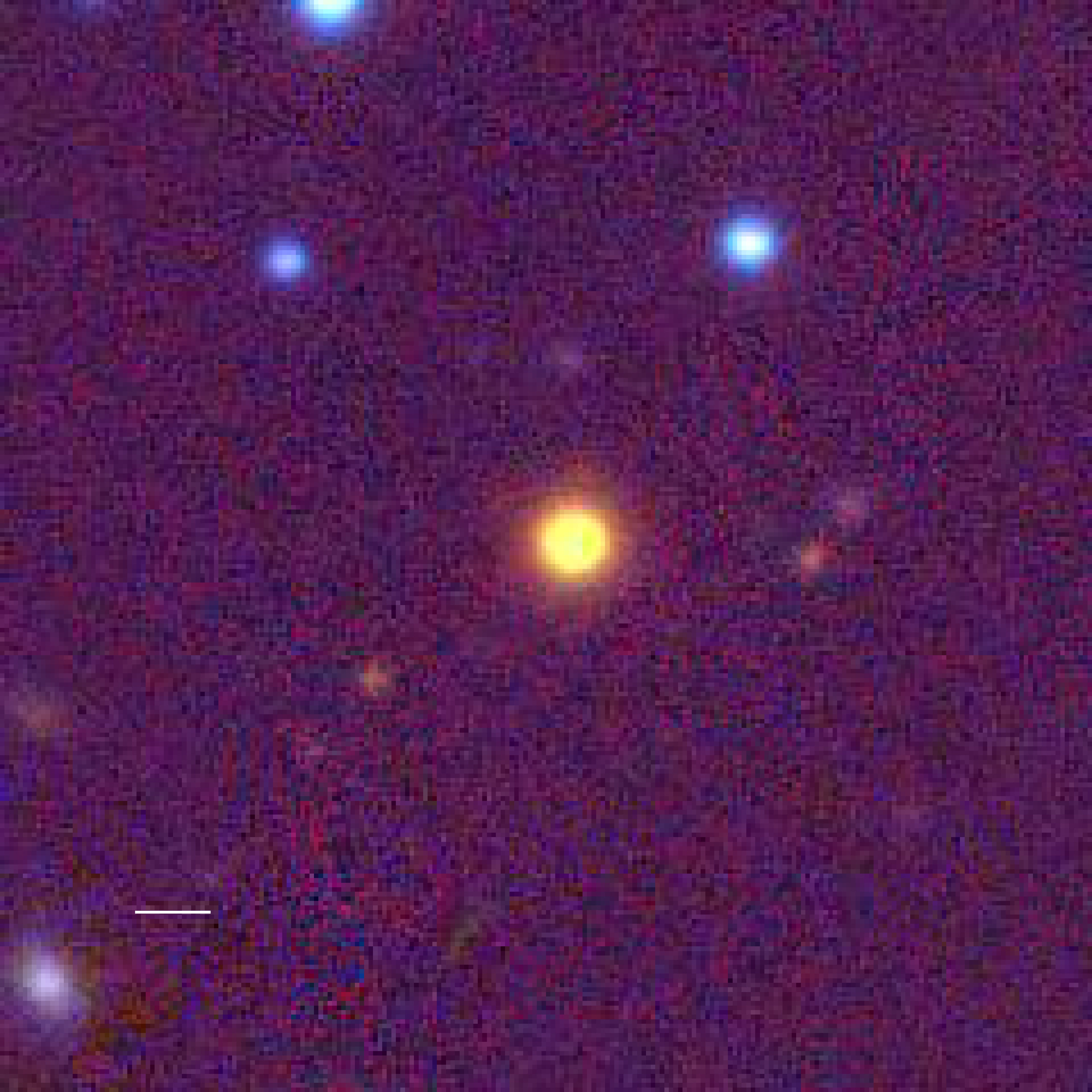}
\caption*{K2-317 (EPIC 249557502)}
\end{subfigure}

\begin{subfigure}{6cm}
\centering\includegraphics[width=5cm]{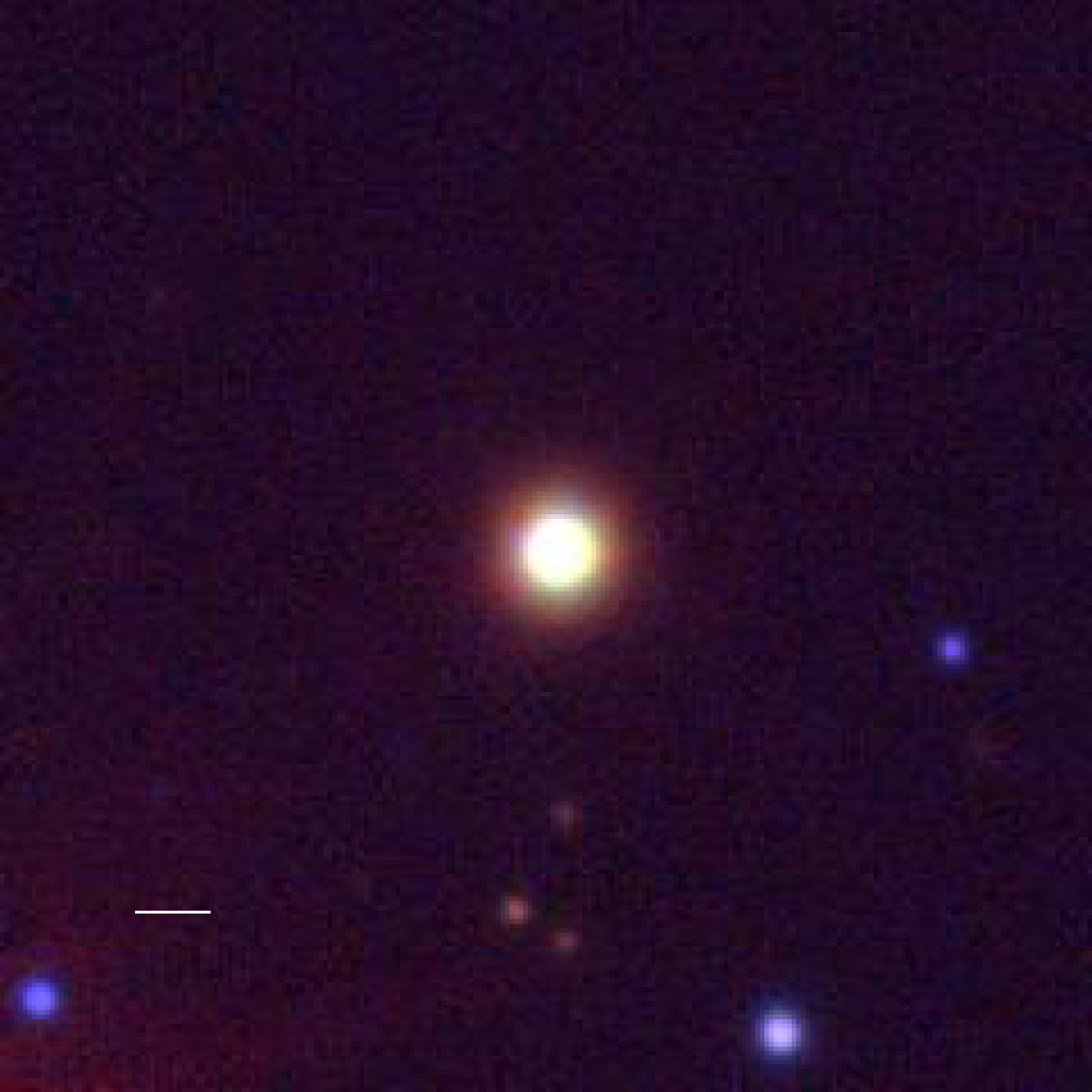}
\caption*{K2-318 (EPIC 249826231)}
\end{subfigure}%
\begin{subfigure}{6cm}
\centering\includegraphics[width=5cm]{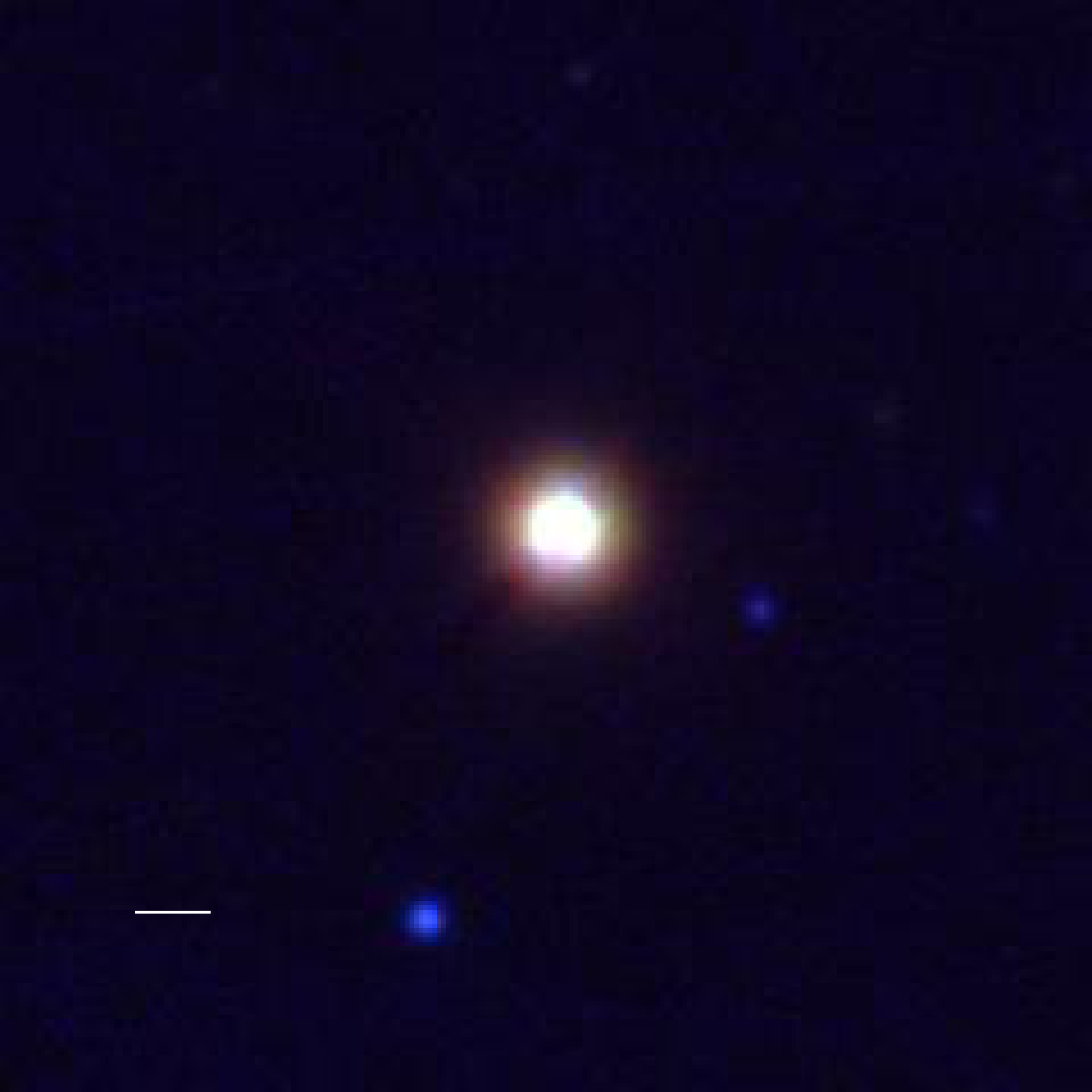}
\caption*{EPIC 250001426}
\end{subfigure}%
\begin{subfigure}{6cm}
\centering\includegraphics[width=5cm]{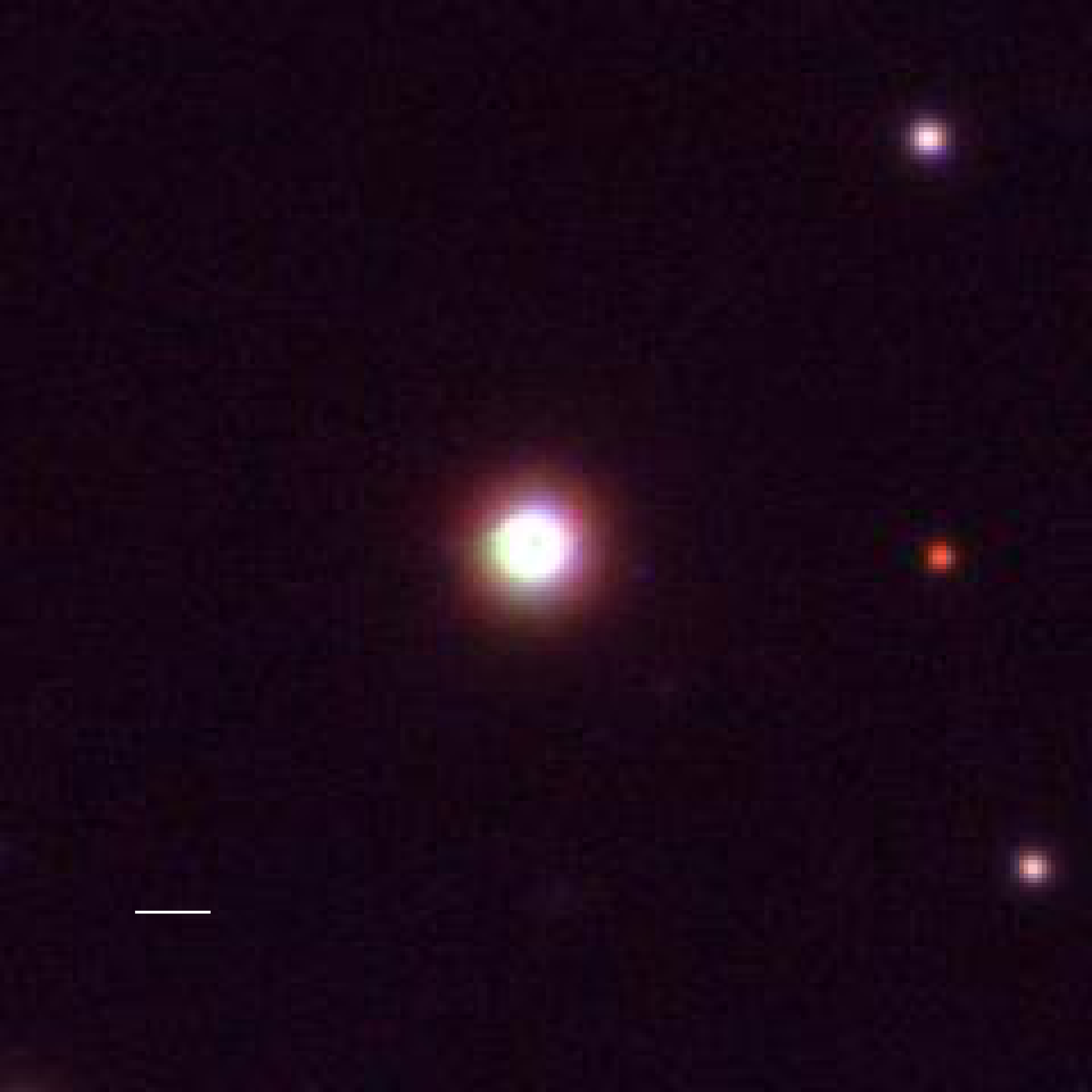}
\caption*{EPIC 250099723}
\end{subfigure}%

\caption{Panstarrs y/i/g stacked images of planet candidate hosts from campaign 15. Image size is 60.00 arc seconds.North up, East left. Straight white lines indicate the 3.98 arc seconds size of the Kepler pixel.}
\label{fig:pans_1}
\end{figure*}

\begin{figure*}
\centering

\begin{subfigure}{6cm}
\centering\includegraphics[width=5cm]{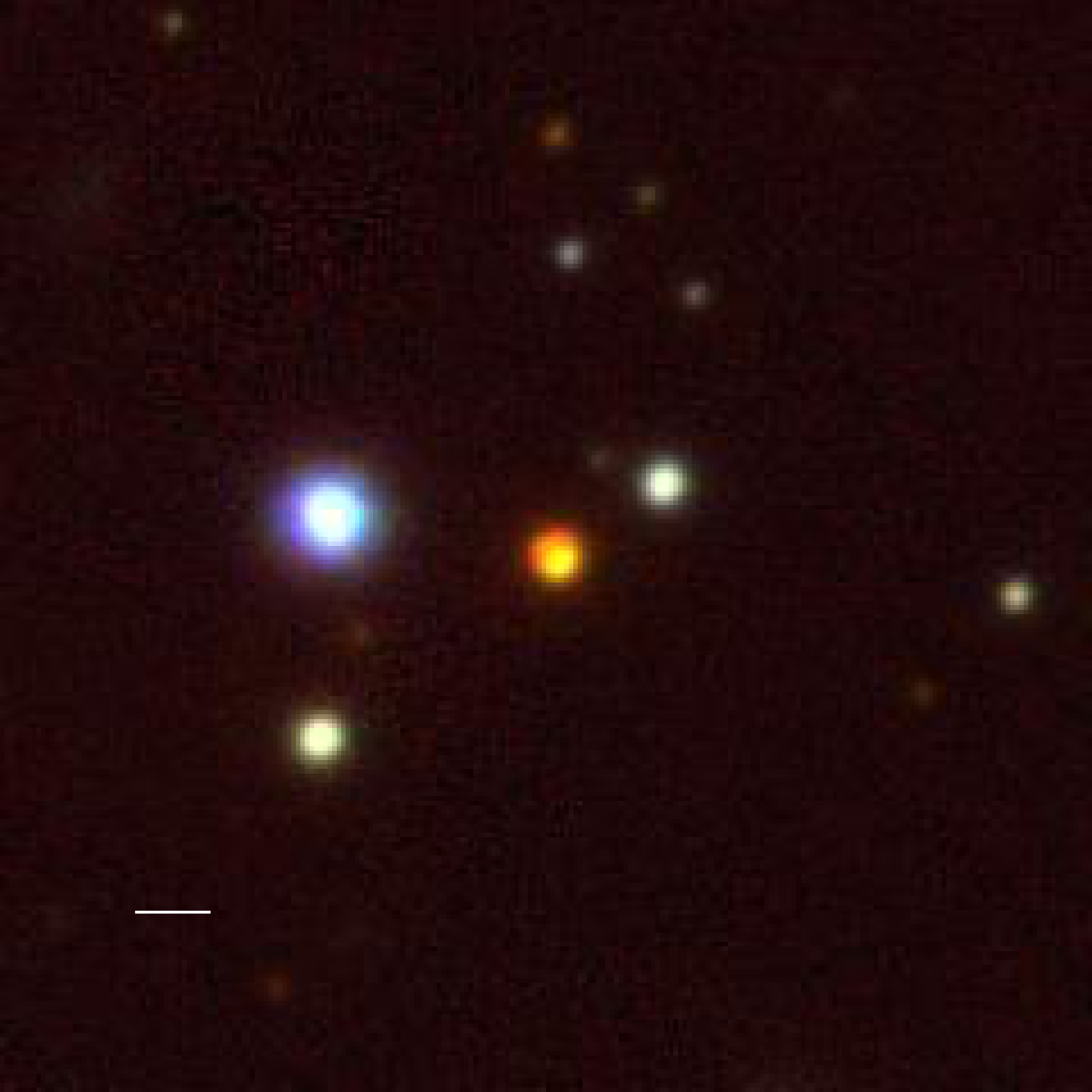}
\caption*{EPIC 246909566}
\end{subfigure}%

\caption{Panstarrs y/i/g stacked images of planet candidate hosts from campaign 13. Image size is 60.00 arc seconds. North up, East left. Straight white line indicate the 3.98 arc seconds size of the Kepler pixel.}
 
\end{figure*}

\begin{figure*}
\centering

\begin{subfigure}{6cm}
\centering\includegraphics[width=5cm]{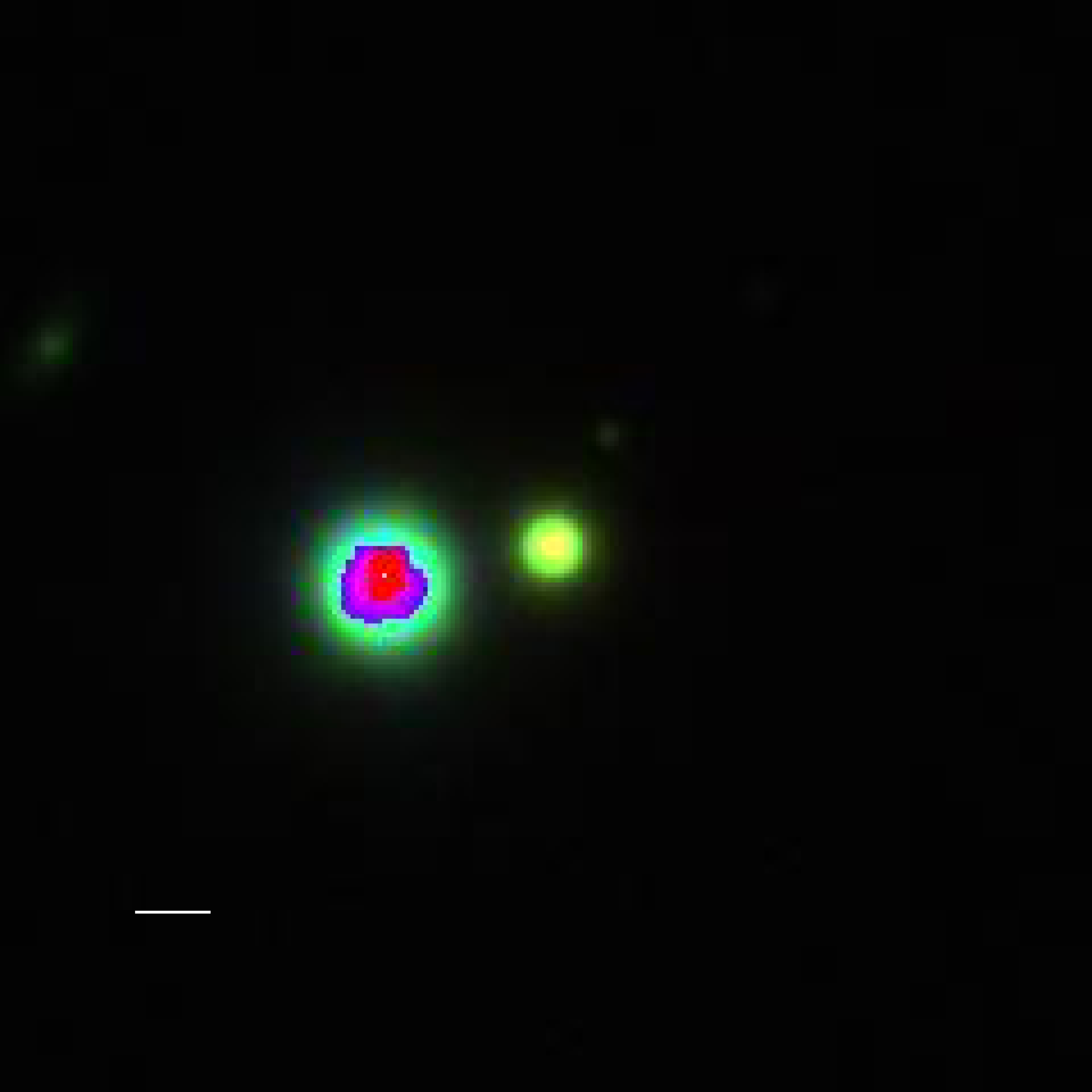}
\caption*{EPIC 201663913}
\end{subfigure}%
\begin{subfigure}{6cm}
\centering\includegraphics[width=5cm]{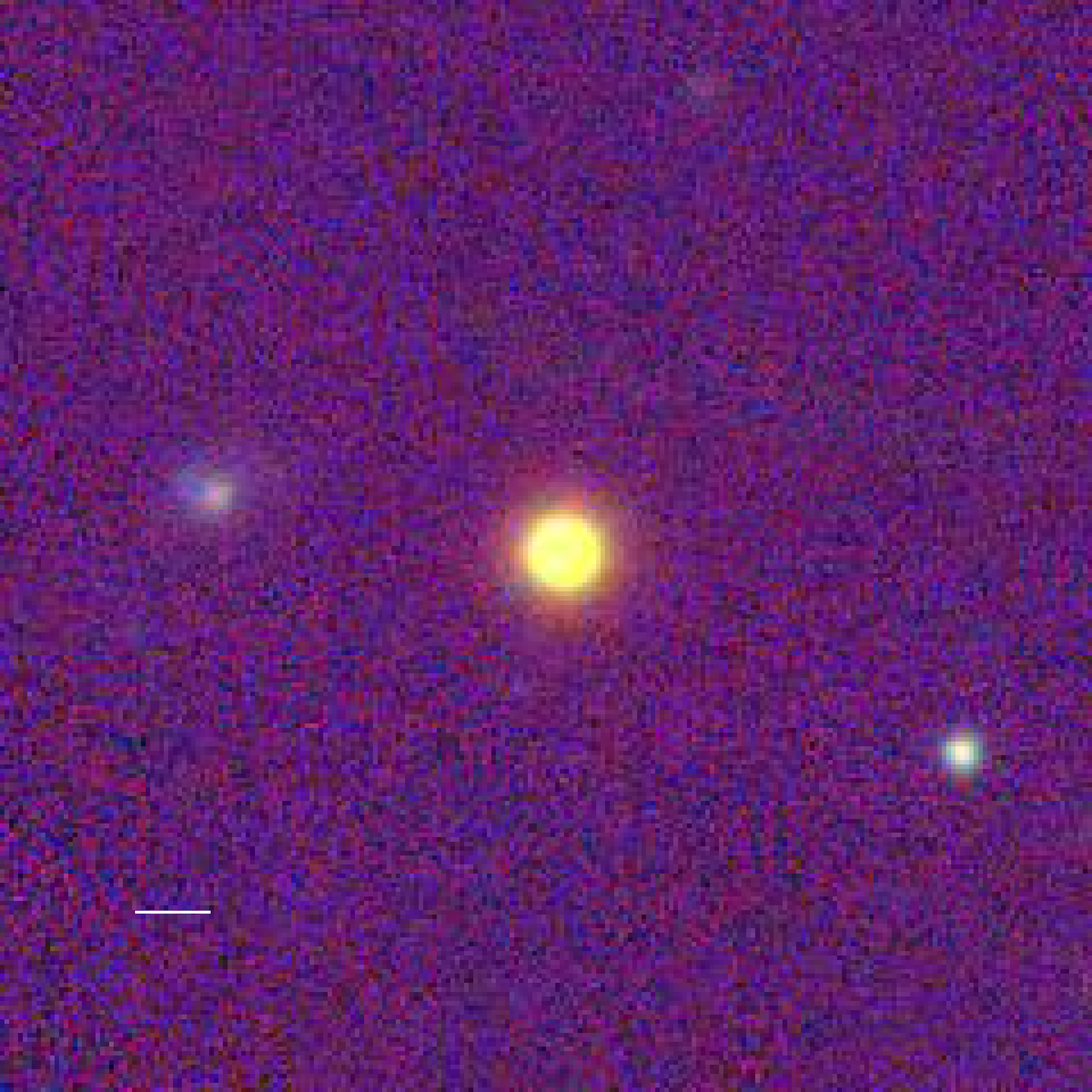}
\caption*{K2-320 (EPIC 201796690)}
\end{subfigure}%
\begin{subfigure}{6cm}
\centering\includegraphics[width=5cm]{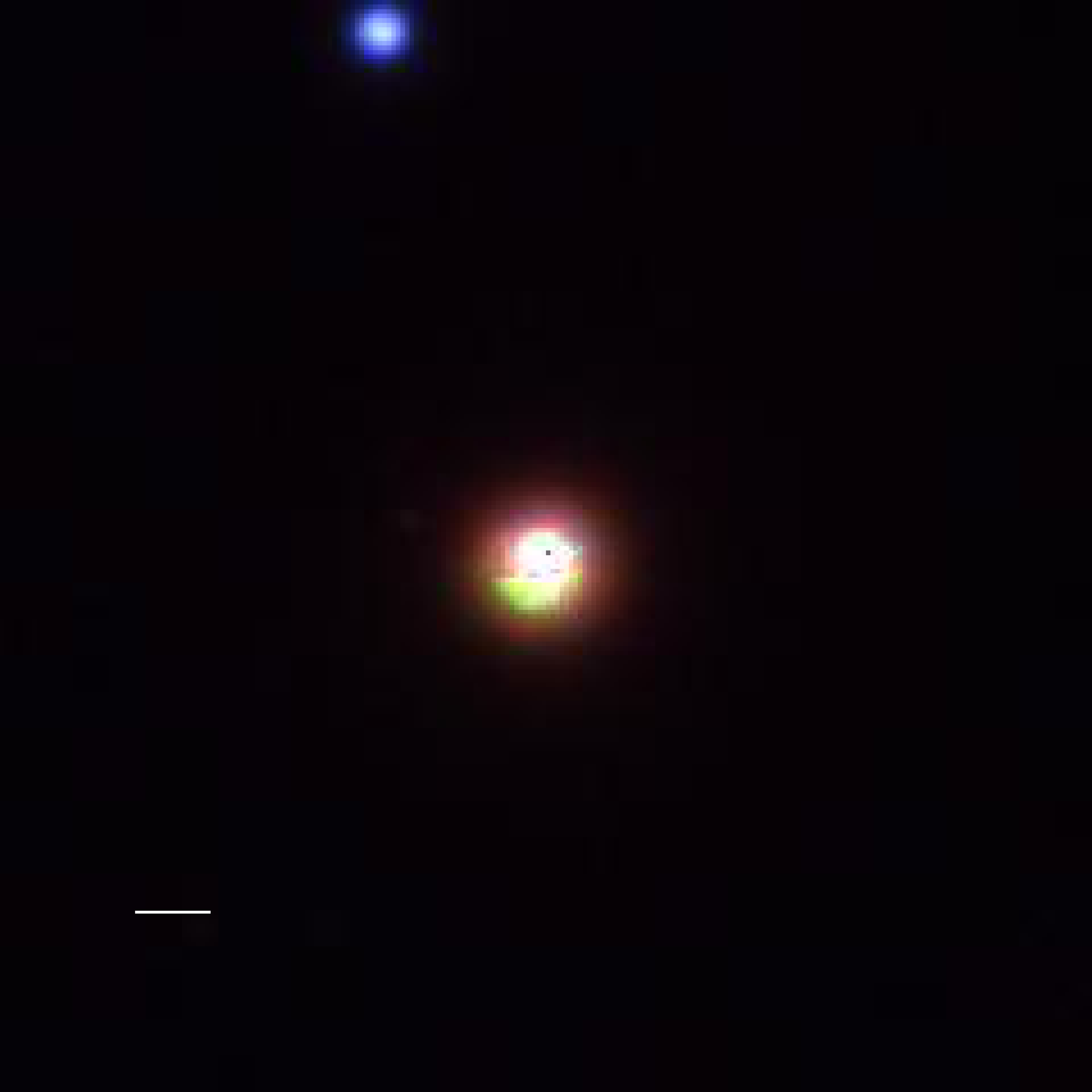}
\caption*{K2-321 (EPIC 248480671)}
\end{subfigure}%

\begin{subfigure}{6cm}
\centering\includegraphics[width=5cm]{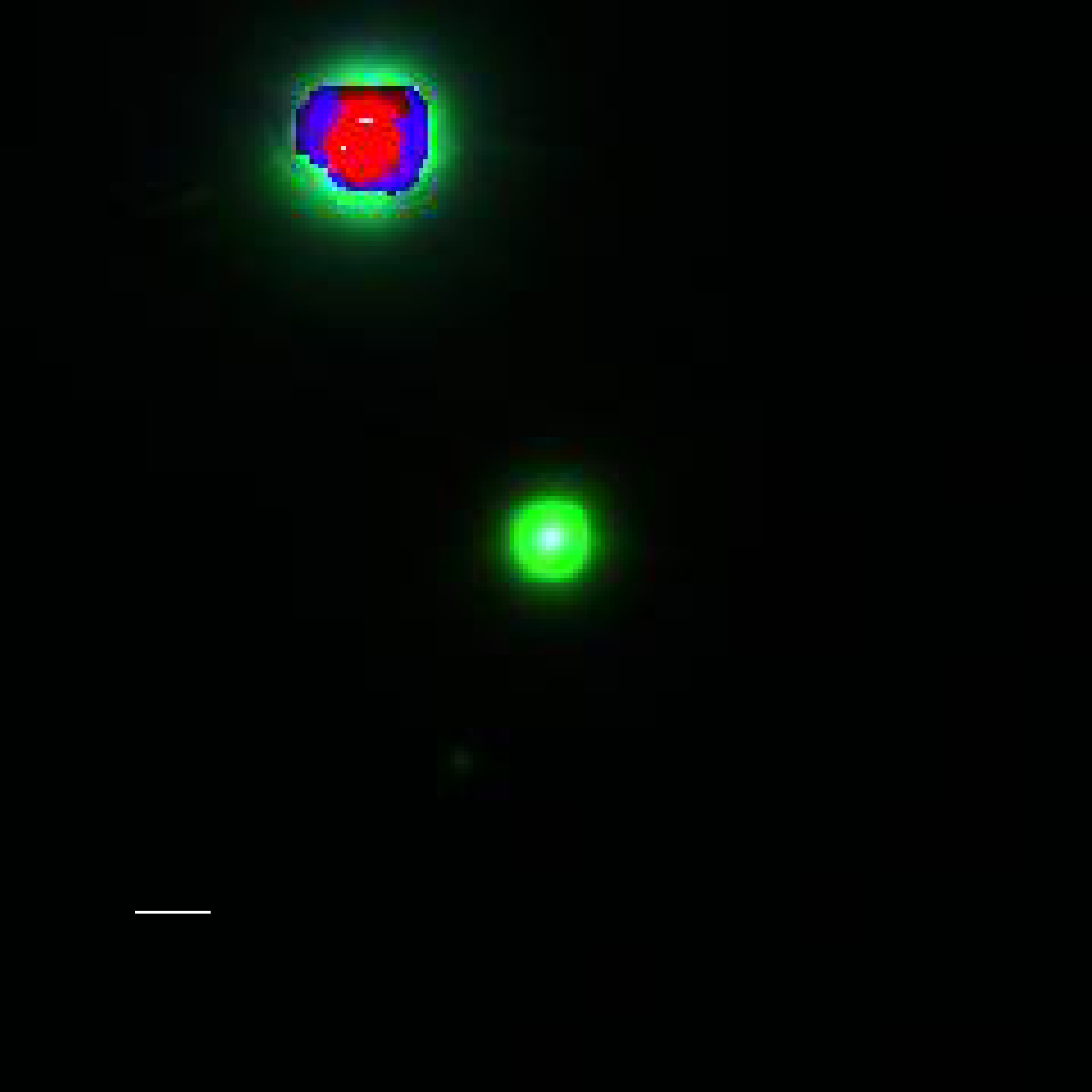}
\caption*{K2-322 (EPIC 248558190)}
\end{subfigure}%
\begin{subfigure}{6cm}
\centering\includegraphics[width=5cm]{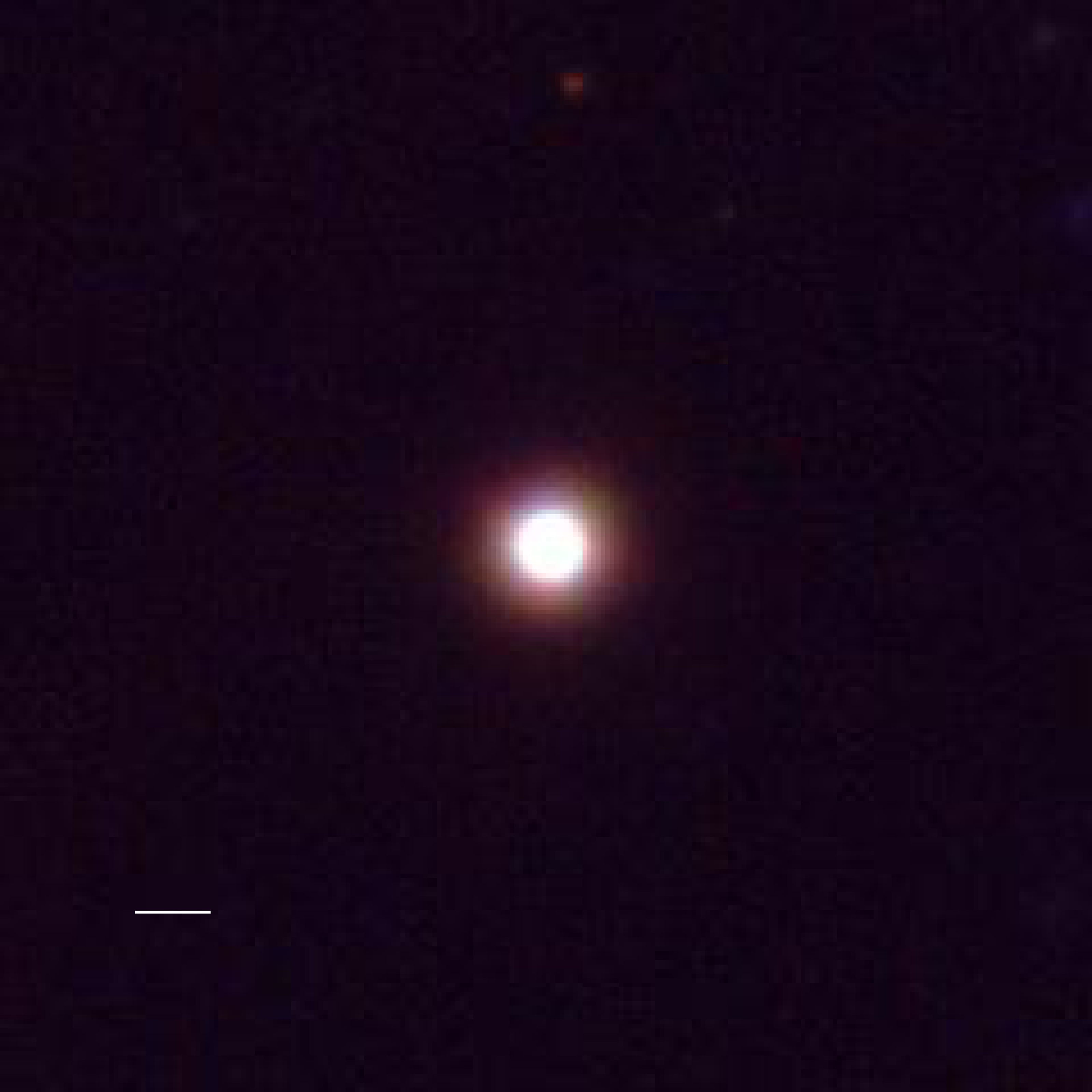}
\caption*{K2-323 (EPIC 248616368)}
\end{subfigure}%
\begin{subfigure}{6cm}
\centering\includegraphics[width=5cm]{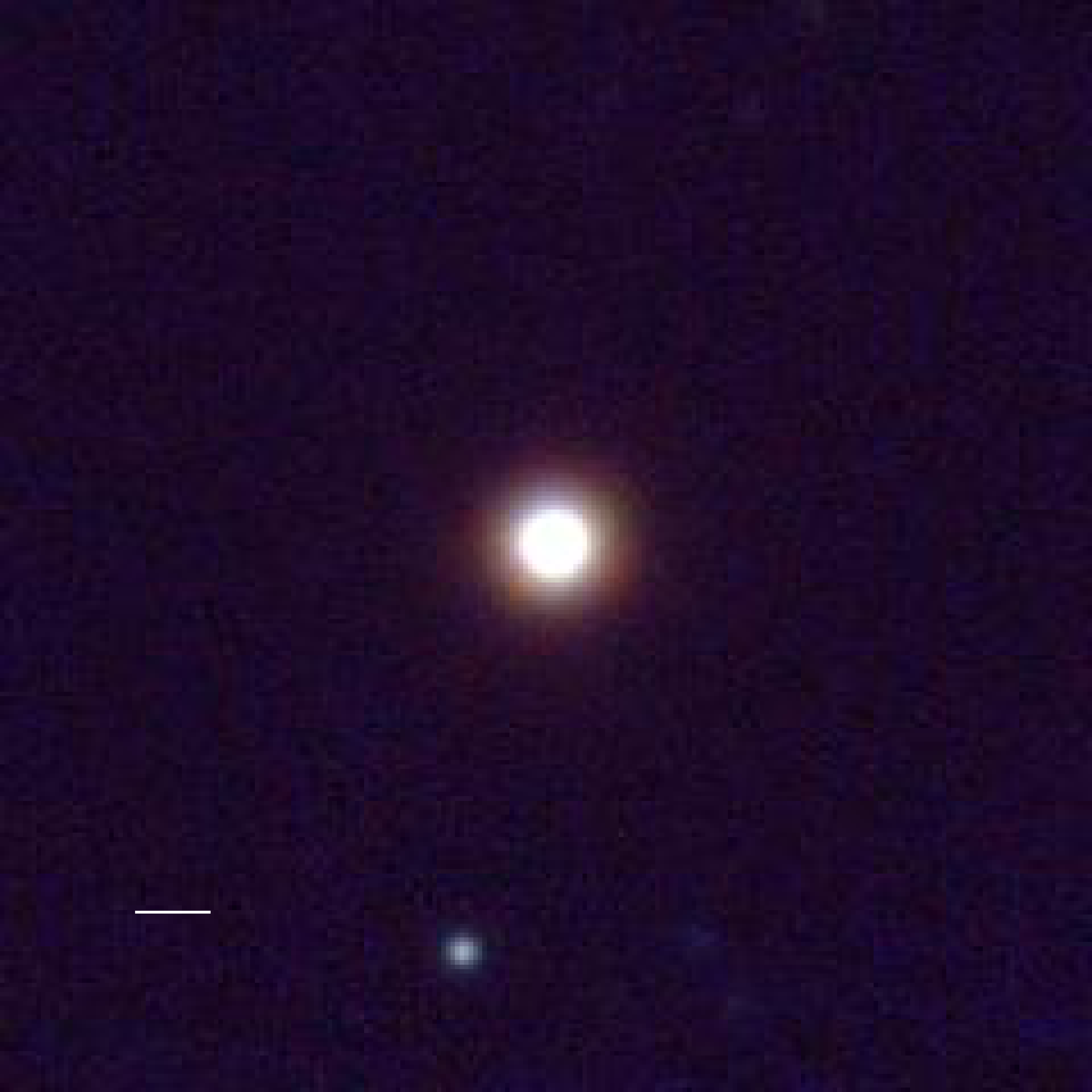}
\caption*{K2-324 (EPIC 248639308)}
\end{subfigure}%

\begin{subfigure}{6cm}
\centering\includegraphics[width=5cm]{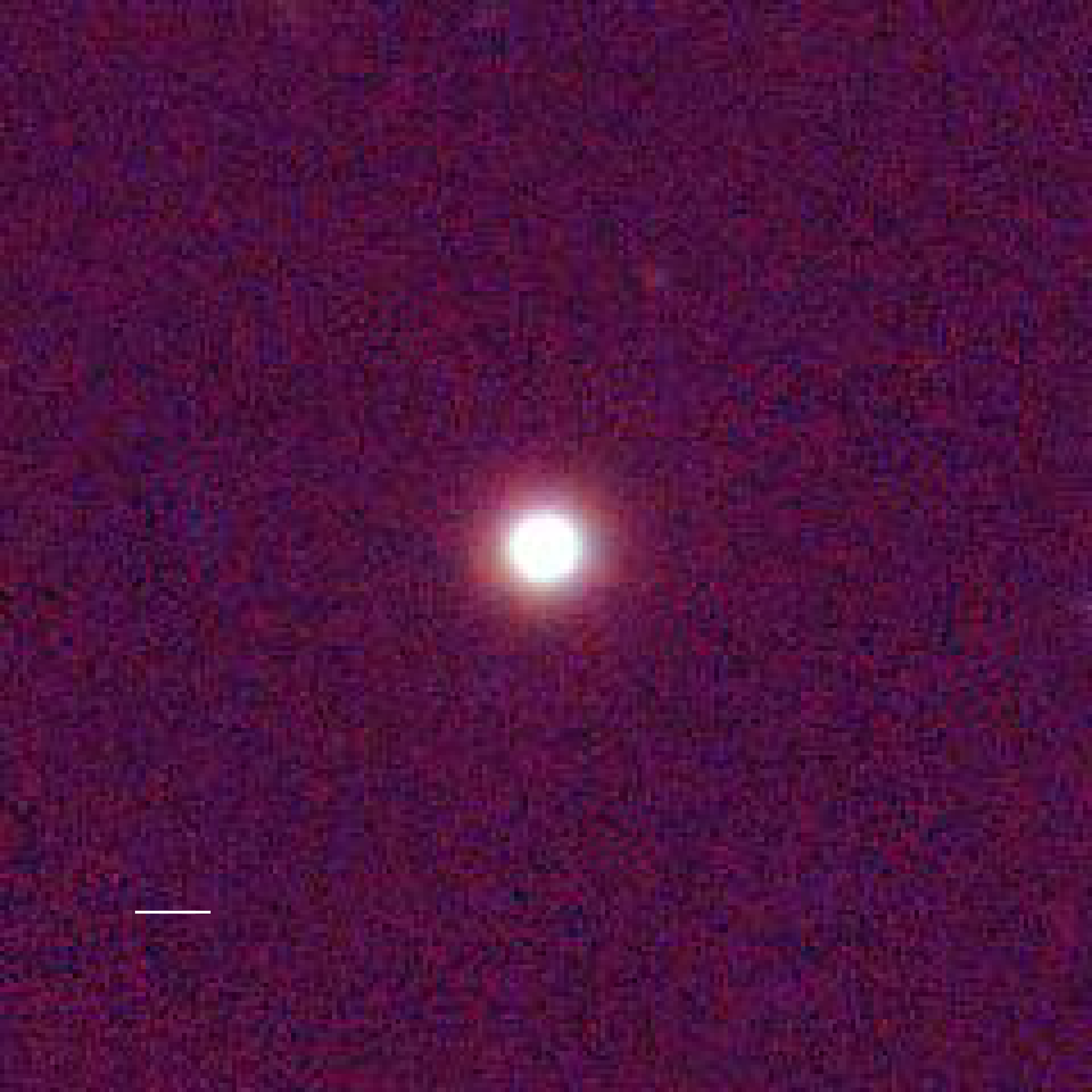}
\caption*{EPIC 248775938}
\end{subfigure}
\begin{subfigure}{6cm}
\centering\includegraphics[width=5cm]{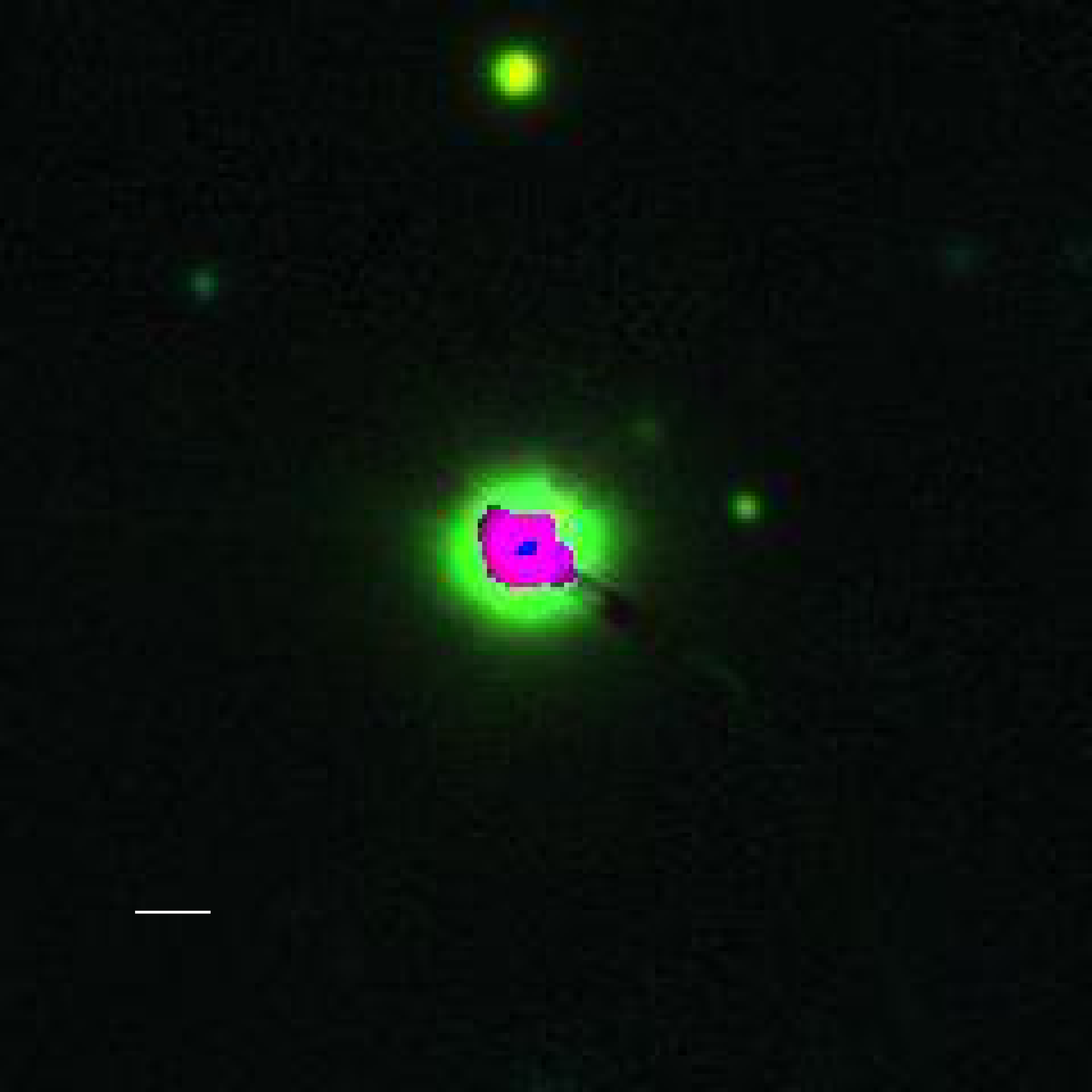}
\caption*{EPIC 248782482}
\end{subfigure}

\caption{Panstarrs y/i/g stacked images of planet candidate hosts from campaign 14. Image size is 60.00 arc seconds. North up, East left. Straight white lines indicate the 3.98 arc seconds size of the Kepler pixel.}
 
\end{figure*}

\begin{figure*}
\centering

\begin{subfigure}{6cm}
\centering\includegraphics[width=5cm]{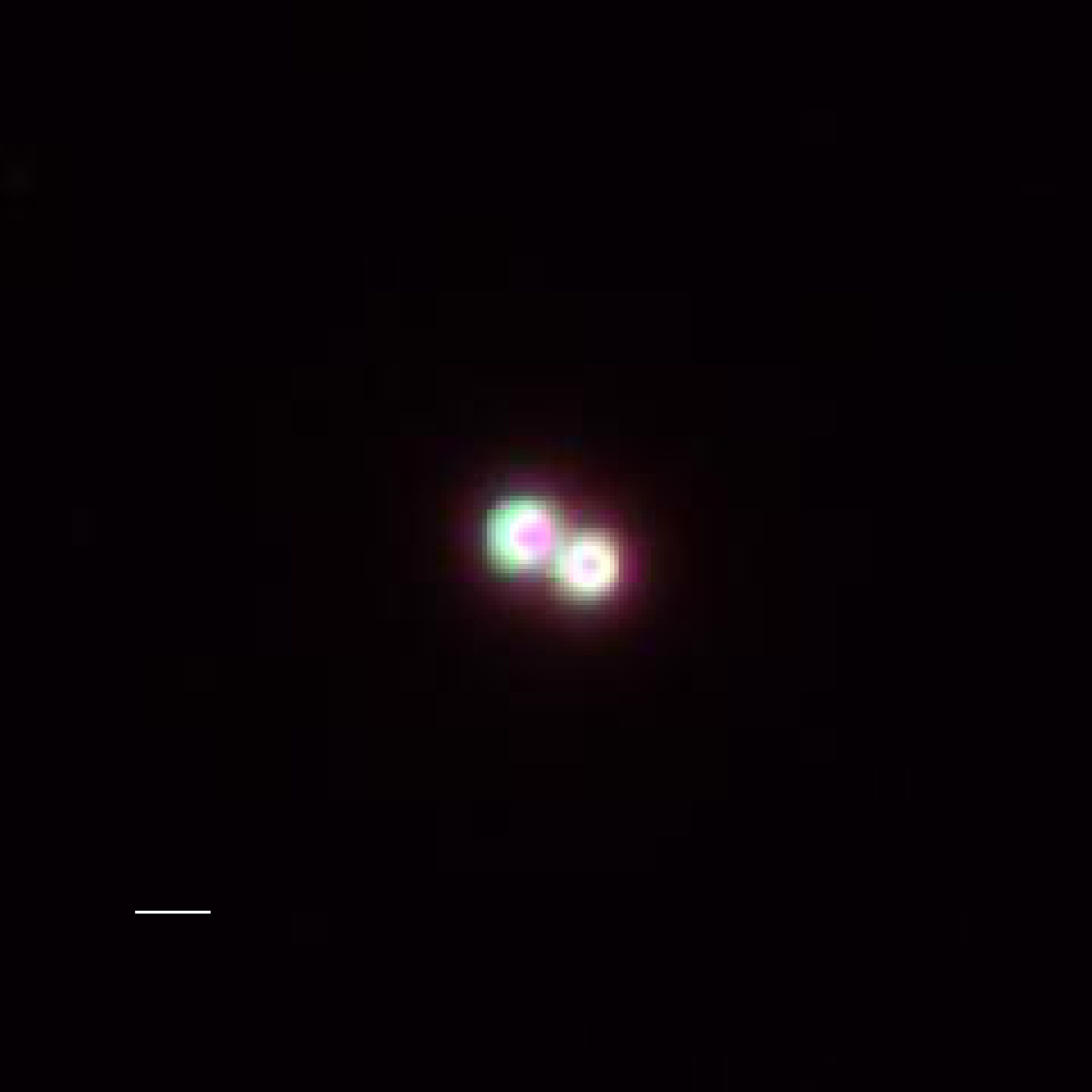}
\caption*{EPIC 2485944983}
\end{subfigure}%
\begin{subfigure}{6cm}
\centering\includegraphics[width=5cm]{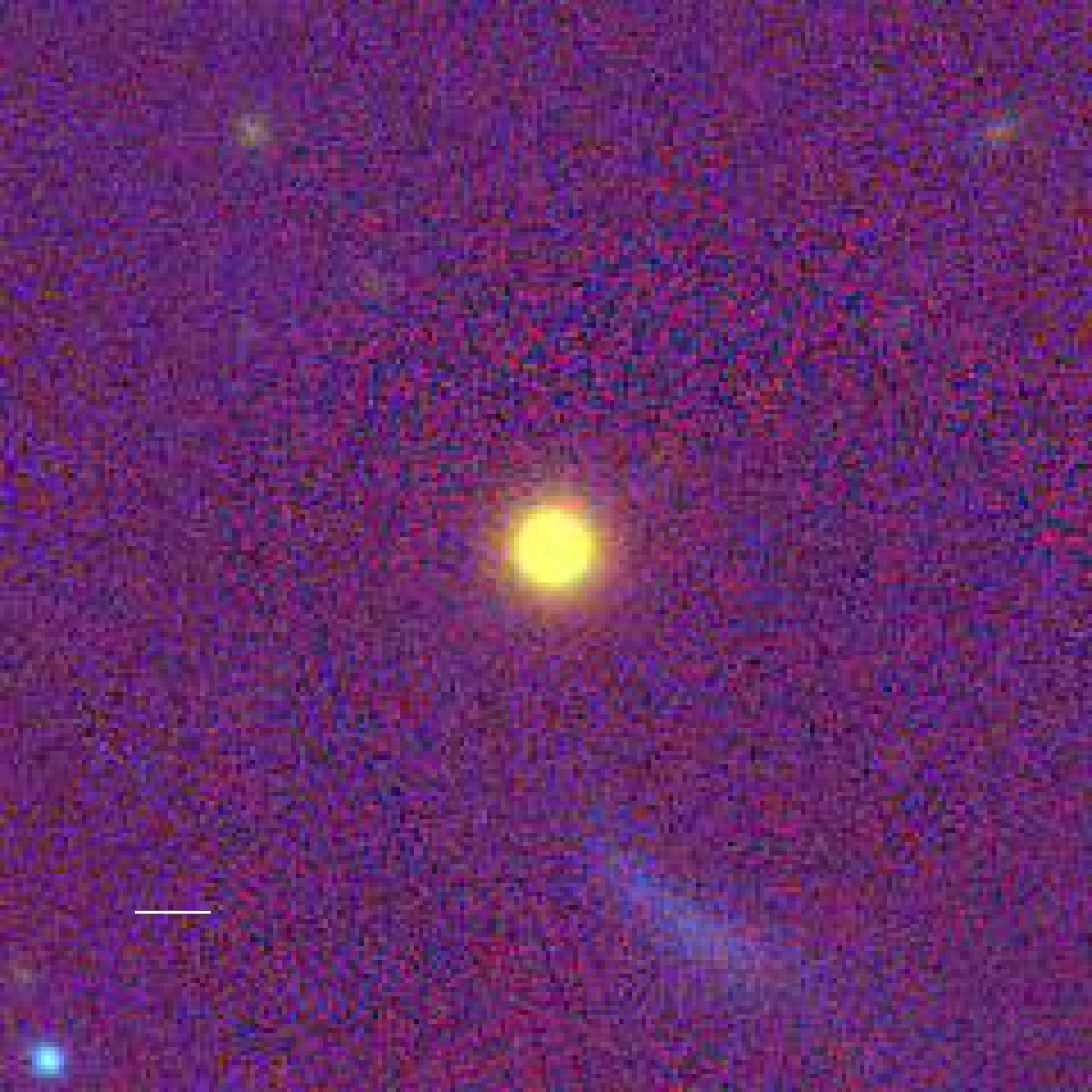}
\caption*{K2-325 (EPIC 246074965)}
\end{subfigure}%
\begin{subfigure}{6cm}
\centering\includegraphics[width=5cm]{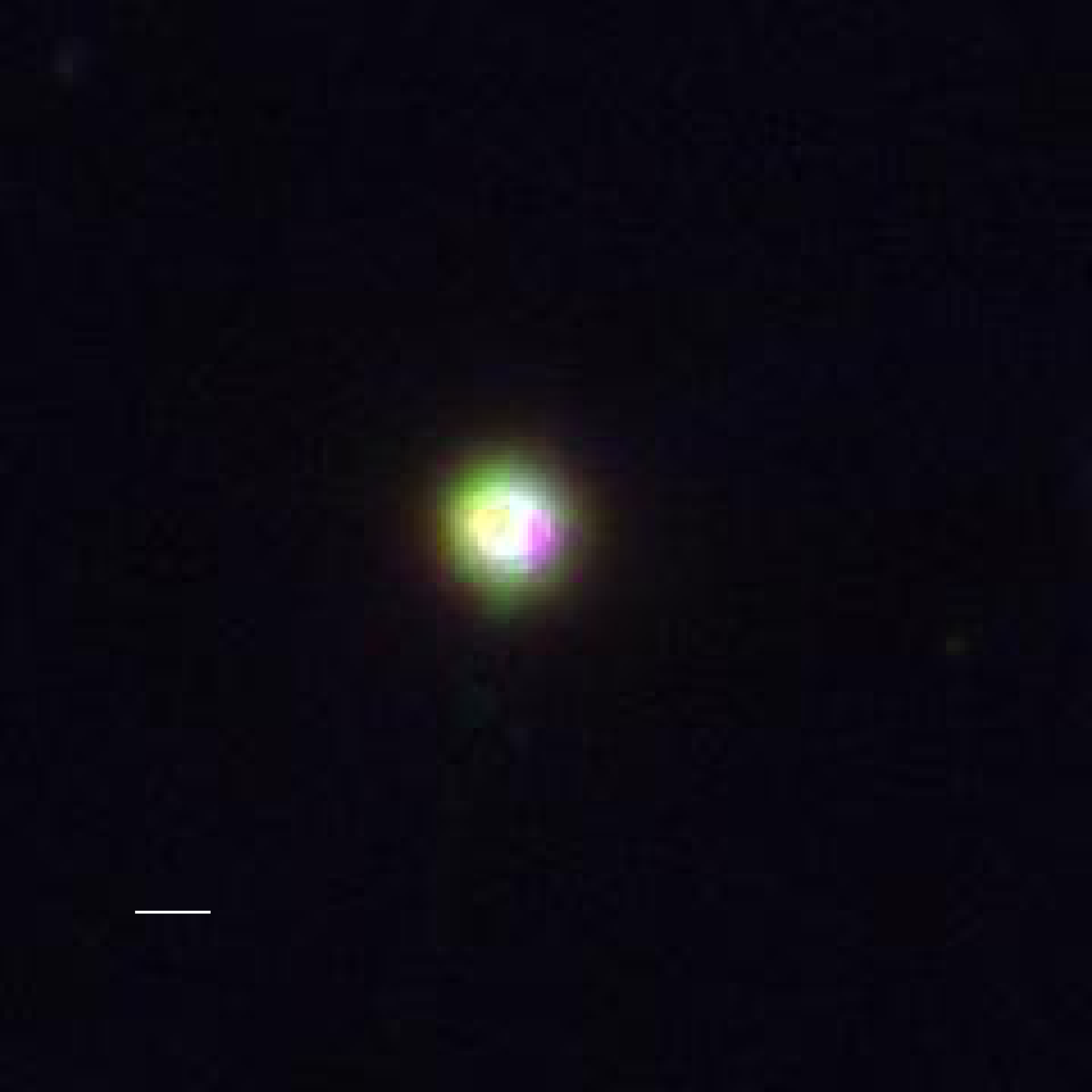}
\caption*{EPIC 246163416}
\end{subfigure}%

\begin{subfigure}{6cm}
\centering\includegraphics[width=5cm]{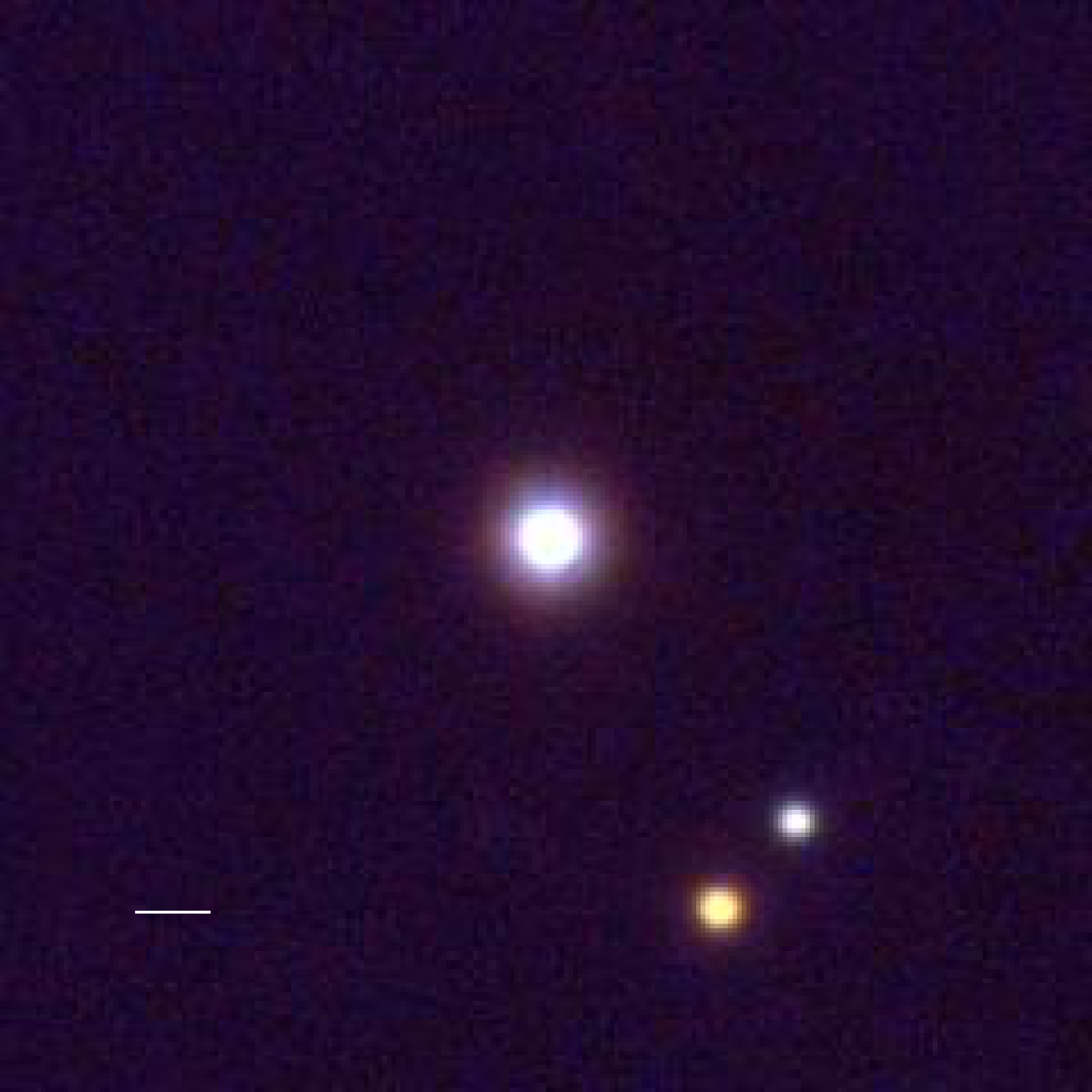}
\caption*{EPIC 246313886}
\end{subfigure}%
\begin{subfigure}{6cm}
\centering\includegraphics[width=5cm]{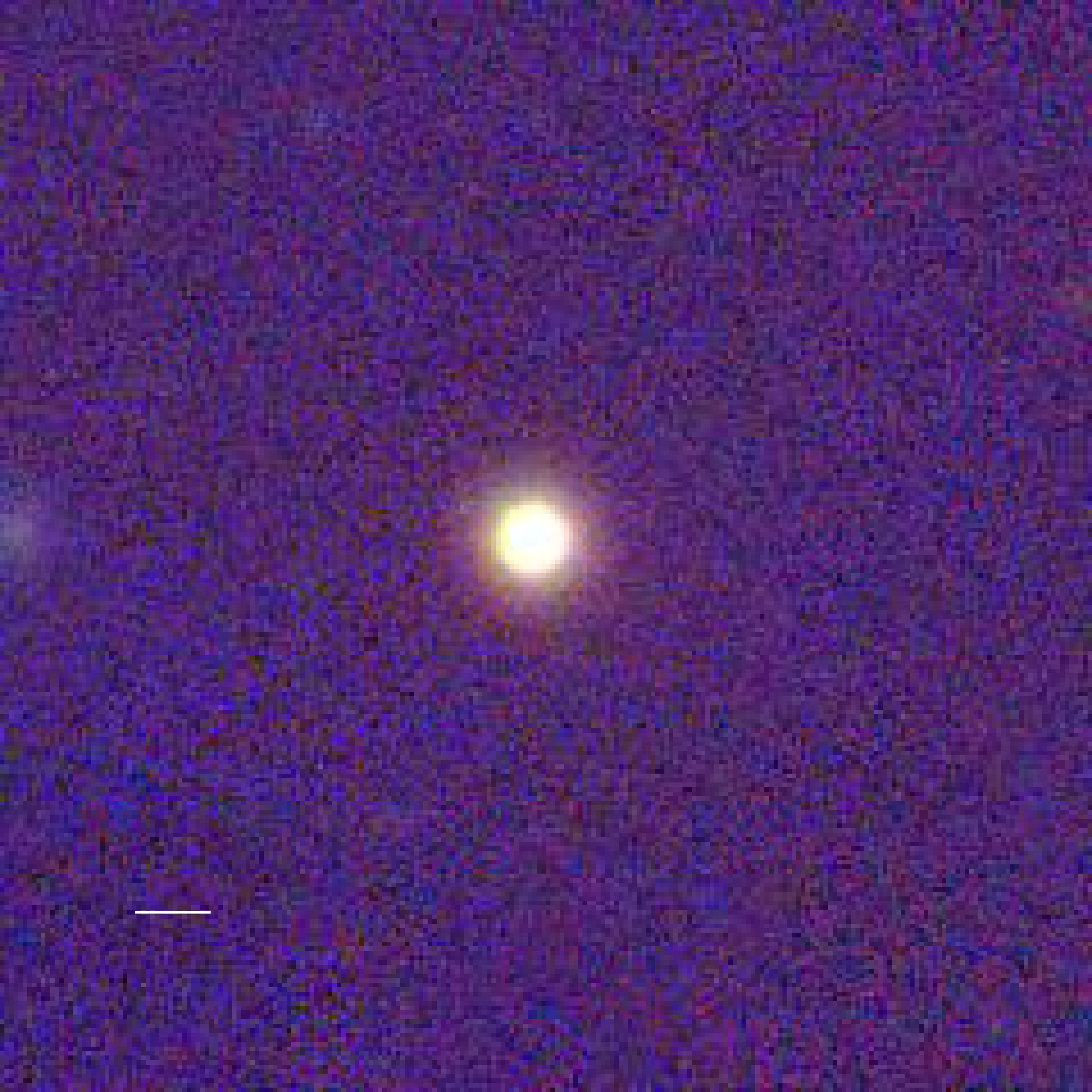}
\caption*{EPIC 246331347}
\end{subfigure}%
\begin{subfigure}{6cm}
\centering\includegraphics[width=5cm]{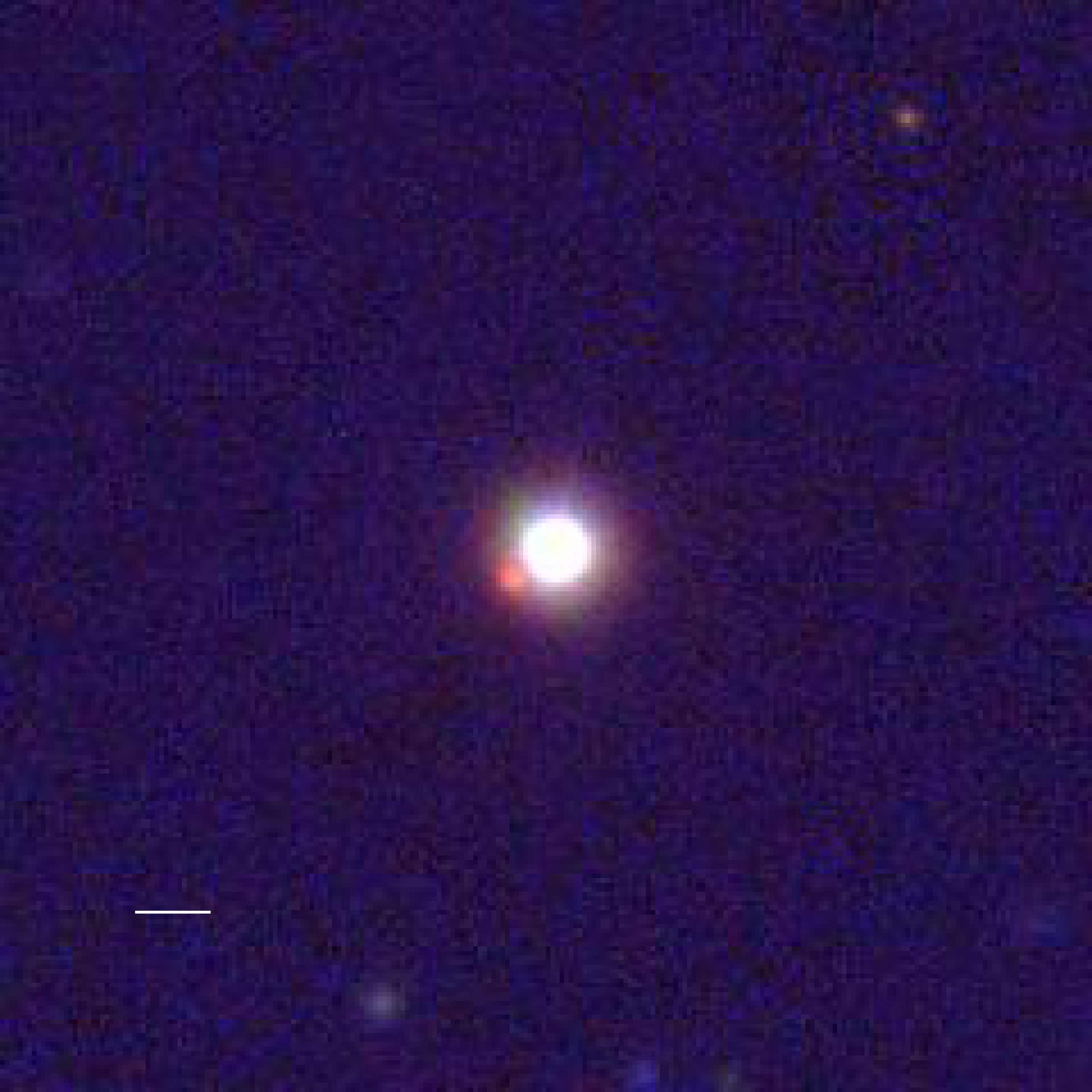}
\caption*{EPIC 246331418}
\end{subfigure}%

\begin{subfigure}{6cm}
\centering\includegraphics[width=5cm]{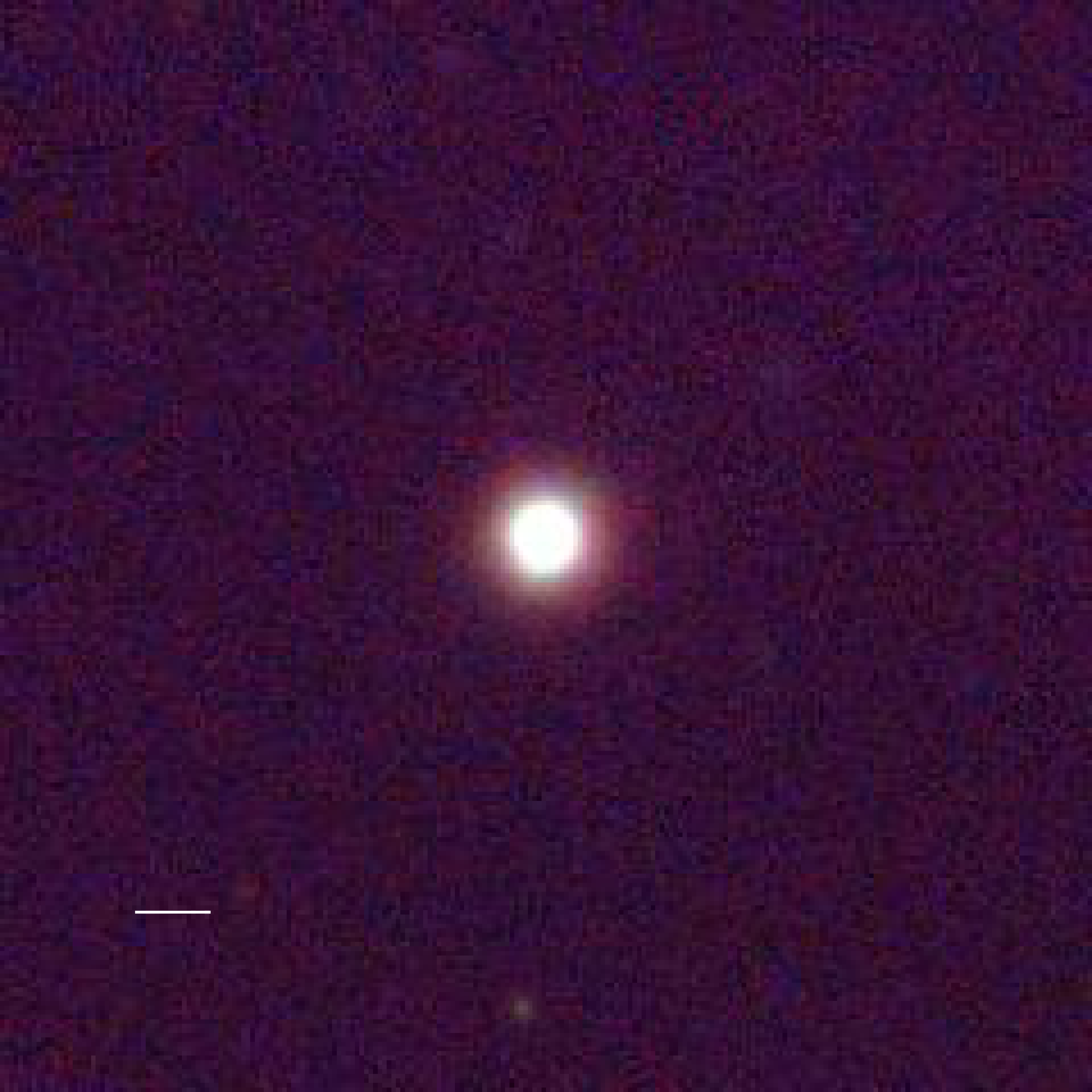}
\caption*{K2-326 (EPIC 246472939)}
\end{subfigure}%

\caption{Panstarrs y/i/g stacked images of planet candidate hosts from campaign 12. Image size is 60.00 arc seconds. North up, East left. Straight white lines indicate the 3.98 arc seconds size of the Kepler pixel.}
\label{ap:pans_last}
\end{figure*}



\bsp	
\label{lastpage}
\end{document}